\documentclass[review,11pt]{elsarticle}
\usepackage[top=1in, bottom=1in, left=1in, right=1in]{geometry}
\def\comment#1{}
\usepackage{graphicx,epsfig,amsfonts,amsmath,amssymb,lscape,color}
\usepackage{algorithm}
\usepackage{algorithmic}

\usepackage{amssymb}
\DeclareSymbolFont{grb}{OML}{cmm}{b}{it}
\DeclareMathSymbol{\alphab}{\mathord}{grb}{"0B}
\DeclareMathSymbol{\betab}{\mathord}{grb}{"0C}
\DeclareMathSymbol{\gammab}{\mathord}{grb}{"0D}
\DeclareMathSymbol{\deltab}{\mathord}{grb}{"0E}
\DeclareMathSymbol{\epsilonb}{\mathord}{grb}{"0F}
\DeclareMathSymbol{\zetab}{\mathord}{grb}{"10}
\DeclareMathSymbol{\etab}{\mathord}{grb}{"11}
\DeclareMathSymbol{\thetab}{\mathord}{grb}{"12}
\DeclareMathSymbol{\kappab}{\mathord}{grb}{"14}
\DeclareMathSymbol{\lambdab}{\mathord}{grb}{"15}
\DeclareMathSymbol{\mub}{\mathord}{grb}{"16}
\DeclareMathSymbol{\nub}{\mathord}{grb}{"17}
\DeclareMathSymbol{\rhob}{\mathord}{grb}{"1A}
\DeclareMathSymbol{\sigmab}{\mathord}{grb}{"1B}
\DeclareMathSymbol{\taub}{\mathord}{grb}{"1C}
\DeclareMathSymbol{\phib}{\mathord}{grb}{"1E}
\DeclareMathSymbol{\psib}{\mathord}{grb}{"20}
\DeclareMathSymbol{\omegab}{\mathord}{grb}{"21}
\DeclareMathSymbol{\epsilonb}{\mathord}{grb}{"22}
\DeclareMathSymbol{\varphib}{\mathord}{grb}{"27}

\biboptions{sort&compress}

\journal{Signal Processing}

\begin{document}
\begin{frontmatter}

\title{Coherent Sources
Direction Finding and Polarization Estimation
with Various Compositions of
Spatially Spread Polarized Antenna Arrays}

\author{Xin Yuan}
\ead{xin.yuan@duke.edu}
\address{Department of Electrical and Computer Engineering, \\Duke University, Durham, NC 27708, USA.}

\begin{abstract}
Various compositions of sparsely polarized antenna arrays are proposed in this paper to estimate the direction-of-arrivals (DOAs) and polarizations of multiple coherent sources.
These polarized antenna arrays are composed of one of the following five sparsely-spread sub-array geometries:
1) four spatially-spread dipoles with three orthogonal orientations,
2) four spatially-spread loops with three orthogonal orientations,
3) three spatially-spread dipoles and three spatially-spread loops with orthogonal orientations,
4) three collocated dipole-loop pairs with orthogonal orientations, and
5) a collocated dipole-triad and a collocated loop-triad.
All the dipoles/loops/pairs/triads in each sub-array can also be sparsely spaced with the inter-antenna spacing far {\em larger} than a half-wavelength.
Only {\em one dimensional} spatial-smoothing is used in the proposed algorithm to derive the {\em two-dimensional} DOAs and polarizations of multiple cross-correlated signals.
From the simulation results, the sparse array composed of dipole-triads and loop-triads is recommended to construct a large aperture array, while
the sparse arrays composed of only dipoles or only loops are recommended to efficiently reduce the mutual coupling across the antennas.
Practical applications include distributed arrays and passive radar systems.
\end{abstract}

\begin{keyword}
Antenna array mutual coupling,
antenna arrays,
aperture antennas,
array signal processing,
direction of arrival estimation,
polarization.
\end{keyword}

\end{frontmatter}

\section{Introduction}
This paper proposes various compositions of sparse arrays to estimate the elevation-azimuth angles and polarizations of coherent sources.
These sparse arrays are composed of spatially-spread dipoles or loops.
First, the dipoles/loops with various orientations form a sub-array with the inter-antenna spacing far larger than a half-wavelength.
Second, these sub-arrays are uniformly spaced with a fixed distant, which can be far larger than a half-wavelength, too.
We term these sparse arrays, the ``bi-sparse arrays".
A novel algorithm is developed to estimate the direction-of-arrivals and polarizations of coherent sources using these bi-sparse arrays.

For coherent sources direction-finding, various spatial-smoothing (SS) algorithms have been investigated, e.g: \cite{ShanASSPT0885,PillaiASSPT0189,HuaSPT0992,WangAEST0498,WilliamsASSPT0488,LiSPT1292}.
The essence of the SS algorithm is to separate the total array into equivalent sub-arrays by adopting the invariance of the uniform linear array
in order to restore the rank of the covariance matrix of the received data-vector, after which the conventional eigen-structure algorithms can be used.
The advantage of the spatial-smoothing algorithm is that it can resolve arbitrary number of coherent sources only if the array has sufficient sensors.
The disadvantage of this algorithm is that it will decrease the array aperture.

For the direction finding and polarization estimation of incident source, various polarized antennas has been used \cite{LiSPT1292,LiAPT0393,NehoraiSPT0294,WongKT_APT1097,LuoEURASIP0512}, such as dipole and/or loop pairs \cite{YuanAPT0512}, dipole or loop triads \cite{YuanSJ0612,XuIET-MAP0612}, dipole and loop quads \cite{YuanAWPL2012}, and
the six-component electromagnetic vector-sensor \cite{NehoraiSPT0294,WongKT_APT0500,
WongKT_APT0800,
Zoltowski1SPT0800,
Zoltowski2SPT0800,YuanSPT0312}.
All of them are composed of dipoles and/or loops with various orientations to measure the electric field and/or the magnetic field of the signal.
Extensively investigated in the literature is the six-component electromagnetic vector-sensor.

\subsection{Review of Electromagnetic Vector and Polarization}

\begin{figure}[htbp]
\centering
  \includegraphics[scale=0.8]{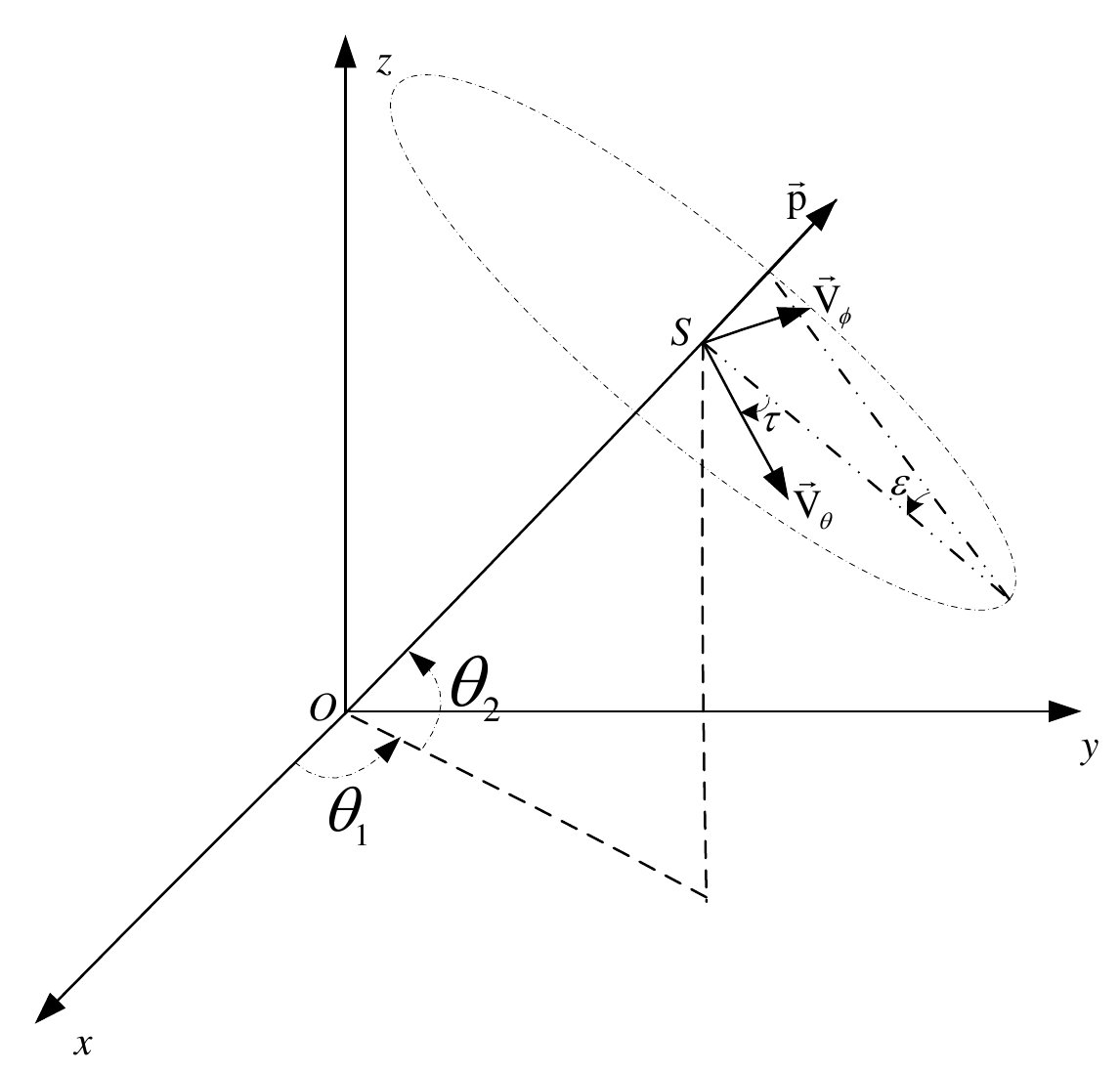}
    \caption{Electromagnetic vector and polarization ellipse. The receiver (polarized antenna arrays) is located at the origin $O$.
    $S$ is the source and $\{\theta_1\in[0,2\pi), \theta_2 \in[-\pi/2,\pi/2]\}$ denotes the azimuth-angle and elevation-angle.
    $\vec{\rm P}$ denotes the Poynting vector of the electromagnetic wave. $\vec{\rm V}_{\theta}$ and $\vec{\rm V}_{\phi}$ symbolize the horizontal and vertical components of the electromagnetic wave.}
    \label{EMV}
\end{figure}

\begin{figure}[htbp]
\centering
  \includegraphics[scale=1.0]{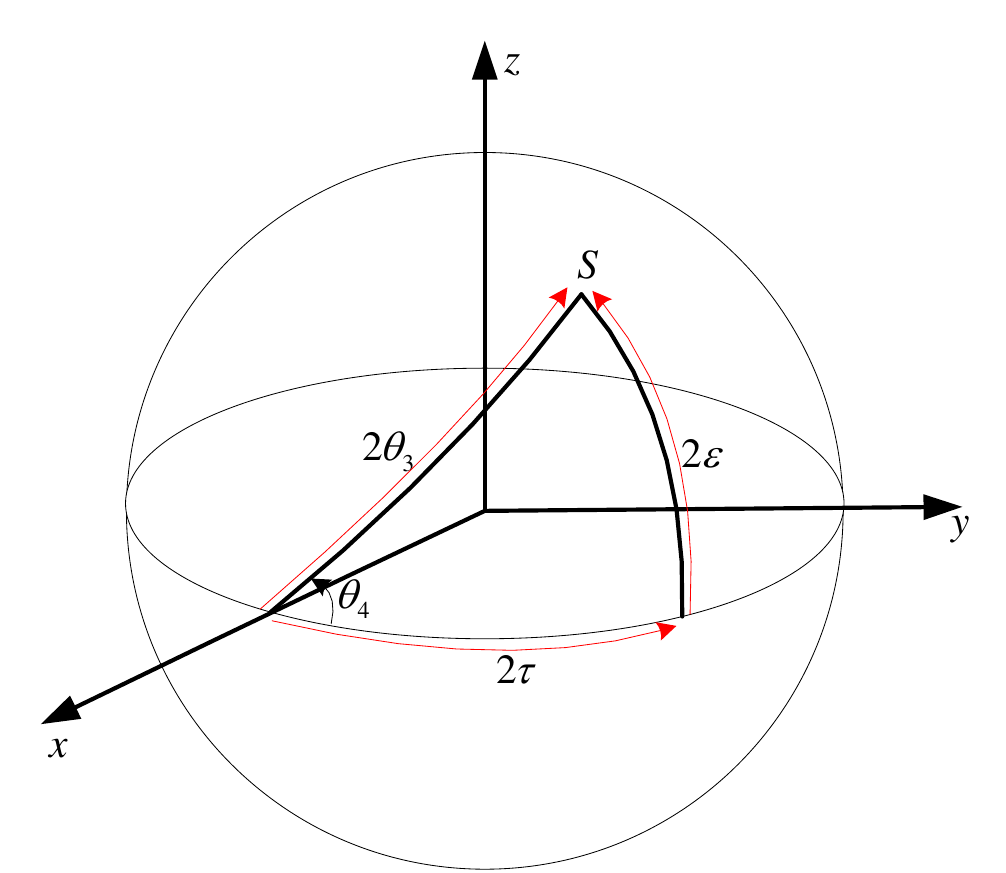}
    \caption{The Poincare sphere. $\{\tau,\epsilon\}$ are the ellipticity and the orientation angles of the electromagnetic wave, and $\{\theta_3,\theta_4\}$ denote the auxiliary polarization angle and polarization phase difference of the electromagnetic wave.}
    \label{Poin}
\end{figure}

Consider a far-field source $S$, depicted in Figure \ref{EMV}, emitting a completely polarized transverse electromagnetic wave,
which can be characterized by the
electric-field vector ${\bf e}$ and the magnetic-field vector ${\bf h}$.
The polarization of the transverse electromagnetic wave can be characterized by $\{\epsilon, \tau\}$ defined in the polarization ellipse in Figure \ref{EMV}.
For the convenience of representation, with the relationship between the  polarization parameters $\{\theta_3,\theta_4\}$ and $\{\epsilon, \tau\}$ shown in Figure \ref{Poin} \cite{DeschampsIRE51},
\footnote{If the electromagnetic wave is linearly polarized, $\theta_4=0$.
If the electromagnetic wave is circularly polarized, $\theta_3=45^\circ$ and $\theta_4=\pm 90^\circ$
($+$ for left circularly polarized and $-$ for right circularly polarized)
\cite{DeschampsIRE51}.},
${\bf e}$ and ${\bf h}$ can be expressed in spherical coordinates as shown in Figure \ref{EMV} as \cite{LiAPT0592,NehoraiSPT0294,WongKT_APT1097}:
\begin{eqnarray}
{\bf e} &=& \sin\theta_3 e^{j\theta_4} \vec{\bf v}_{\phi} + \cos \theta_3 \vec{\bf v}_{\theta}, \\
{\bf h} &=& {\cal Z}_o\left( \cos \theta_3 \vec{\bf v}_{\phi} + \sin\theta_3 e^{j\theta_4}\vec{\bf v}_{\theta}\right),
\end{eqnarray}
where
${\cal Z}_o$ is the intrinsic impedance of the transmission medium,
$\vec{\bf v}_m$ is a unit vector along the direction of $m$ (see Figure \ref{EMV}).
The state of light polarization correspond to various points on the surface of the Poincare sphere in Figure \ref{Poin}.
Poincare used the circular polarization basis, with the latitude representing the relative magnitudes for the left and right circularly polarized components and the longitude representing the relative phase difference between left and right circularly polarized components

Equivalently, in Cartesian coordinates after normalization, ${\bf e}$ and ${\bf h}$ can be re-expressed as \cite{LiAPT0592,NehoraiSPT0294,WongKT_APT1097}
\begin{eqnarray}
{\bf e} &=& (\cos\theta_1\sin\theta_2\sin\theta_3 e^{j\theta_4}- \sin\theta_1\cos\theta_3)\vec{\bf v}_x \notag\\
&& + (\sin\theta_1\sin\theta_2\sin\theta_3 e^{j\theta_4}+ \cos\theta_1\cos\theta_3)\vec{\bf v}_y \notag\\
&&-\cos\theta_2\sin\theta_3 e^{j\theta_4}\vec{\bf v}_z,   \\
{\bf h} &=&(-\sin\theta_1\sin\theta_3 e^{j\theta_4}-\cos\theta_1\sin\theta_2\cos\theta_3))\vec{\bf v}_x \notag\\
&& (\cos\theta_1\sin\theta_3 e^{j\theta_4}- \sin\theta_1\sin\theta_2\cos\theta_3)\vec{\bf v}_y \notag\\
&&+\cos\theta_2\cos\theta_3  \vec{\bf v}_z,
\end{eqnarray}
where $\{\theta_1\in[0,2\pi), \theta_2 \in[-\pi/2,\pi/2]\}$ denotes the azimuth-angle and elevation-angle of the incident source.

\subsection{Coherent Sources Direction-Finding with Electromagnetic Vector-Sensors}
An electromagnetic vector-sensor is composed of a triad-triad plus a loop-triad.
A dipole-triad \cite{WongKT_AEST0401,AuYeungAEST0109} comprises three orthogonally-collocated dipoles which are used to measure the three components of the signal's electric field.
A loop-triad \cite{WongKT_AEST0401,AuYeungAEST0109} comprises three orthogonally-collocated loops which are used to measure the three components of the signal's magnetic field.
The electromagnetic vector-sensor can resolve both the polarization and the direction-of-arrival (DOA) differences of the source \cite{LiAPT0393,NehoraiSPT0294}.
The electromagnetic vector-sensor (array) has been investigated extensively for direction finding and polarization estimation:
\cite{NehoraiSPT0294,
LiAPT0393,
HochwaldSPT0895,
WongKT_APT1097,
HoSPT1097,
HoSPT1099,
WongKT_APT0500,
WongKT_APT0800,
Zoltowski1SPT0800,
Zoltowski2SPT0800,
WongKT_AEST0101,
KoAEST0702,
MironSPT0406,
HurtadoAEST0407,
LeBihanSPT0907,
XuAEST1008,
HurtadoSPM0109,
SunCISP09,
XiaoSPT0209,
GongSP0509,
SunAPSEC10,SunCISP10,
GongSP0411,
GongIET-SP0411,
GuoSPT0711,
YuanSPT0312,
LuoEURASIP0512,
YuanICASSP2012}.
However, much of the literature models the sources as uncorrelated signals. The algorithms proposed therein will degrade adversely when the incident sources are correlated in practical applications \cite{ShanASSPT0885,LiSPT1292,RahamimSPT1104}.

For the coherent sources direction-finding with electromagnetic vector-sensor arrays,
Rahamim {\em et al.} \cite{RahamimSPT1104} proposed the polarization-smoothing (PS) algorithm,
based on the unique collocated geometry of the electromagnetic vector-sensor.
This PS algorithm restores the rank of the covariance matrix by adding the covariance matrices of the data sets collected by different antennas in the vector-sensor.
Compared with the spatial-smoothing algorithm described above, the polarization-smoothing algorithm will not decrease the array aperture.
Some improved polarization-smoothing algorithms were developed in \cite{XuIET-RSN0607,HeAEST0110,GongAEST0711}.
However, the polarization-smoothing algorithm needs iterative search to estimate the DOAs of the sources.
Particularly, for the two-dimensional DOA estimation, this iterative search will have a high computation workload.
Furthermore, because of this polarization-smoothing, the signal subspace of the covariance matrix will not provide the steering vectors of the sources.
This polarization-smoothing approach makes the vector-sensor lose the {\em vector} property.

The spatial-smoothing algorithm has been used in the electromagnetic vector-sensor arrays to estimate the DOAs and the polarizations of the coherent sources
in \cite{XuJCIC0504,LiuJEIT1010,LiuEURASIP2011}.
Because of the unique geometry of the electromagnetic vector-sensor, when the spatial-smoothing algorithm is adopted, the inter vector-sensor spacing in the arrays can be larger than a half-wavelength, and a sparse array is thus formed to enhance the direction finding resolution.
Though the SS algorithm will decrease the array-aperture, the large inter vector-sensor spacing in the sparse array will compensate for the array aperture decrease.
However, since the mutual coupling across the six antennas in the electromagnetic vector-sensor is a serious problem \cite{WongKT_SPT0111,,YuanDLTss}, all the above studies are difficult to apply in practical applications.
In order to overcome the above problem, sparse arrays are proposed in this work to bridge the gap in the literature.

Another important research direction is the identifiability of these polarized antenna arrays \cite{TanSP93,HoSP95,HochwaldTSP0196,TanTSP1296,HoTAP1198,XuTAES1008}, which investigates the linear dependence of steering vectors of these antenna arrays. For the bi-sparse arrays  proposed in this paper, the identifiability is under development.


\subsection{Contributions of This Work}
This paper proposes a novel algorithm to estimate the DOAs and the polarizations of the coherent sources, based on:
1) a sparsely distributed array geometry composed of dipoles or loops, and
2) the spatial-smoothing algorithm.

In order to reduce the mutual coupling across the six antennas in the electromagnetic vector-sensor,
a spatially-spread electromagnetic vector-sensor composed of three orthogonally oriented dipoles and three orthogonally oriented loops is proposed in \cite{WongKT_SPT0111}.
However, in practical applications, the responses of the dipole and the loop vary from each other \cite{LiAPT0393}.
Hence, the mixed use of dipoles and loops is not a good choice for engineers.
Unlike the spatially-spread dipoles and loops used in \cite{WongKT_SPT0111,YuanSSP2011,YuanAWPL2013}, this paper proposes bi-sparse arrays composed of only dipoles or only loops.
Because the dipoles or loops are spatially spread, the arriving angles and polarizations of the sources can not be estimated by the formulas derived in
\cite{WongKT_AEST0401,YuanSJ0612} for the collocated triads.
In order to eliminate the phase factor arising from the inter-sensor spacing, an additional dipole/loop is employed in the sub-array with the same orientation as one of the dipoles/loops in the spatially-spread triads. Thus, the sub-array is a spatially-spread dipole-quad or loop-quad.
Following this, the spatially-spread dipole-quads or loop-quads are uniformly spaced on a line to construct a sparsely linear array.
Finally, a bi-sparse array composed of spatially-spread dipoles or loops is formed.
Based on these bi-sparse arrays, a new spatial-smoothing algorithm is proposed, which can offer the {\em closed-form} estimates of the DOA and polarization of the sources.
Only {\em one} dimensional spatial-smoothing is necessary to derive the two dimensional DOAs and polarizations.
The inter-sensor spacing phase factors of the sparse array are used to improve the direction-finding resolutions, and at the same time to compensate for the array aperture decrease in the spatial-smoothing algorithm.

The specific contributions of this work are summarized as follows:
\begin{enumerate}
\item[1)] All the inter-sensor spacings in the bi-sparse arrays are far larger than a half-wavelength. This will offer the following three advantages:
a) The inter-sensor spacing in the bi-sparse array can be used to improve the DOA estimation accuracy;
b) The mutual coupling among the antennas is reduced;
c) The inter-sensor spacing will compensate for the array aperture decrease in the SS algorithm.
\item[2)] Only one dimensional SS is utilized to provide the two dimensional DOA and polarization estimation, and it thus requires no pair algorithm.
\item[3)] The closed-form estimation is obtained to avoid the iterative search.
\item[4)] The hardware cost is reduced since only dipoles or loops is utilized.
\item[5)] The proposed algorithm is extended to the following geometries:
a) the vector-sensor array composed of the non-collocated electromagnetic vector-sensor proposed in \cite{WongKT_SPT0111},
b) the vector-sensor array composed of the spatially-spread dipole-loop pairs with orthogonal orientations, and
c) the vector-sensor array composed of the spatially-spread dipole-triads and loop-triads proposed in \cite{WongKT_AEST0401}.
\end{enumerate}
It is worth noting that the proposed algorithm requires that the sub-arrays are uniformly spaced, which is not required in the PS algorithm.

This SS algorithm was used in the rectangular array composed of  {\em collocated} electromagnetic vector-sensors to estimate the DOA of coherent sources by the propagator method in \cite{LiuJEIT1010}. A planar-plus-an-isolated array geometry composed of electromagnetic vector-sensors and a pressure sensor was used in \cite{HeIET-RSN0509} to estimate the DOA of coherent sources.
In contrast, this paper uses the bi-sparse array composed of spatially-spread dipoles or loops.
The mutual coupling is thus reduced and the described contributions can be achieved.
Unlike \cite{LiuJEIT1010}, the derived algorithm in the present work will be based on the eigen-decomposition.\footnote{The propagator method \cite{MarcosSP0395} can also be utilized in the proposed array geometry in this paper.
For the comparison of the propagator method and the ESPRIT\cite{RoyASSPT0789}-based algorithm used for the vector sensor array, please refer to \cite{HeDSP0509}.}
The steering-vectors of the sources are derived from the signal sub-space of the covariance matrix of the collected data-set.
The inter-sensor spacing phase factors in the sparse array is utilized to offer the {\em fine but ambiguous} estimates of the direction-cosines.
Then the disambiguation algorithm is used to derive the {\em fine and unambiguous} estimates of the direction-cosines to enhance the estimation accuracy.
Similar scenario has been used in \cite{YuanAEST0712} under a sparsely-distributed acoustic vector-sensor array scenario.

\subsection{Organization of This Paper}
The remainder of this paper is organized as follows:
Section \ref{Sec:geo} provides the array geometry used in this work.
Section \ref{Sec:ag} presents the algorithm to derive the closed-form estimation of DOAs and polarizations of multiple coherent sources.
Sections \ref{Sec:EMVSss}-\ref{Sec:DLTss} extend the proposed algorithm to the other three array geometries.
Section \ref{Sec:CRB} analyzes the Cram\'{e}r-Rao bounds of the parameter estimation.
Section \ref{Sec:sim} shows the simulation results to verify the performance of the proposed algorithm.
Section \ref{Sec:con} concludes the whole paper.

\section{Array Geometry}
\label{Sec:geo}

\begin{figure}
\centering
\begin{minipage}{5in}
\centering
\includegraphics[height=8cm,width=12cm]{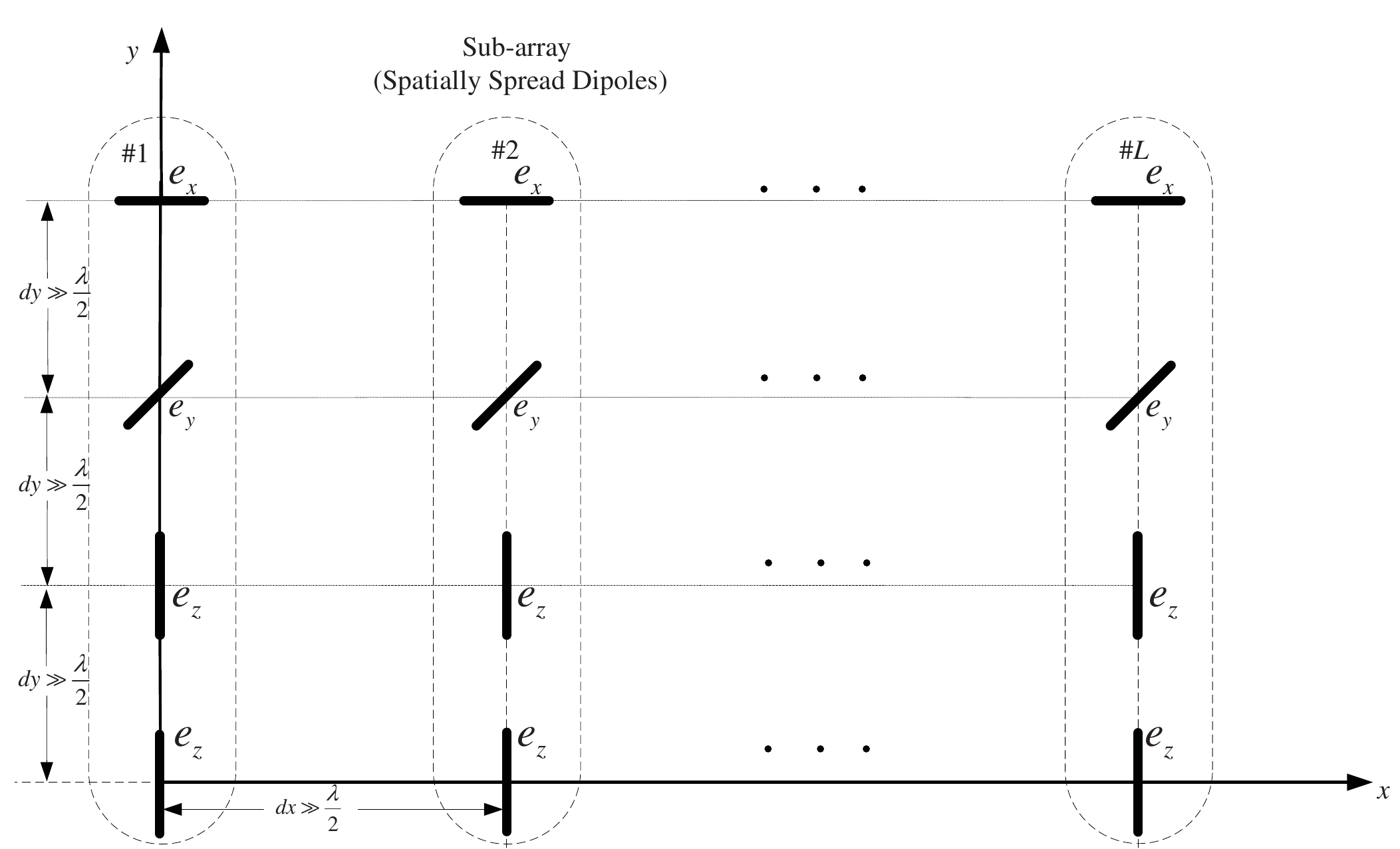}
\caption{A possible geometry of the bi-sparse array composed of spatially spaced dipoles. The inter-dipole spacings among the dipoles are all beyond $\lambda/2$.}
\label{DT-SS}
\end{minipage}
    \hfill
\begin{minipage}{5in}
\includegraphics[height=8cm,width=12cm]{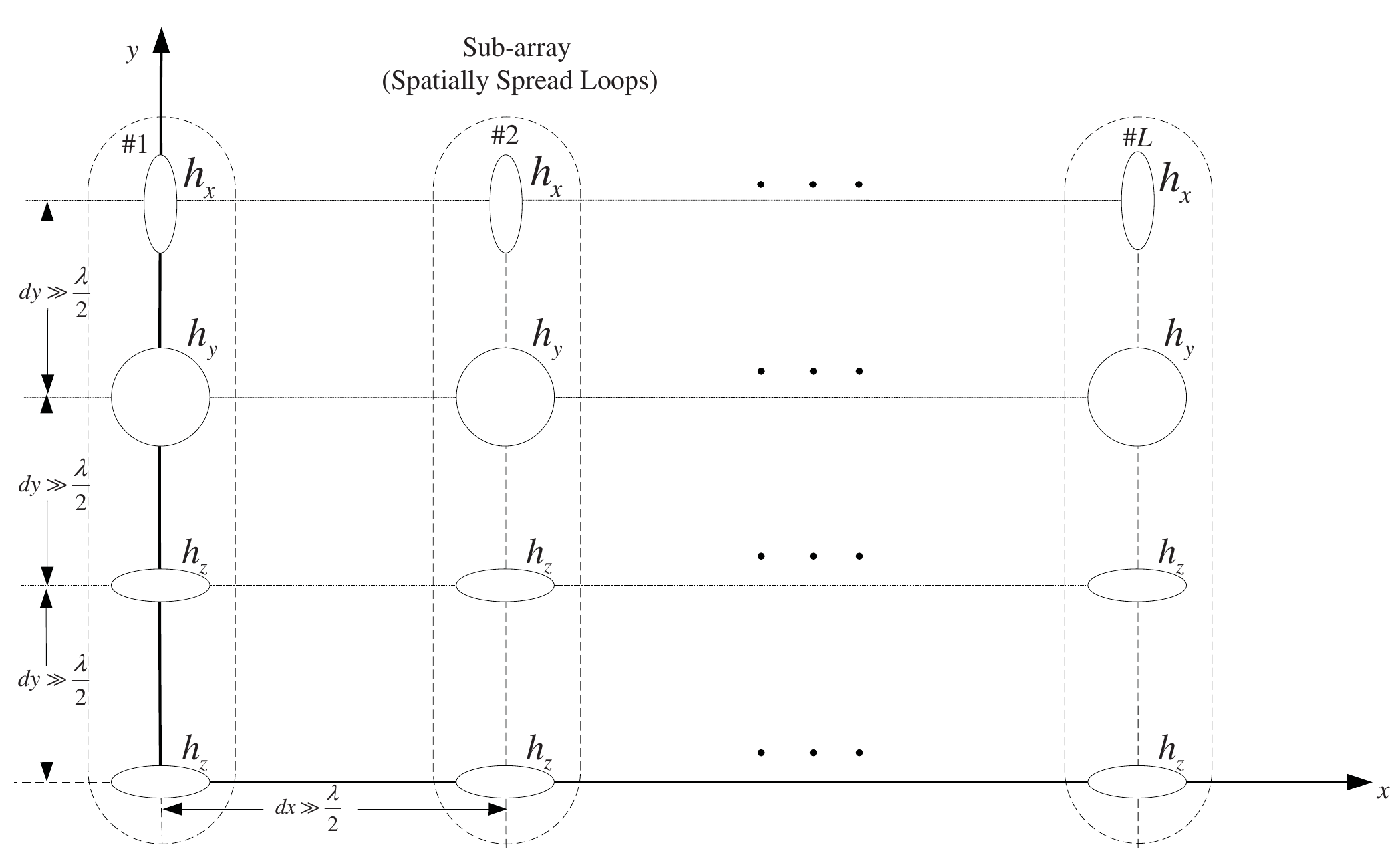}
\centering
\caption{A possible geometry of the bi-sparse array composed of spatially spaced loops. The inter-loop spacings among the loops are all beyond $\lambda/2$.}
\label{LT-SS}
\end{minipage}
\end{figure}

Figures \ref{DT-SS}-\ref{LT-SS} depicts the bi-sparse array-geometries proposed in this paper.
The bi-sparse array composed of dipoles is demonstrated in Figure \ref{DT-SS}, and the bi-sparse array composed of loops is demonstrated in Figure \ref{LT-SS}.
Four dipoles or loops are displaced along the $y$-axis with a distance $d_y\gg\lambda/2$, where $\lambda$ denotes the wavelength of the signal.
These four dipoles or loops comprise a sub-array, and $L$ such sub-arrays are spaced along the $x$-axis with a distance $d_x\gg\lambda/2$.

In a multiple-source scenario with $K$ sources, the responses of the dipoles along each axis for the $k$th signal are \cite{NehoraiSPT0294,LiAPT0393}:
\begin{eqnarray}
{\bf e}_k &\stackrel{\rm def}{=} & \left[\begin{array}{l}   e_{x,k}  \\
                          e_{y,k}  \\
                          e_{z,k}  \end{array}\right] 
\stackrel{\rm def}{=}
\left[\begin{array}{c}
                    \cos\theta_{1,k}\sin\theta_{2,k}\sin\theta_{3,k} e^{j\theta_{4,k}}- \sin\theta_{1,k}\cos\theta_{3,k} \\
\sin\theta_{1,k}\sin\theta_{2,k}\sin\theta_{3,k} e^{j\theta_{4,k}}+ \cos\theta_{1,k}\cos\theta_{3,k}\\
-\cos\theta_{2,k}\sin\theta_{3,k} e^{j\theta_{4,k}}
                   \end{array} \right], \label{eq:aDT}
\end{eqnarray}
where $\{\theta_{1,k}\in[0,2\pi), \theta_{2,k}\in[-\pi/2,\pi/2]\}$ are the azimuth-angle and elevation-angle of the $k$th source,
and $\{\theta_{3,k}\in[0,\pi/2],\theta_{4,k}\in[-\pi,\pi)\}$ denote the auxiliary polarization angle and polarization phase difference of the $k$th incident signal, respectively (equating to $\{\gamma,\eta\}$ in \cite{LiAPT0393}).
The responses of the loops along each axis for the $k$th signal are \cite{NehoraiSPT0294,LiAPT0393}:
\begin{eqnarray}
{\bf h}_k &  \stackrel{\rm def}{=} &
 \left[\begin{array}{l}   h_{x,k}  \\
                          h_{y,k}  \\
                          h_{z,k} \end{array}\right]
\stackrel{\rm def}{=}
\left[\begin{array}{c}
-\sin\theta_{1,k}\sin\theta_{3,k} e^{j\theta_{4,k}}-\cos\theta_{1,k}\sin\theta_{2,k}\cos\theta_{3,k}\\
\cos\theta_{1,k}\sin\theta_{3,k} e^{j\theta_{4,k}}- \sin\theta_{1,k}\sin\theta_{2,k}\cos\theta_{3,k}\\
\cos\theta_{2,k}\cos\theta_{3,k}
                   \end{array} \right].\label{eq:aLT}
\end{eqnarray}

Since the dipoles or loops in Figures \ref{DT-SS}-\ref{LT-SS} are spatially-spread, the inter-sensor phase factors will be introduced in the array-manifold.
The array-manifold of the sub-array with four sparsely spaced dipoles in Figure \ref{DT-SS} corresponding to $k$th source is:
\begin{eqnarray}
{\bf a}_{{\rm sub},k}&\stackrel{\rm def}{=}&
\left[\begin{array}{r}
e^{-j\frac{2\pi u_{y,k}}{\lambda_k}3d_y } e_x \\
e^{-j\frac{2\pi u_{y,k}}{\lambda_k}2d_y } e_y \\
e^{-j\frac{2\pi u_{y,k}}{\lambda_k}d_y } e_z \\
e_z
\end{array}\right], \label{eq:aDTss}
\end{eqnarray}
where $u_{y,k}\stackrel{\rm def}{=}\cos\theta_{2,k}\sin\theta_{1,k}$ is the direction-cosine of the $k$th source align to $y$-axis.
The array-manifold of the sub-array with four sparsely spaced loops in Figure \ref{LT-SS} corresponding to $k$th source is:
\begin{eqnarray}
{\bf a}_{{\rm sub},k}&\stackrel{\rm def}{=}&
\left[\begin{array}{r}
e^{-j\frac{2\pi u_{y,k}}{\lambda_k}3d_y } h_x \\
e^{-j\frac{2\pi u_{y,k}}{\lambda_k}2d_y } h_y \\
e^{-j\frac{2\pi u_{y,k}}{\lambda_k}d_y } h_z \\
h_z
\end{array}\right]. \label{eq:aLTss}
\end{eqnarray}
In these two bi-sparse arrays, only dipoles or loops are used, and so the ``vector-cross-product" algorithm \cite{NehoraiSPT0294} can not be adopted. Thus a new algorithm will be derived below.\footnote{The two array geometries in Figures \ref{DT-SS}-\ref{LT-SS} employ only dipoles or only loops and they just show two possible permutations of the array configurations. There are still many other different possible permutations to form the bi-sparse array.
In each sub-array, two $e_z$ dipoles or two $h_z$ loops are used.
The additional dipole/loop with the same orientation with the other dipole/loop is used to decouple the inter-antenna phase factor and it can be replaced by the dipole/loop with the other two orientations, which means the dipole/loop lies on the $x$-axis, $e_z/h_z$ can be replaced by $e_x/h_x$ or $e_y/h_y$.
}

The array-manifold of the
demonstrated array in Figures \ref{DT-SS}-\ref{LT-SS} for $k$th source is thus a $4L \times1$ vector:
\begin{eqnarray} \label{eq:bfa}
{\bf a}_k&=& {\bf a}_{{\rm sub},k}\otimes {\bf q}_{x,k} ,
\end{eqnarray}
where $\otimes$ denotes the Kronecker-product operator,
\begin{eqnarray}
{\bf q}_{x,k}=  \left[1, e^{-j\frac{2\pi}{\lambda}d_x u_{x,k}}, \cdots, e^{-j\frac{2\pi}{\lambda}(L-1)d_x u_{x,k}} \right]^T,
\end{eqnarray}
with $^T$ symbolizing the transposition, and
$u_{x,k}\stackrel{\rm def}{=}\cos\theta_{2,k}\cos\theta_{1,k}$ is the direction-cosine of the $k$th source align to $x$-axis.

\section{Proposed Algorithm}
\label{Sec:ag}
In a $K$-source scenario, the data set measured at time $t$ by the array in Section \ref{Sec:geo} is:
\begin{eqnarray} \label{eq:bfyt}
{\bf y}(t) &=& \sum^K_{k=1}{\bf a}_k s_k (t) + {\bf n}(t),
\end{eqnarray}
where ${\bf a}_k$ is the steering-vector of the $k$th source as shown in (\ref{eq:bfa}), ${\bf n}(t)$ is the additive Gaussian-distributed complex noise,
and $s_k(t)$ is the $k$th signal.
In the coherent sources scenario, $s_k(t) = \eta_k s_1(t), \forall k=2,\cdots,K$, where $\eta_k$ is a complex number denoting the correlation between the sources.\footnote{The spatial smoothing algorithm has been used in \cite{XuJCIC0504,LiuEURASIP2011,LiuJEIT1010} with the six-component electromagnetic vector sensor array. Originality of this paper is to adopt the spatial smoothing algorithm to the spatially spaced polarized vector sensor array, instead of the collocated
electromagnetic vector sensor array, which introduces a lot of mutual coupling.
The mutual coupling is thus reduced because of the large spacing between the adjacent antennas.
The fine estimates of the direction-cosine are obtained from the inter-sensor spacing within one sub-array, not from the spacing between the electromagnetic vector sensors as in \cite{Zoltowski1SPT0800,Zoltowski2SPT0800}.}

The $4L\times 1$ steering vector ${\bf a}_k$ in (\ref{eq:bfyt}) can be expressed as a $4\times L$ matrix as:
\begin{eqnarray}
{\bf A}_k &=& \left[{\bf a}_{{\rm sub},k}, ~ {\bf a}_{{\rm sub},k} e^{-j\frac{2\pi}{\lambda}d_x u_{x,k}}, \dots, {\bf a}_{{\rm sub},k} e^{-j\frac{2\pi}{\lambda}(L-1)d_x u_{x,k}} \right] \nonumber\\
&=&\left[\begin{array}{rrrr}
e^{-j\frac{2\pi u_{y,k}}{\lambda_k}3d_y } h_x , & e^{-j\frac{2\pi u_{y,k}}{\lambda_k}3d_y } h_x e^{-j\frac{2\pi}{\lambda}d_x u_{x,k}}, &\dots, & e^{-j\frac{2\pi u_{y,k}}{\lambda_k}3d_y } h_x e^{-j\frac{2\pi}{\lambda}(L-1)d_x u_{x,k}} \\
e^{-j\frac{2\pi u_{y,k}}{\lambda_k}2d_y } h_y , & e^{-j\frac{2\pi u_{y,k}}{\lambda_k}2d_y } h_y e^{-j\frac{2\pi}{\lambda}d_x u_{x,k}}, &\dots, & e^{-j\frac{2\pi u_{y,k}}{\lambda_k}2d_y } h_y e^{-j\frac{2\pi}{\lambda}(L-1)d_x u_{x,k}} \\
e^{-j\frac{2\pi u_{y,k}}{\lambda_k}d_y } h_z , & e^{-j\frac{2\pi u_{y,k}}{\lambda_k}d_y } h_z e^{-j\frac{2\pi}{\lambda}d_x u_{x,k}}, &\dots, & e^{-j\frac{2\pi u_{y,k}}{\lambda_k}d_y } h_z e^{-j\frac{2\pi}{\lambda}(L-1)d_x u_{x,k}} \\
h_z , & h_z e^{-j\frac{2\pi}{\lambda}d_x u_{x,k}}, &\dots, &  h_z e^{-j\frac{2\pi}{\lambda}(L-1)d_x u_{x,k}}
\end{array}\right]
\end{eqnarray}
Following this, the  measurement data set in (\ref{eq:bfyt}) can also be represented to a $4\times L$ matrix form:
\begin{eqnarray} \label{eq:bfYt}
{\bf Y}(t) &\stackrel{\rm def}{=}& \sum^K_{k=1}{\bf A}_k s_k(t) + {\bf N}(t) \\
&=& \sum^K_{k=1} \left\{\left[{\bf a}_{{\rm sub},k}, {\bf a}_{{\rm sub},k} e^{-j\frac{2\pi}{\lambda}d_x u_{x,k}}, \cdots,  {\bf a}_{{\rm sub},k} e^{-j\frac{2\pi}{\lambda}(L-1)d_x u_{x,k}}  \right] s_k (t)\right\}  + {\bf N}(t), \notag
\end{eqnarray}
where  $u_{x,k}=\cos\theta_{2,k}\cos\theta_{1,k}$ is the direction-cosine of the $k$th source align to $x$-axis.
In a multiple source scenario with correlated sources, the covariance matrices of (\ref{eq:bfyt}) and (\ref{eq:bfYt}) will be rank-deficient. The following will develop a method to solve this problem.
The SS algorithm \cite{ShanASSPT0885} will be adapted to the data-set in (\ref{eq:bfYt}).

\subsection{Spatial-Smoothing Through Matrix Enhancement}
With the similar approach in \cite{HuaSPT0992}, we can restore the rank of the covariance matrix of (\ref{eq:bfYt}) by matrix enhancement.
The $4\times L$ matrix in (\ref{eq:bfYt}) can be partitioned into $P$ overlapped $4\times (L-P+1)$ sub-matrices:
\begin{eqnarray}
{\bf Y}_p(t) &\stackrel{\rm def}{=} & {\bf Y}(t) {\bf J}_p,  \forall p=1,2,\cdots P,
\end{eqnarray}
where ${\bf J}_p$ is an $L\times (L-P+1)$ selection matrix, and
\begin{eqnarray}
{\bf J}_p &=&  \left[\begin{array}{l}{\bf 0}_{(p-1)\times(L-P+1)} \\
------\\
{\bf I}_{(L-P+1)}\\
------\\
{\bf 0}_{(P-p+1)\times(L-P+1)} \\
\end{array}\right],
\end{eqnarray}
where ${\bf 0}_{a\times b}$ denotes an $a\times b$ matrix with all entries equaling to zero,
and ${\bf I}_i$ is an $i\times i$ identity matrix.

In order to simplify the exposition, the following derivation will set $P=2$.
Then:
\begin{eqnarray}
{\bf Y}_1 (t)&=& \sum^K_{k=1} \left[{\bf a}_{{\rm sub},k}, {\bf a}_{{\rm sub},k} e^{-j\frac{2\pi}{\lambda}d_x u_{x,k}}, \cdots,  
{\bf a}_{{\rm sub},k} e^{-j\frac{2\pi}{\lambda}(L-2)d_x u_{x,k}}  \right] s_k (t)  + {\bf N}_1(t),  \\
{\bf Y}_2 (t)&=& \sum^K_{k=1} \left[{\bf a}_{{\rm sub},k} e^{-j\frac{2\pi}{\lambda}d_x u_{x,k}}, {\bf a}_{{\rm sub},k} e^{-j\frac{2\pi}{\lambda}2d_x u_{x,k}}, \cdots,   
{\bf a}_{{\rm sub},k} e^{-j\frac{2\pi}{\lambda}(L-1)d_x u_{x,k}}  \right] s_k (t)  + {\bf N}_2(t),
\end{eqnarray}
where ${\bf N}_1(t) ={\bf N}(t) {\bf J}_1,{\bf N}_2(t) ={\bf N}(t) {\bf J}_2 $.

Construct the following $8\times (L-1)$ enhancement matrix:
\begin{eqnarray} \label{eq:Z}
{\bf Z}(t) &\stackrel{\rm def}{=}& \left[\begin{array}{c}
{\bf Y}_1 (t)\\
{\bf Y}_2 (t)
\end{array}\right].
\end{eqnarray}

\subsection{DOA and Polarization Estimation}
\label{Sec:CU}
The ${\bf Z}(t)$ can now be used to estimate the DOAs and polarizations of the coherent sources.
Consider there are $M$ time samples collected at uniform time-slots $t_1,t_2,\cdots,t_M$. Compute the covariance matrix of ${\bf Z}$ by:
\begin{eqnarray} \label{eq:bfR}
{\bf R} &=& \frac{1}{M}\sum^M_{m=1}{\bf Z}(t_m) {\bf Z}^H(t_m),
\end{eqnarray}
where $^H$ is the Hermitian operator.
The steering vectors corresponding to ${\bf R}$ in (\ref{eq:bfR}) are:
\begin{eqnarray}
{\bf A} &\stackrel{\rm def}{=}& \left[\begin{array}{c}
{\bf A}_1 \\
{\bf A}_2
\end{array}\right] = \left[\begin{array}{l}
{\bf A}_1 \\
{\bf A}_1 {\bf \Phi}_x
\end{array}\right],
\end{eqnarray}
where ${\bf A}_1\stackrel{\rm def}{=}\left[{\bf a}_{{\rm sub},1}, {\bf a}_{{\rm sub},2}, \cdots, {\bf a}_{{\rm sub},K}\right]$,
and ${\bf \Phi}_x ={\rm diag}\left[e^{-j\frac{2\pi}{\lambda_1}u_{x,1} d_x}, e^{-j\frac{2\pi}{\lambda_2}u_{x,2} d_x}, \cdots, e^{-j\frac{2\pi}{\lambda_K}u_{x,K} d_x}\right]$.

Perform the eigen-decomposition of the covariance matrix ${\bf R}$:
\begin{eqnarray}
{\bf R}  &=& {\bf E}_s {\Lambda}_s  {\bf E}^H_s + {\bf E}_n {\Lambda}_n  {\bf E}^H_n,
\end{eqnarray}
where ${\bf E}_s$ is the signal subspace composed of the eigen-vectors associated with the $K$ largest eigen-values.

Partition the $8\times K$ signal subspace ${\bf E}_s$ into two $4\times K$ sub-matrices, ${\bf E}_{s,1}, {\bf E}_{s,2}$,
where ${\bf E}_{s,1}$ is composed of the top 4 rows, and ${\bf E}_{s,2}$ is composed of the bottom 4 rows.
In the {\em noiseless} case, ${\bf E}_{s,1}$ and ${\bf E}_{s,2}$ are inter-related with each other by:
\begin{eqnarray}
{\bf E}_{s,2} &=& {\bf E}_{s,1} {\bf \Phi}_x.
\end{eqnarray}
In the {\em noisy} case, this ${\bf \Phi}_x$ can be estimated by \cite{RoyASSPT0789}:
\begin{eqnarray}
{\hat{\bf \Phi}}_x &=& \left({\bf E}^H_{s,1} {\bf E}_{s,1}\right)^{-1} {\bf E}^H_{s,1} {\bf E}_{s,2}.
\end{eqnarray}

Similar to \cite{WongKT_APT1097}, there exists a unique $K\times K$ nonsingular matrix ${\bf T}$ such that \cite{RoyASSPT0789}:
\begin{eqnarray}
{\bf E}_{s,1} &=& {\bf A}_1 {\bf T},\\
{\bf E}_{s,2} &=& {\bf A}_2 {\bf T} = {\bf A}_1 {\bf \Phi}_x {\bf T}.
\end{eqnarray}
In the {\em noisy} case, this ${\bf T}$ can be estimated by performing the eigen-decomposition of $\left({\bf E}^H_{s,1} {\bf E}_{s,1}\right)^{-1} {\bf E}^H_{s,1} {\bf E}_{s,2}$.
${\hat{\bf T}}$ is composed of the eigenvectors, and ${\bf D}={\rm diag} \left[\sigma_1,\sigma_2,\cdots, \sigma_K\right]$ comprises the eigenvalues.

It is worth noting that ${\bf D}$ will offer the {\em fine but ambiguous} estimates of the sources' direction-cosines along the $x$-axis:
\begin{eqnarray} \label{eq:ufine}
\hat{u}_{x,k}^{\rm fine} &=& -\frac{\lambda_k}{2\pi d_x}\angle \sigma_k ,
\end{eqnarray}
where $\angle$ denotes the complex angle of the ensuing number.

The steering vectors of the sources can be estimated by \cite{RoyASSPT0789}:
\begin{eqnarray} \label{eq:hatA1}
\hat{\bf A}_1 &=& {\bf E}_{s,1}  {\hat{\bf T}}^{-1} + {\bf E}_{s,2}  {\hat{\bf T}}^{-1} {\bf D}^{-1} \notag\\
&=& \left[\hat{\bf a}_{{\rm sub},1}, \hat{\bf a}_{{\rm sub},2}, \cdots, \hat{\bf a}_{{\rm sub},K}\right].
\end{eqnarray}

Note that $\hat{\bf A}_1$ will offer the {\em coarse} estimates of the arriving angles.
We can obtain from $\hat{\bf A}_1 $ that $\hat{\bf a}_{{\rm sub},k}= c{\bf a}_{{\rm sub},k}$, where $c$ is an unknown complex number.
From $\hat{\bf a}_{{\rm sub},k}= c{\bf a}_{{\rm sub},k}$ and (\ref{eq:aDTss})-(\ref{eq:aLTss}), the {\em fine but ambiguous} estimates of the sources' direction-cosines along the $y$-axis::
\begin{eqnarray} \label{eq:vfine}
{\hat u}^{\rm fine}_{y,k} &=& \frac{\lambda_k}{2\pi}\frac{1}{d_y}\angle\left(\frac{[{\hat{\bf a}}_{{\rm sub},k}]_4}{[{\hat{\bf a}}_{{\rm sub},k}]_3}\right).
\end{eqnarray}
Define:
\begin{eqnarray}
{\bf d}&=& \left[\begin{array}{c}
\frac{[{\hat{\bf a}}_{{\rm sub},k}]_1}{[{\hat{\bf a}}_{{\rm sub},k}]_3} e^{j\frac{2\pi {\hat u}^{\rm fine}_{y,k}}{\lambda_k}2d_y }\\
\frac{[{\hat{\bf a}}_{{\rm sub},k}]_2}{[{\hat{\bf a}}_{{\rm sub},k}]_3} e^{j\frac{2\pi {\hat u}^{\rm fine}_{y,k}}{\lambda_k}d_y }
\end{array}\right] 
=\left\{\begin{array}{cc}\left[\frac{e_x}{e_z},  \frac{e_y}{e_z}\right]^T & \mbox{for the array in Figure \ref{DT-SS}},\\
\left[\frac{h_x}{h_z},  \frac{h_y}{h_z}\right]^T & \mbox{for the array in Figure \ref{LT-SS}}.
\end{array}\right.
\end{eqnarray}
where $^T$ denotes the transposition.

From the equations derived in \cite{YuanSPT0312,YuanSJ0612}, we can get:
\begin{eqnarray}
\hat{\theta}^{\rm coarse}_{1,k} &=& \left\{\begin{array}{l}\tan^{-1}\left(\frac{-{\mathfrak Im}\{[{\bf d}]_1\}}{{\mathfrak Im}\{[{\bf d}]_2\}}\right), ~~~~~~~~ {\mbox{if}}\hspace{0.1in}  ({\mathfrak Im}\{[{\bf d}]_2\}\sin\theta_{4,k})\ge0 \\
 \tan^{-1}\left(\frac{-{\mathfrak Im}\{[{\bf d}]_1\}}{{\mathfrak Im}\{[{\bf d}]_2\}}\right) +\pi, ~~~{\mbox{if}} \hspace{0.1in}  ({\mathfrak Im}\{[{\bf d}]_2\}\sin\theta_{4,k})<0
\end{array}\right.  \label{eq:phidt}\\
\hat{\theta}^{\rm coarse}_{2,k} &=& \left\{\begin{array}{l}
\tan^{-1}\left(-B_{{\theta}_{2,k}}\right),
\hspace{0.37in}{\mbox{if}} \hspace{0.1in} \left(B_{{\theta}_{2,k}}\right)\le0  \\
\tan^{-1}\left(-B_{{\theta}_{2,k}}\right)+\pi,
\hspace{0.14in}{\mbox{if}} \hspace{0.1in} \left(B_{{\theta}_{2,k}}\right)>0 \end{array}\right. \\
B_{{\theta}_{2,k}}&=& {\mathfrak Re}\{[{\bf d}]_1\}\cos{\hat \theta^{\rm coarse}_{1,k}} + {\mathfrak Re}\{[{\bf d}]_2\}\sin{\hat \theta^{\rm coarse}_{1,k}}\notag\\
\hat{\theta}_{4,k} &=&  -\angle\left([{\bf d}]_1\sin{\hat \theta^{\rm coarse}_{1,k}}-[{\bf d}]_2\cos{\hat \theta^{\rm coarse}_{1,k}}\right) \\
\hat{\theta}_{3,k} &=& \left\{\begin{array}{l}
\cot^{-1}\left(\frac{{\mathfrak Im}\{[{\bf d}]_2\}\cos{\hat\theta}^{\rm coarse}_{2,k}}{\sin{\hat\theta_{4,k}}\cos{\hat\theta}^{\rm coarse}_{1,k}}\right), ~~~\mbox{for the array in Figure \ref{DT-SS}};\\
\tan^{-1}\left(\frac{{\mathfrak Im}\{[{\bf d}]_2\}\cos{\hat\theta}^{\rm coarse}_{2,k}}{\sin{\hat\theta_{4,k}}\cos{\hat\theta}^{\rm coarse}_{1,k}}\right), ~~~\mbox{for the array in Figure \ref{LT-SS}}.
\end{array} \right.
\end{eqnarray}
where ${\mathfrak Re}\{\hspace{0.03in}\}$ and ${\mathfrak Im}\{\hspace{0.03in}\}$ denote the real part and the imaginary part of the entry in $\{\hspace{0.03in}\}$, respectively.
\footnote{Under a finite number of discrete values, we can not estimate the DOA correctly with only dipoles or loops in Figures \ref{DT-SS}-\ref{LT-SS}. For example, under the horizontally linear-polarized case, $\theta_3 = \theta_4 = 0$,
$\{e_x = -\sin \theta_1,~ e_y = \cos\theta_1, ~ e_z = 0,\}$.
In this case, with only the dipoles as in Figure 3a, we can not estimate the elevation-angle $\theta_2$. However, these values occur
with probability zero, anyway. Similar discrete values exist in other array geometries.}
Thus,
\begin{eqnarray}
{\hat u}^{\rm coarse}_{x,k}&=& \cos{\hat\theta}^{\rm coarse}_{2,k}\cos{\hat\theta}^{\rm coarse}_{1,k}, \label{eq:ucoarse}\\
{\hat u}^{\rm coarse}_{y,k}&=& \cos{\hat\theta}^{\rm coarse}_{2,k}\sin{\hat\theta}^{\rm coarse}_{1,k}. \label{eq:vcoarse}
\end{eqnarray}
Following this the disambiguation method can be adopted to derive the final estimates of direction-cosines and so the arriving angles.
Using the coarse estimates of direction-cosines in (\ref{eq:ucoarse})-(\ref{eq:vcoarse}) to disambiguate the fine estimates in
(\ref{eq:ufine}) and (\ref{eq:vfine}) by the method in \cite{Zoltowski1SPT0800,Zoltowski2SPT0800,WongKT_SPT0111}, we can obtain the {\em fine and unambiguous} (final) estimates of direction-cosines, $\left\{{\hat u}_k, {\hat v}_k\right\}$:
\begin{eqnarray}
{\hat u}_{x,k} &=& {\hat u}^{\rm fine}_{x,k} + m^{\circ}_{x,k} \frac{\lambda_k}{d_x},  \label{eq:ufinal}\\
{\hat u}_{y,k} &=& {\hat u}^{\rm fine}_{y,k} + m^{\circ}_{y,k} \frac{\lambda_k}{d_y},  \label{eq:vfinal}
\end{eqnarray}
where $\{m^{\circ}_{x,k} ,m^{\circ}_{y,k} \}$ are two integers that can be determined by ${\hat u}^{\rm coarse}_{x,k},{\hat u}^{\rm coarse}_{y,k}$ \cite{Zoltowski1SPT0800,Zoltowski2SPT0800,WongKT_SPT0111},
\begin{eqnarray}
m_{x,k}^\circ
&=&
  \underset{m_{x,k}}{\operatorname{arg min}}  \left| \hat{u}^{\rm coarse}_{x,k}- \hat{u}^{\rm fine}_{x,y} - m_{x,k} \frac{\lambda}{ d_x}   \right|   \nonumber \\
m_{y,k}^\circ
&=&
\underset{m_{y,k}}{\operatorname{arg min}}   \left| \hat{u}^{\rm coarse}_{y,k}- \hat{u}^{\rm fine}_{y,k} - m_{y,k} \frac{\lambda}{d_y }   \right|   \nonumber
\end{eqnarray}
for
\begin{eqnarray}
 m_{x,k}
&\in&
\left\{   \left\lceil
\frac{d_x}{\lambda}(-1-\hat{u}^{\rm coarse}_{x,k})\right\rceil,
 \left\lfloor
\frac{d_x}{\lambda}(1-\hat{u}^{\rm coarse}_{x,k}) \right\rfloor \right\}, \nonumber  \\
 m_{y,k}
&\in&
  \left\{   \left\lceil
\frac{d_y}{\lambda}(-1-\hat{u}^{\rm coarse}_{y,k})
\right\rceil,
 \left\lfloor
\frac{d_y}{\lambda}(1-\hat{u}^{\rm coarse}_{y,k}) \right\rfloor \right\} . \nonumber
\end{eqnarray}
where $\lceil \alpha \rceil$ refers to the smallest integer not less than $\alpha$, and
$\lfloor \alpha \rfloor$ refers to the largest integer not exceeding $\alpha$.

Lastly, after the unique $\{{\hat u}_{x,k}, {\hat u}_{y,k}\}$  has been obtained, the direction-of-arrival of $k$th incident source $\{\theta_{1,k},\theta_{2,k}\}$ can be estimated by \cite{WongKT_APT1097}:
\begin{eqnarray}
{\hat\theta}_{1,k} &=& \angle\left({\hat u}_{x,k}+j \hspace{0.03in}{\hat u}_{y,k}\right), \label{eq:hath1}\\
{\hat\theta}_{2,k} &=& \arccos\left(\sqrt{{{\hat u}_{y,k}}^2+{{\hat u}_{x,k}}^2}\right). \label{eq:hath2}
\end{eqnarray}

The derivation in this section just shows an example with $P=2$.
In practical applications, this $P$ depends on the source number $K$. The conditions to select $P$ and $L$ are:
\begin{eqnarray} \label{eq:Cond}
4\ge K, \hspace{0.2in}\mbox{and} \hspace{0.2in} (L-P+1)\ge K.
\end{eqnarray}
Thus $2 \le P \le (L-K+1)$.
When $P>2$, there will be several estimates of the sources' steering-vectors. Then the average values can be used to improve the estimation accuracy.
In addition, the Forward/Backward spatial-smoothing technique in \cite{PillaiASSPT0189} can be incorporated with the proposed algorithm to
increase the resolvable source number.

\begin{table}
\caption{Summary of the algorithm}
\centering
\begin{tabular}{cl}
\hline
1. & Determine $P$ and $L$, based on the coherent source-number $K$;   \\
2. & Construct the enhancement matrix ${\bf Z}(t)$ as in (\ref{eq:Z});  \\
3. & Compute the covariance matrix ${\bf R}$ of the matrix  ${\bf Z}$ as in (\ref{eq:bfR});\\
4. & Perform the eigen-decomposition to the covariance matrix ${\bf R}$;  \\
5. & Derive the steering vectors of the sources from the signal-subspace of ${\bf R}$ as in (\ref{eq:hatA1});\\
6. & Compute $\hat{u}_{x,k}^{\rm fine}$ as in (\ref{eq:ufine}); \\
7. & Derive the coarse estimates of arriving angles and then of the direction-cosines, \{$\hat{u}_{x,k}^{\rm coarse},\hat{u}_{y,k}^{\rm coarse}$\};\\
8. & Compute $\hat{u}_{y,k}^{\rm fine}$ as in (\ref{eq:vfine}) from the inter-sensor spacing phase-factors; \\
9. & Derive the {\em final} estimates of direction-cosines by disambiguating 
   the {\em fine} estimates;\\
10. & Compute the DOA of each source from (\ref{eq:hath1})-(\ref{eq:hath2}) and then
   the polarization parameters.\\
\hline
\end{tabular}
\label{table:sum}
\end{table}
Table \ref{table:sum} summarizes the ten steps to estimate the DOAs and the polarizations of the coherent-sources.
Note that the estimates of the direction-cosines are automatically paired for each source.
Compared with the PS algorithm, the proposed approach is based on the SS algorithm and it requires that the sub-arrays in the array-geometry are uniformly spaced.
However, if the PS algorithm is used to the array-geometry in this paper, only one-dimensional direction-cosine can be estimated.
Furthermore, the PS algorithm needs iterative search and the inter-sensor spacing should be less than a half-wavelength.
In contrast, the proposed algorithm 1) requires no iterative search, 2) is adopted in the bi-sparse array to reduce the mutual coupling, and 3) offers two-dimensional direction-cosines with only one dimensional spatial-smoothing.

It is worth nothing that from (\ref{eq:ufinal})-(\ref{eq:vfinal}), if ${\hat u}_{x,i}={\hat u}_{x,j} + n \frac{\lambda_{i,j}}{d_x}$, where $i, j$ denote the $i$th source and $j$th source, $n$ is an integer and $\lambda_{i,j}$ denotes the wavelength of the coherent sources ($i$th source and $j$th source), (\ref{eq:ufinal})-(\ref{eq:vfinal}) will not present the right estimates of ${\hat u}_{x,i}, {\hat u}_{x,j}$. In this source scenario, the covariance matrix ${\bf R}$ of ${\bf Z}$ in (\ref{eq:bfR}) will be rank-deficient, and following this the SS algorithm will break down.

\section{Extend the Proposed Algorithm to the Spatially-Spread Electromagnetic Vector Sensor Array}
\label{Sec:EMVSss}

\begin{figure}[ht!]
\centering
\includegraphics[scale=0.8]{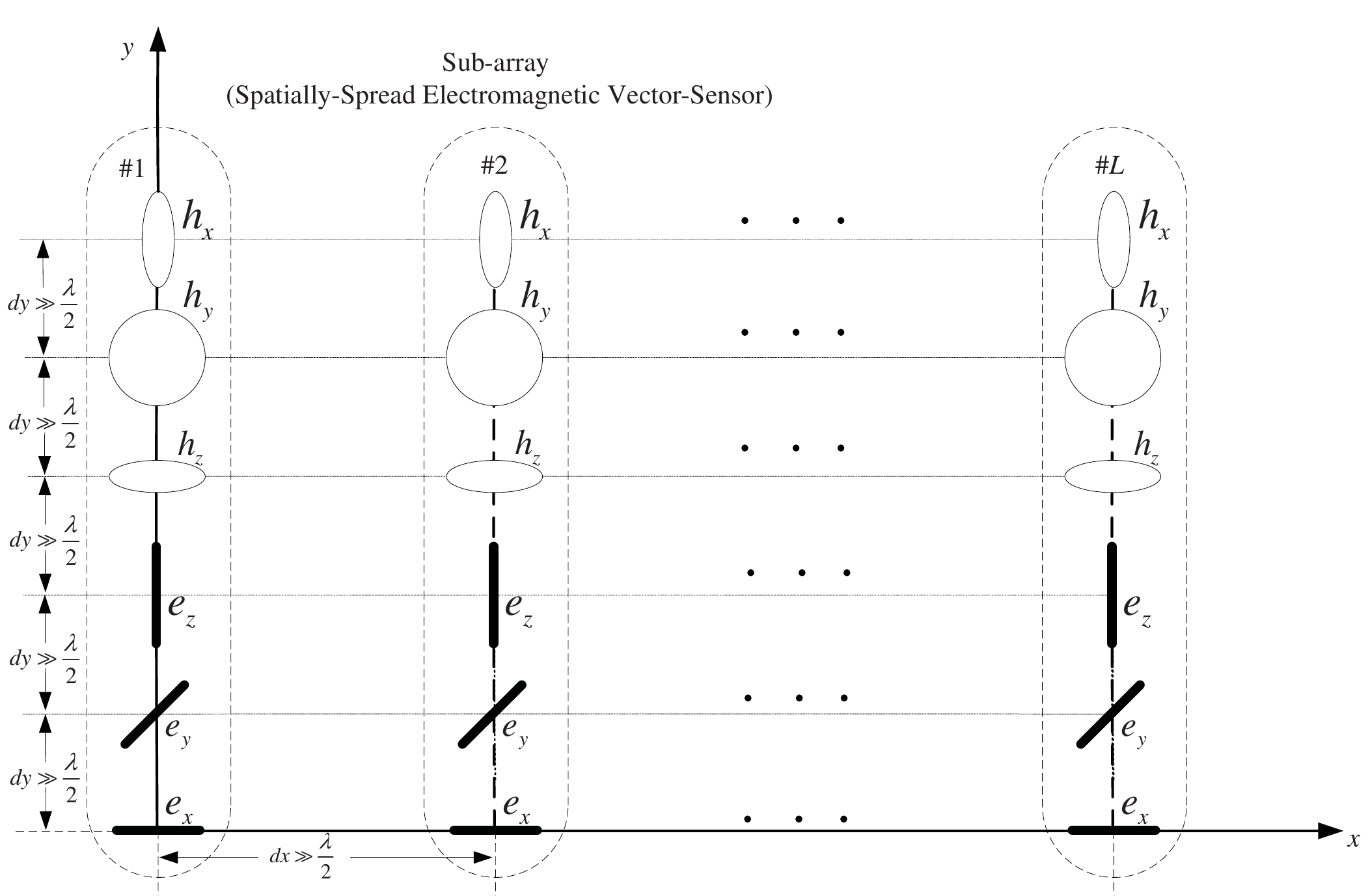}
\caption{The geometry of the bi-sparse array composed of spatially-spread electromagnetic vector-sensors \cite{SeeSPA03,SeeSSP03,WongKT_SPT0111}. The inter-sensor spacings among the dipoles/loops are all beyond $\lambda/2$.}
\label{EMVS-SS}
\end{figure}

The proposed algorithm can be adopted to the spatially-spread electromagnetic vector-sensors proposed in \cite{SeeSPA03,SeeSSP03,WongKT_SPT0111}.
Please refer to Figure \ref{EMVS-SS} for the array geometry composed of spatially-spread electromagnetic vector-sensors (EMVSs) in \cite{WongKT_SPT0111}, and the direction-finding algorithm proposed therein can be incorporated with the proposed algorithm in the present paper to resolve coherent sources.
Let the inter-sensor spacing among the dipoles and loops in the sub-array (the spatially-spread electromagnetic vector-sensor) be equal to $d_y$.
\footnote{The inter-sensor spacings can be different in the sub-array of Figure \ref{EMVS-SS}, only if to satisfy the conditions in \cite{WongKT_SPT0111}.}
The array-manifold of the sub-array corresponding to $k$th source is:
\begin{eqnarray}
{\bf a}_{{\rm sub},k}&\stackrel{\rm def}{=}&
\left[\begin{array}{r}
{\tilde {\bf e}}_k\\
{\tilde {\bf h}}_k \end{array}\right]\stackrel{\rm def}{=}\left[\begin{array}{r}
e_{x,k}\\
e^{-j\frac{2\pi u_{y,k}}{\lambda_k}d_y } e_{y,k} \\
e^{-j\frac{2\pi u_{y,k}}{\lambda_k}2d_y} e_{z,k} \\
e^{-j\frac{2\pi u_{y,k}}{\lambda_k}5d_y } h_{x,k} \\
e^{-j\frac{2\pi u_{y,k}}{\lambda_k}4d_y } h_{y,k} \\
e^{-j\frac{2\pi u_{y,k}}{\lambda_k}3d_y } h_{z,k}
\end{array}\right].
\end{eqnarray}
With the similar algorithm investigated in Section \ref{Sec:ag}, we can obtain $\hat{\bf a}_{{\rm sub},k}=c {\bf a}_{{\rm sub},k}$.
It follows that ${\hat{\tilde {\bf e}}}_k=c {\tilde {\bf e}}_k, {\hat{\tilde {\bf h}}}_k = c {\tilde {\bf h}}_k$.
The ``vector-cross-product" \cite{NehoraiSPT0294,WongKT_SPT0111} will be:
\begin{eqnarray}
{\hat{\tilde{\bf u}}}_k &=& \frac{{\hat{\tilde {\bf e}}}_k \times {\hat{\tilde {\bf h}}}^*_k}{\|{\hat{\tilde {\bf e}}}_k \times {\hat{\tilde {\bf h}}}^*_k\|}
= e^{j\frac{2\pi u_{y,k}}{\lambda_k}2d_y }\left[\begin{array}{r}
u_{x,k} \\
e^{j\frac{2\pi u_{y,k}}{\lambda_k}d_y } u_{y,k} \\
e^{j\frac{2\pi u_{y,k}}{\lambda_k}2d_y} u_{z,k}
\end{array}\right]
=\left[\begin{array}{l}
\cos\theta_{2,k}\cos\theta_{1,k}\hspace{0.05in} e^{j\frac{2\pi u_{y,k}}{\lambda_k}2d_y }\\
\cos\theta_{2,k}\sin\theta_{1,k}\hspace{0.05in}e^{j\frac{2\pi u_{y,k}}{\lambda_k}3d_y } \\
\sin\theta_{2,k}\hspace{0.53in}e^{j\frac{2\pi u_{y,k}}{\lambda_k}4d_y}
\end{array}\right]. \label{eq:vcp-ss}
\end{eqnarray}
where $^{*}$ denotes complex conjugation,
$\times$ symbolizes the vector cross-product operator,
$\|\cdot \|$ represents the Frobenius norm of the element inside $\|\hspace{0.05in}\|$,
and $\{u_{x,k},u_{y,k},u_{z,k}\}$ are the direction-cosines of the $k$th source align to $x$-axis, $y$-axis, $z$-axis respectively.

Adopt the proposed algorithm in Section \ref{Sec:ag} and from (\ref{eq:vcp-ss}),
\begin{enumerate}
\item[1)]
If $\theta_{1,k}\in[0,\pi]$, which means $u_{y,k}\ge0$, then:
\begin{eqnarray}
\acute{\bf u}_k &=& {\hat{\tilde{\bf u}}}_k e^{-j \angle{[{\hat{\tilde{\bf u}}}_k]_2}} =
\left[\begin{array}{l}
{\hat u}^{\rm coarse}_{x,k}  e^{-j\frac{2\pi u_{y,k}}{\lambda_k}d_y } \\
{\hat u}^{\rm coarse}_{y,k}   \\
{\hat u}_{z,k}  e^{j\frac{2\pi u_{y,k}}{\lambda_k}d_y }
\end{array}\right],\\
{\hat u}^{\rm coarse}_{y,k} &=& [\acute{\bf u}_k]_2, \\
{\hat u}^{\rm coarse}_{x,k} &=& [\acute{\bf u}_k]_1 e^{j\frac{2\pi {\hat u}^{\rm coarse}_{y,k}}{\lambda_k}d_y },\\
{\hat u}^{\rm fine}_{y,k} &=&  \frac{\lambda_k}{2\pi}\frac{1}{3d_y}\angle{[{\hat{\tilde{\bf u}}}_k]_2}.
\end{eqnarray}
\item[2)]
If $\theta_{1,k}\in(\pi,2\pi]$, which means $u_{y,k}<0$, then:
\begin{eqnarray}
\acute{\bf u}_k &=& -{\hat{\tilde{\bf u}}}_k e^{-j \angle{[{\hat{\tilde{\bf u}}}_k]_2}} =
\left[\begin{array}{l}
{\hat u}^{\rm coarse}_{x,k}  e^{-j\frac{2\pi {u}_{y,k}}{\lambda_k}d_y } \\
{\hat u}^{\rm coarse}_{y,k}   \\
{\hat u}_{z,k}  e^{j\frac{2\pi u_{y,k}}{\lambda_k}d_y }
\end{array}\right],\\
{\hat u}^{\rm coarse}_{y,k} &=& [\acute{\bf u}_k]_2, \\
{\hat u}^{\rm coarse}_{x,k} &=& [\acute{\bf u}_k]_1 e^{j\frac{2\pi {\hat u}^{\rm coarse}_{y,k}}{\lambda_k}d_y },\\
{\hat u}^{\rm fine}_{y,k} &=&  \frac{\lambda_k}{2\pi}\frac{1}{3d_y}\left(\angle[{\hat{\tilde{\bf u}}}_k]_2+\pi \right).
\end{eqnarray}
\end{enumerate}
The ${\hat u}^{\rm fine}_{x,k}$ can be obtained by the same method in Section \ref{Sec:ag} as shown in (\ref{eq:ufine}).
It follows that the same disambiguation approach can be adopted to get the final estimates of the direction-cosines and then the direction-of-arrivals.
After this, the polarizations of the sources can be estimated by the formulas derived in \cite{YuanAPT0512}.

In this array geometry and the following two array geometries, the conditions in (\ref{eq:Cond}) will be:
\begin{eqnarray}
6\ge K, \hspace{0.2in}\mbox{and} \hspace{0.2in} (L-P+1)\ge K,
\end{eqnarray}
since there are six antennas in each sub-array.

\section{Extend the Proposed Algorithm to the Sparsely Spaced Dipole-Loop Pairs}
\label{Sec:Pairss}

\begin{figure}[ht!]
\centering
\includegraphics[scale=0.8]{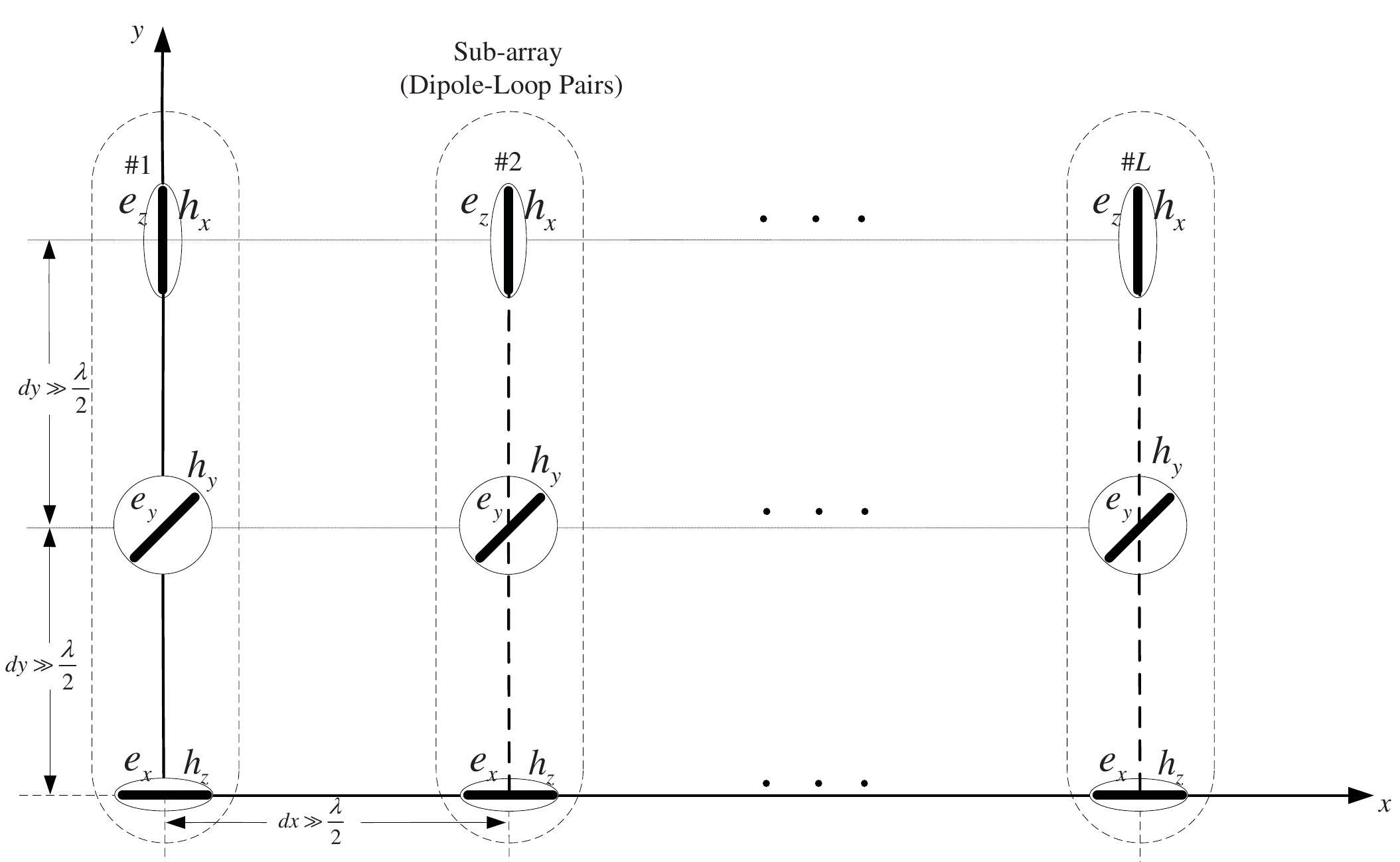}
\caption{The geometry of the bi-sparse array composed of sparsely spaced dipole-loop pairs. The inter-pair spacings among the pairs are all beyond $\lambda/2$.}
\label{pair-SS}
\end{figure}

The proposed algorithm can be adopted to the sparsely spaced dipole-loop pairs \cite{YuanAPT0512} as shown in Figure \ref{pair-SS}. The new bi-sparse array geometry in Figure \ref{pair-SS} is composed of dipole-loop pairs of orthogonal orientations. 
\footnote{Please note that this geometry is first proposed in this paper and it can be seen as a special case of the non-collocated electromagnetic vector-sensor proposed in \cite{WongKT_SPT0111}. The two inter-pair spacings can be different in the sub array of Figure \ref{pair-SS}.}
The array-manifold of the sub-array composed of dipole-loop pairs corresponding to $k$th source is:
\begin{eqnarray}
{\bf a}_{{\rm sub},k}&\stackrel{\rm def}{=}&
\left[\begin{array}{r}
{\tilde {\bf e}}_k\\
{\tilde {\bf h}}_k \end{array}\right]\stackrel{\rm def}{=}\left[\begin{array}{r}
e_{x,k}\\
e^{-j\frac{2\pi u_{y,k}}{\lambda_k}d_y } e_{y,k} \\
e^{-j\frac{2\pi u_{y,k}}{\lambda_k}2d_y} e_{z,k} \\
e^{-j\frac{2\pi u_{y,k}}{\lambda_k}2d_y } h_{x,k} \\
e^{-j\frac{2\pi u_{y,k}}{\lambda_k}d_y } h_{y,k} \\
h_{z,k}
\end{array}\right].
\end{eqnarray}
From the proposed algorithm in Section \ref{Sec:ag}, we can obtain $\hat{\bf a}_{{\rm sub},k}=c {\bf a}_{{\rm sub},k}$, where $c$ is an unknown complex number.
It follows that ${\hat{\tilde {\bf e}}}_k=c {\tilde {\bf e}}_k, {\hat{\tilde {\bf h}}}_k = c {\tilde {\bf h}}_k$.
The vector-cross-product result will be:
\begin{eqnarray}
{\hat{\tilde{\bf u}}}_k &=& \frac{{\hat{\tilde {\bf e}}}_k \times {\hat{\tilde {\bf h}}}^*_k}{\|{\hat{\tilde {\bf e}}}_k \times {\hat{\tilde {\bf h}}}^*_k\|}
= \left[\begin{array}{r}
e^{-j\frac{2\pi u_{y,k}}{\lambda_k}d_y }u_{x,k} \\
 u_{y,k} \\
e^{j\frac{2\pi u_{y,k}}{\lambda_k}d_y} u_{z,k}
\end{array}\right]. \label{eq:vcp-pair}
\end{eqnarray}
Then:
\begin{eqnarray}
{\hat u}^{\rm coarse}_{y,k} &=& [{\hat{\tilde{\bf u}}}_k]_2,\\
{\hat u}^{\rm coarse}_{x,k} &=& [{\hat{\tilde{\bf u}}}_k]_1 e^{j\frac{2\pi u^{\rm coarse}_{y,k}}{\lambda_k}d_y }
\end{eqnarray}
\begin{enumerate}
\item[1)]
If $\theta_{2,k}\in[0,\pi/2]$,
\begin{eqnarray}
{\hat u}^{\rm fine}_{y,k} &=&   \frac{\lambda_k}{2 \pi}\frac{1}{d_y}\angle[{\hat{\tilde{\bf u}}}_k]_3.
\end{eqnarray}
\item[2)]
If $\theta_{2,k}\in[-\pi/2, 0)$,
\begin{eqnarray}
{\hat u}^{\rm fine}_{y,k} &=&   \frac{\lambda_k}{2 \pi}\frac{1}{d_y}\left(\angle[{\hat{\tilde{\bf u}}}_k]_3+\pi \right).
\end{eqnarray}
\end{enumerate}
The ${\hat u}^{\rm fine}_{x,k}$ can be obtained by the same method in Section \ref{Sec:ag} as in (\ref{eq:ufine}).
It follows that the same disambiguation approach can be adopted to get the final estimates of the direction-cosines and then the direction-of-arrivals and the polarizations.

\section{Extend the Proposed Algorithm to the Sparsely Spaced Dipole-Triads or Loop-Triads}
\label{Sec:DLTss}

\begin{figure}
\centering
\includegraphics[scale=0.9]{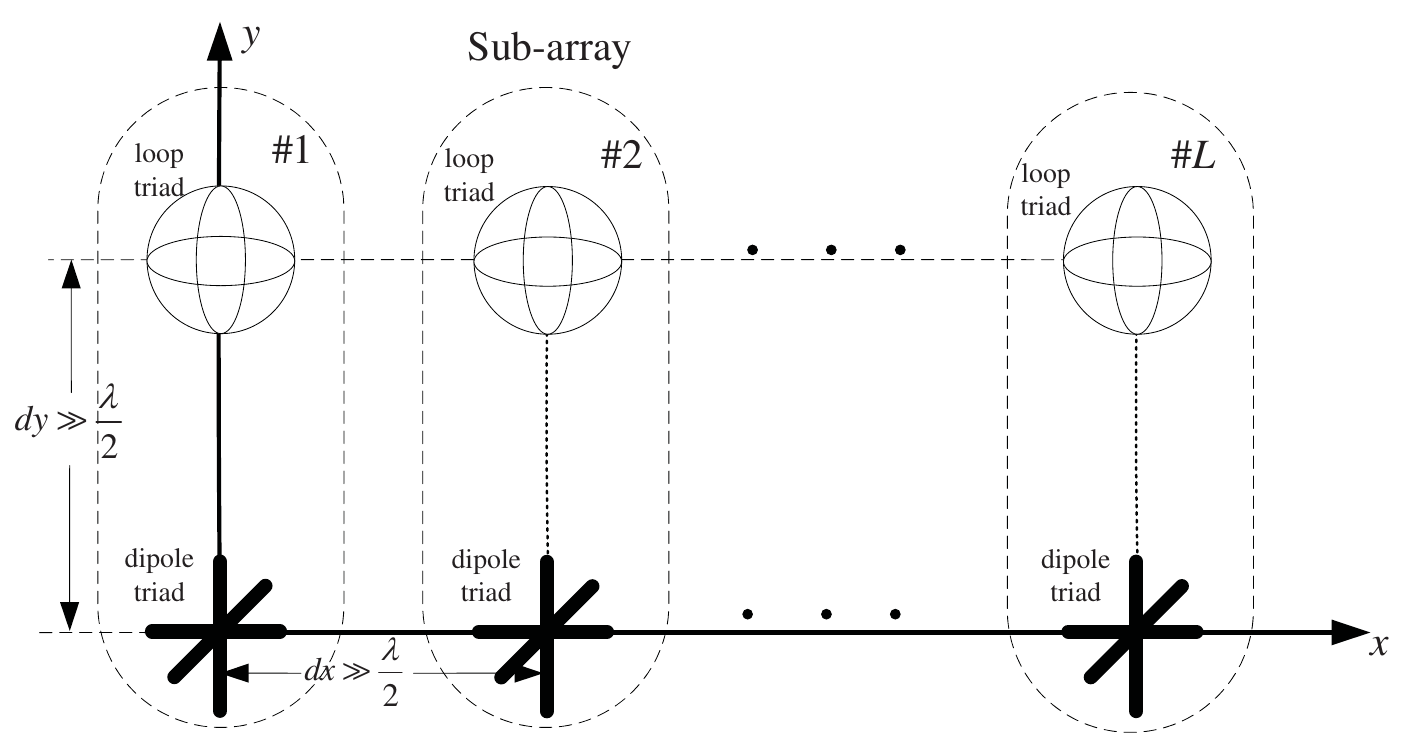}
\caption{The geometry of the bi-sparse triad array. One dipole-triad and one loop-triad comprise one sub-array. The inter-triad spacing $d_x\gg\frac{\lambda}{2}$, $d_y\gg\frac{\lambda}{2}$.}
\label{DLT-Sparse}
\end{figure}

Figure \ref{DLT-Sparse} depicts the bi-sparse array-geometry composed of dipole-triads and loop-triads.
The dipole-triad and loop-triad are displaced along the $y$-axis with a distance $d_y\gg\lambda/2$.
The two triads comprises a sub-array \cite{WongKT_AEST0401}, and $L$ sub-arrays are spaced along the $x$-axis with a distance $d_x\gg\lambda/2$.
The array-manifold of the sub-array corresponding to $k$th source in Figure \ref{DLT-Sparse} is thus \cite{WongKT_AEST0401}:
\begin{eqnarray} \label{eq:asub}
{\bf a}_{{\rm sub},k}&\stackrel{\rm def}{=}&\left[\begin{array}{r}
{\bf e}_k\\
q_{y,k}{\bf h}_k
\end{array}\right],
\end{eqnarray}
where $q_{y,k}=e^{-j\frac{2\pi}{\lambda_k} u_{y,k} d_y}$.
From (\ref{eq:asub}), using the vector-cross-product \cite{WongKT_AEST0401}:
\begin{eqnarray} \label{eq:vcpq}
\tilde{\bf u}_k &=&\frac{{\bf e}_k \times (q_{y,k}{\bf h}_k)^{*}}{\left\|
                 {\bf e}_k \times (q_{y,k}{\bf h}_k)^{*} \right\|}
\hspace{0.08in}=\hspace{0.08in}q_{y,k}^*{\bf u}_k
=
q_{y,k}^*\left[\begin{array}{c}
u_{x,k}\\
u_{y,k}\\
u_{z,k}
\end{array}\right].
\end{eqnarray}

From the proposed algorithm in Section \ref{Sec:ag}, we can obtain $\hat{\bf a}_{{\rm sub},k}=c {\bf a}_{{\rm sub},k}$..
It follows that ${\hat{ {\bf e}}}_k=c {{\bf e}}_k, {\hat{{\bf h}}}_k = c {{\bf h}}_k$.
From the vector-cross product result, we can get:
\begin{eqnarray} \label{eq:vcpcoar}
{\hat{\tilde{\bf u}}}_k &=&\frac{\hat{\bf e}_k \times \hat{\bf h}_k^{*}}{\left\|
                 \hat{\bf e}_k \times \hat{\bf h}_k^{*} \right\|}
= \hat{q}_{y,k}^*
\left[\begin{array}{c}
\hat{u}_{x,k}\\
\hat{u}_{y,k}\\
\hat{u}_{z,k}
\end{array}\right]=\hat{q}_{y,k}^* \hat{{\bf u}}_k.
\end{eqnarray}
Note that (\ref{eq:vcpcoar}) will offer both the {\em coarse} estimates of the sources' direction-cosines $\hat{{\bf u}}_k$, and the
{\em fine} estimates of the sources' direction-cosines along the $y$-axis, $\hat{u}_{y,k}^{\rm fine}$, from $\hat{q}_{y,k}$.
Separately consider the following two cases:
\begin{enumerate}
\item[1)]
If $\theta_{2,k}\in[0,\pi/2]$, which means $u_{z,k}\ge0$, then:
\begin{eqnarray}
{\hat {\bf u}_k}&=&{\hat{\tilde {\bf u}}}_k e^{-j\angle[{\hat{\tilde {\bf u}}}_k]_3}=\left[\begin{array}{c}
{\hat u}^{\rm coarse}_{x,k}\\
{\hat u}^{\rm coarse}_{y,k}\\
{\hat u}_{z,k}  \end{array}\right],\\
{\hat u}^{\rm fine}_{y,k} &=&   \frac{\lambda_k}{2 \pi}\frac{1}{d_y}\angle[{\hat{\tilde {\bf u}}}_k]_3. \label{eq:vfine1}
\end{eqnarray}
\item[2)]
If $\theta_{2,k}\in[-\pi/2, 0)$, which means $u_{z,k}\le0$, then:
\begin{eqnarray}
{\hat {\bf u}_k}&=&-{\hat{\tilde {\bf u}}}_k e^{-j\angle[{\hat{\tilde {\bf u}}}_k]_3}=\left[\begin{array}{c}
{\hat u}^{\rm coarse}_{x,k}\\
{\hat u}^{\rm coarse}_{y,k}\\
{\hat u}_{z,k}  \end{array}\right],\\
{\hat u}^{\rm fine}_{y,k} &=&  \frac{\lambda_k}{2 \pi}\frac{1}{d_y}\left(\angle[{\hat{\tilde {\bf u}}}_k]_3+\pi \right). \label{eq:vfine2}
\end{eqnarray}
\end{enumerate}
It follows that:
\begin{eqnarray}
{\hat u}^{\rm coarse}_{x,k} &=& [{\hat {\bf u}_k}]_1, \\
{\hat u}^{\rm coarse}_{y,k} &=& [{\hat {\bf u}_k}]_2.
\end{eqnarray}
The ${\hat u}^{\rm fine}_{x,k}$ can be obtained by the same computation as in (\ref{eq:ufine}).
It follows that the same disambiguation approach can be adopted to get the final estimates of the direction-cosines and then the direction-of-arrivals and the polarizations.

\begin{figure}
\centering
\includegraphics[scale=0.9]{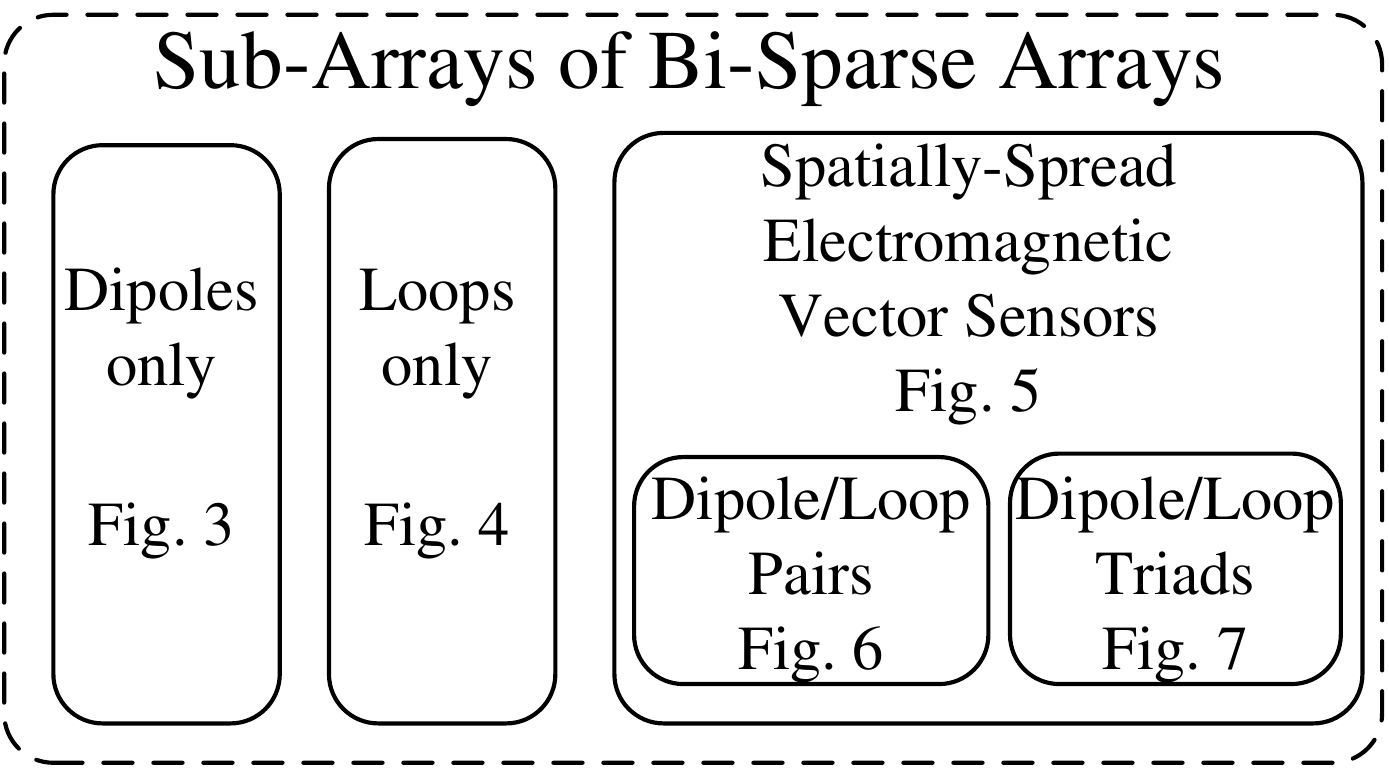}
\caption{Depiction of the relationship for the proposed bi-sparse arrays based on the  used sub-arrays. Note that the dipole/loop pair array in Figure \ref{pair-SS} and dipole/loop triad array in Figure \ref{DLT-Sparse} can be seen as special case of the spatially-spread electromagnetic vector-sensors in Figure \ref{EMVS-SS} by collocating some sensors.}
\label{Scheme}
\end{figure}

Based on the sub-arrays used in the proposed bi-sparse arrays, we depict the relationship of these arrays in Figure \ref{Scheme}. Note that the dipole/loop triads/pairs can be seen as special cases of the spatially-spread electromagnetic vector-sensors (SS-EMVS) in Figure \ref{EMVS-SS} \cite{WongKT_SPT0111}.

\section{Cram\'{e}r-Rao Bound Derivation}\label{Sec:CRB}
A far-field unit-power pure-tone is used in this section for the following Cram\'{e}r-Rao bound (CRB) derivation.
Provided that $s(t)= e^{j(2\pi f_0 t + \epsilon)}$ with a prior-known frequency $f_0$ and a prior-known initial phase $\epsilon$,
with $N$ snapshots uniformly sampled at time-slots $\{t=t_1,t_2,\cdots,t_N\}$, we have:
\begin{eqnarray}
{\bf s} &=& \left[e^{j(2\pi f_0 t_1 + \epsilon)}, e^{j(2\pi f_0 t_2 + \epsilon)},\cdots, e^{j(2\pi f_0 t_N + \epsilon)}\right]^T.
\end{eqnarray}
The above sequence is received by the proposed arrays, corrupted with additive noise ${\bf n}(t)$, which is assumed to be zero-mean Gaussian, with its diagonal covariance matrix ${\bf \Gamma}_0={\rm diag}[\sigma^2,\dots,\sigma^2]$, where $\sigma^2$ refers to the prior-known noise variance at each constituent antenna, ${\bf z}(t)= {\bf a} s(t) + {\bf n}(t)$.
For the arrays in Figures \ref{DT-SS}-\ref{LT-SS}, the following $4NL \times 1$ data-vector could be acquired
\begin{eqnarray}\label{eq:bfZ}
{\zetab}&=&\left[{\bf z}^T(t_1),{\bf z}^T(t_2),\cdots, {\bf z}^T(t_N)\right]^T \nonumber\\
&=&\underbrace{{\bf s}\otimes {\bf a}}_{\stackrel{\rm def}{=}{\alphab}} + \underbrace{\left[{\bf n}^T(t_1),{\bf n}^T(t_2),\cdots, {\bf n}^T(t_N),\right]^T}_{\stackrel{\rm def}{=}{\betab}},
\end{eqnarray}
where $\otimes$ denotes the Kronecker product, ${\betab}$ is the noise vector with a covariance matrix ${\bf \Gamma}={\bf \Gamma}_0 \otimes {\bf I}_{N}$, with ${\bf I}_{N}$ denoting an $N\times N$ identity matrix.
Hence, ${\zetab}$ is a complex Gaussian distributed process with mean ${\alphab}$ and a covariance matrix ${\bf \Gamma}$.
Let
\begin{eqnarray}
\psib&\stackrel{\rm def}{=}&[\theta_1,\theta_2,\theta_3,\theta_4]^T
\end{eqnarray}
refer to the vector comprising all the concerned unknown parameters.  We could derive all the elements of the $4\times 4$ Fisher Information Matrix (FIM) by \cite{VanTrees02}:
\begin{eqnarray} \label{eq:Jij}
J_{[\psib]_i,[\psib]_j}&=& 2{\rm Re} \left[\left(\frac{\partial {\alphab}}{\partial [\psib]_i} \right)^H
{\bf \Gamma}^{-1}
                \left(\frac{\partial {\alphab}}{\partial [\psib]_j }\right)\right],
                ~\forall i,j=1,2,3,4,
\end{eqnarray}
where $J_{[\psib]_i,[\psib]_j}$ refers to the $(i,j)$th entry of the FIM, and ${\rm Re}[.]$ denotes the real-value part of the entry inside $[\hspace{0.05in}]$.
Then the Cram\'{e}r-Rao bounds of $\psib$ equal:
\begin{eqnarray} \label{eq:CRBall}
\mbox{CRB}([\psib]_i)&=& \left[{\bf J}^{-1}\right]_{i,i}, \hspace{0.1in} \forall i=1,2,3,4.
\end{eqnarray}

The $4\times 4$ Fisher Information Matrix can be expressed as:
\begin{eqnarray}
{\bf J}&=& \left[\begin{array}{cccc}
J_{\theta_1,\theta_1} & J_{\theta_1,\theta_2} & J_{\theta_1,\theta_3} & J_{\theta_1,\theta_4} \\
J_{\theta_2,\theta_1} & J_{\theta_2,\theta_2} & J_{\theta_2,\theta_3} & J_{\theta_2,\theta_4} \\
J_{\theta_3,\theta_1} & J_{\theta_3,\theta_2} & J_{\theta_3,\theta_3} & J_{\theta_3,\theta_4} \\
J_{\theta_4,\theta_1} & J_{\theta_4,\theta_2} & J_{\theta_4,\theta_3} & J_{\theta_4,\theta_4} \\
\end{array}\right].
\end{eqnarray}
The Cram\'{e}r-Rao bounds for each parameters can be obtained straightforwardly from (\ref{eq:CRBall}) after we get the values of ${\bf J}$ by (\ref{eq:Jij}).
Similarly, for all the other geometries, the Cram\'{e}r-Rao bounds can be derived\footnote{For the Gaussian source scenario, please refer to the derivation in \cite{YuanIET-RSN0712}.}.


\section{Monte Carlo Simulation}
\label{Sec:sim}

\begin{figure*}
\centering
\begin{minipage}{3in}
\includegraphics[height=6cm,width=8.0cm]{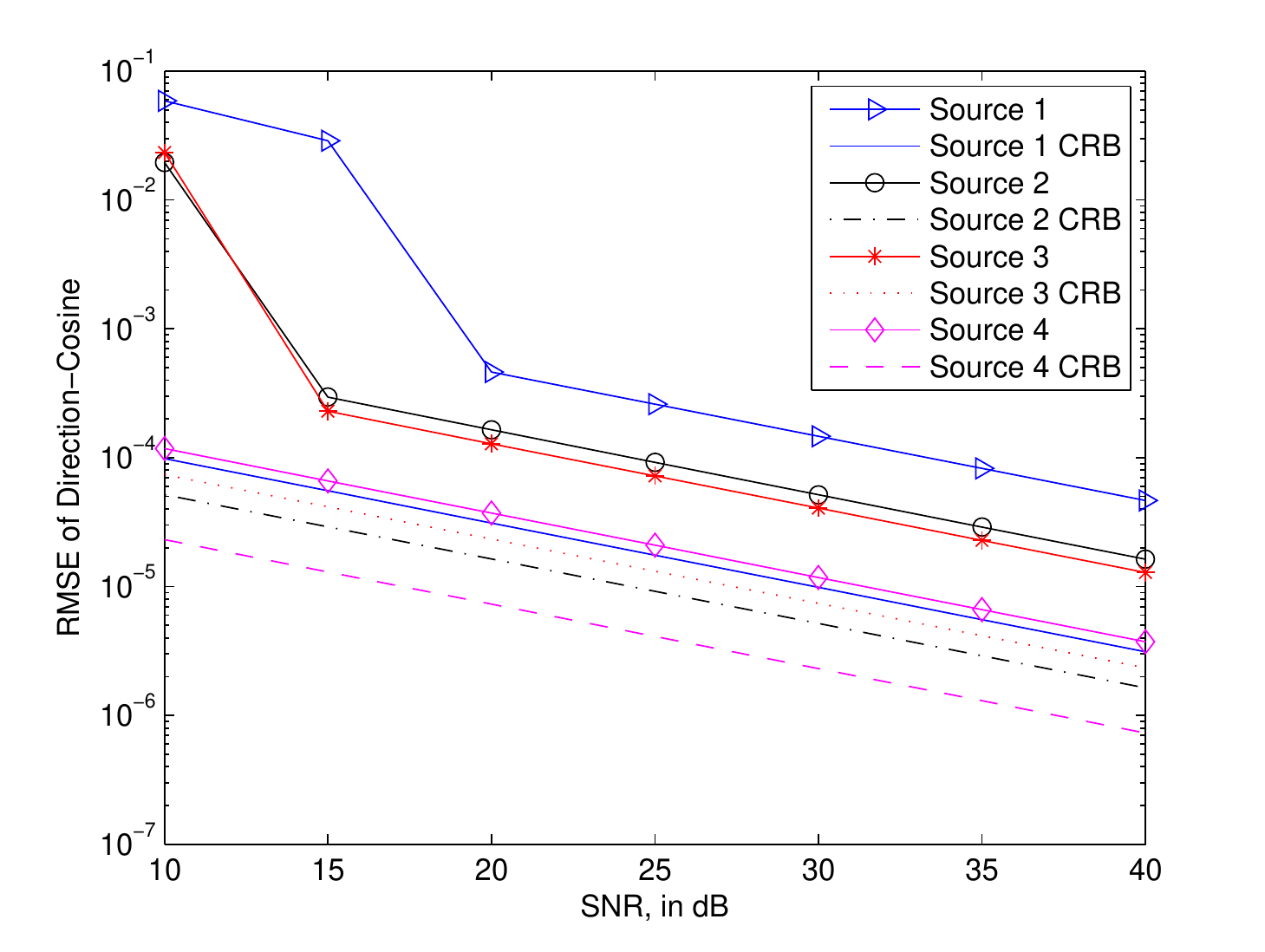}
\caption{Estimation RMSEs of direction-cosines versus SNR in a four-source scenario with the array geometry in Figure \ref{DT-SS}.}
\label{RMSESNRdtss}
\end{minipage}
    \hfill
\begin{minipage}{3in}
\includegraphics[height=6cm,width=8.0cm]{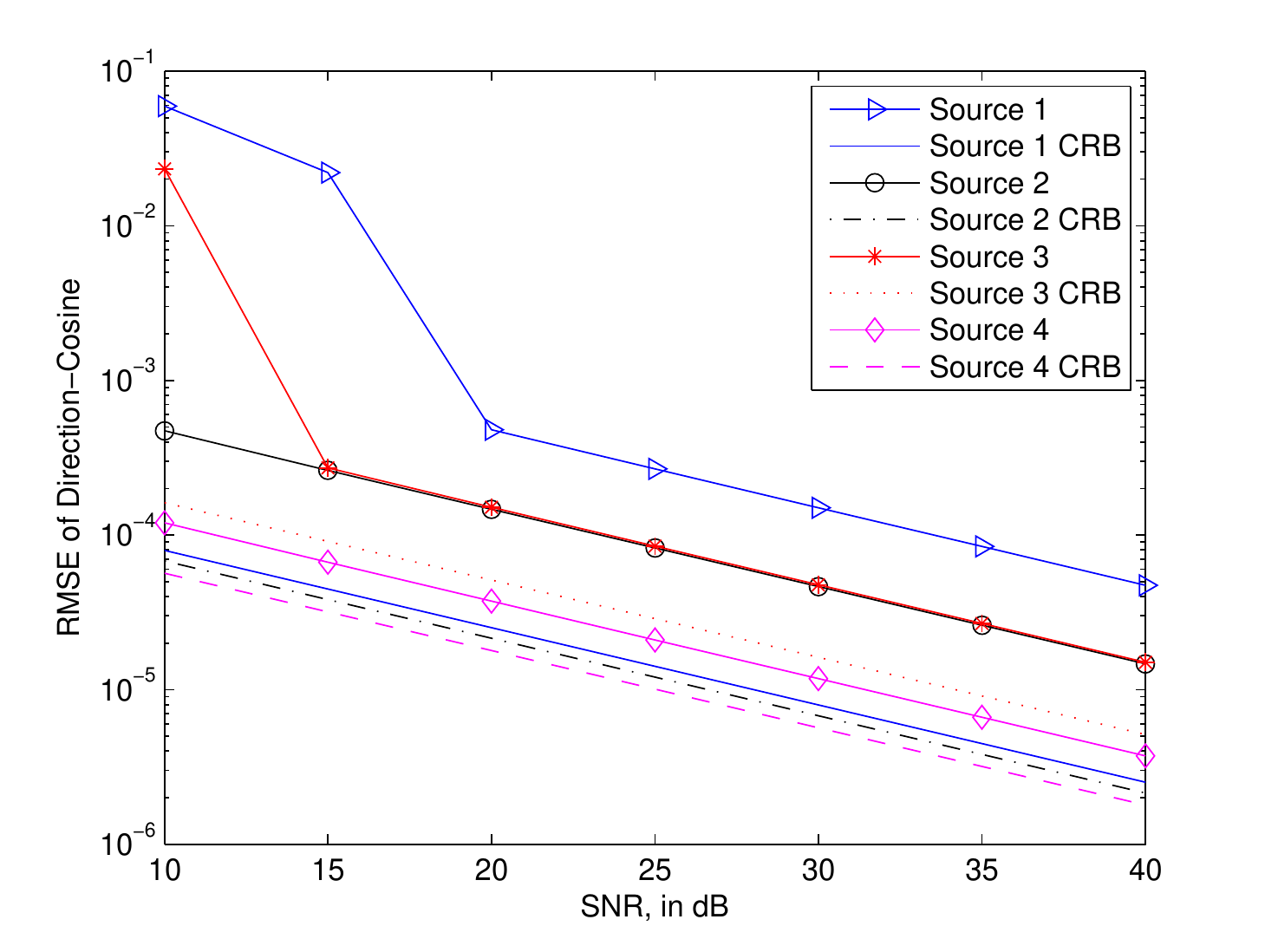}
\caption{Estimation RMSEs of direction-cosines versus SNR in a four-source scenario with the array geometry in Figure \ref{LT-SS}.}
\label{RMSESNRltss}
\end{minipage}
 \hfill
\begin{minipage}{3in}
\includegraphics[height=6cm,width=8.0cm]{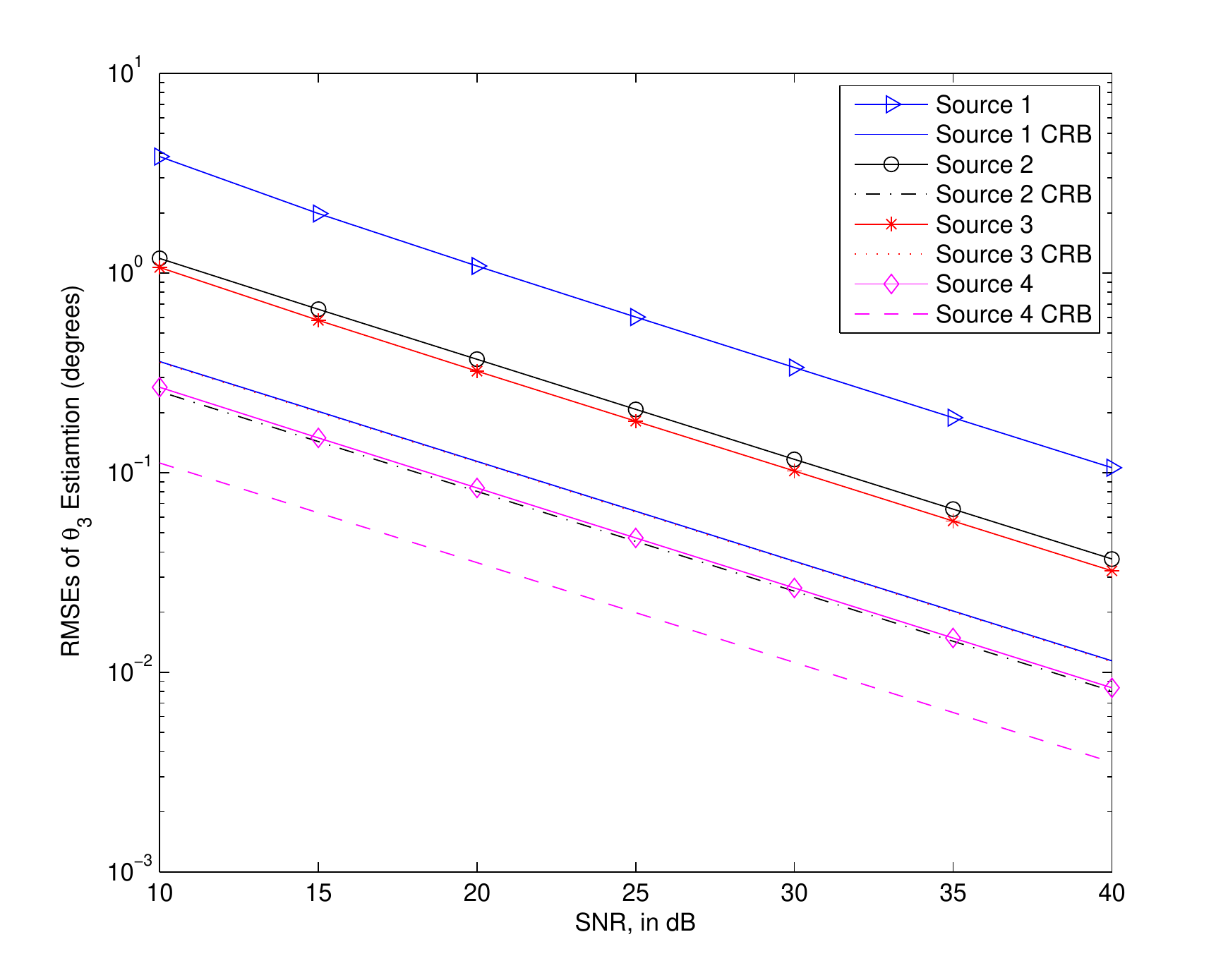}
\caption{Estimation RMSEs of $\{\theta_{3,1},\cdots,\theta_{3,4}\}$ versus SNR in a four-source scenario with the array geometry in Figure \ref{DT-SS}.}
\label{RMSEgamma}
\end{minipage}
 \hfill
\begin{minipage}{3in}
\includegraphics[height=6cm,width=8.0cm]{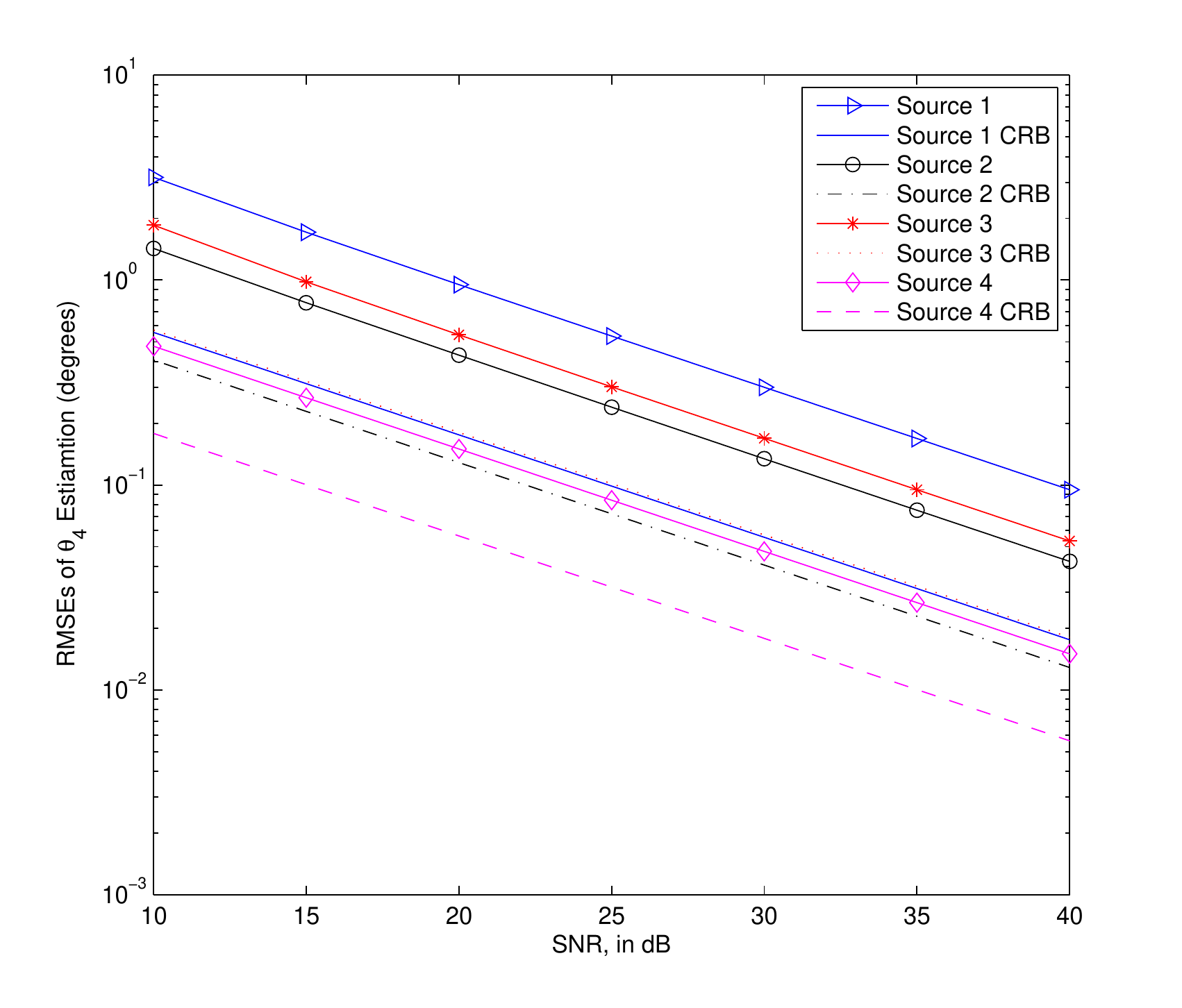}
\caption{Estimation RMSEs of $\{\theta_{4,1},\cdots,\theta_{4,4}\}$ versus SNR in a four-source scenario with the array geometry in Figure \ref{DT-SS}.}
\label{RMSEeta}
\end{minipage}
\label{RMSE-dltss}
\end{figure*}

\begin{figure}
\centering
\includegraphics[height=6cm,width=8.0cm]{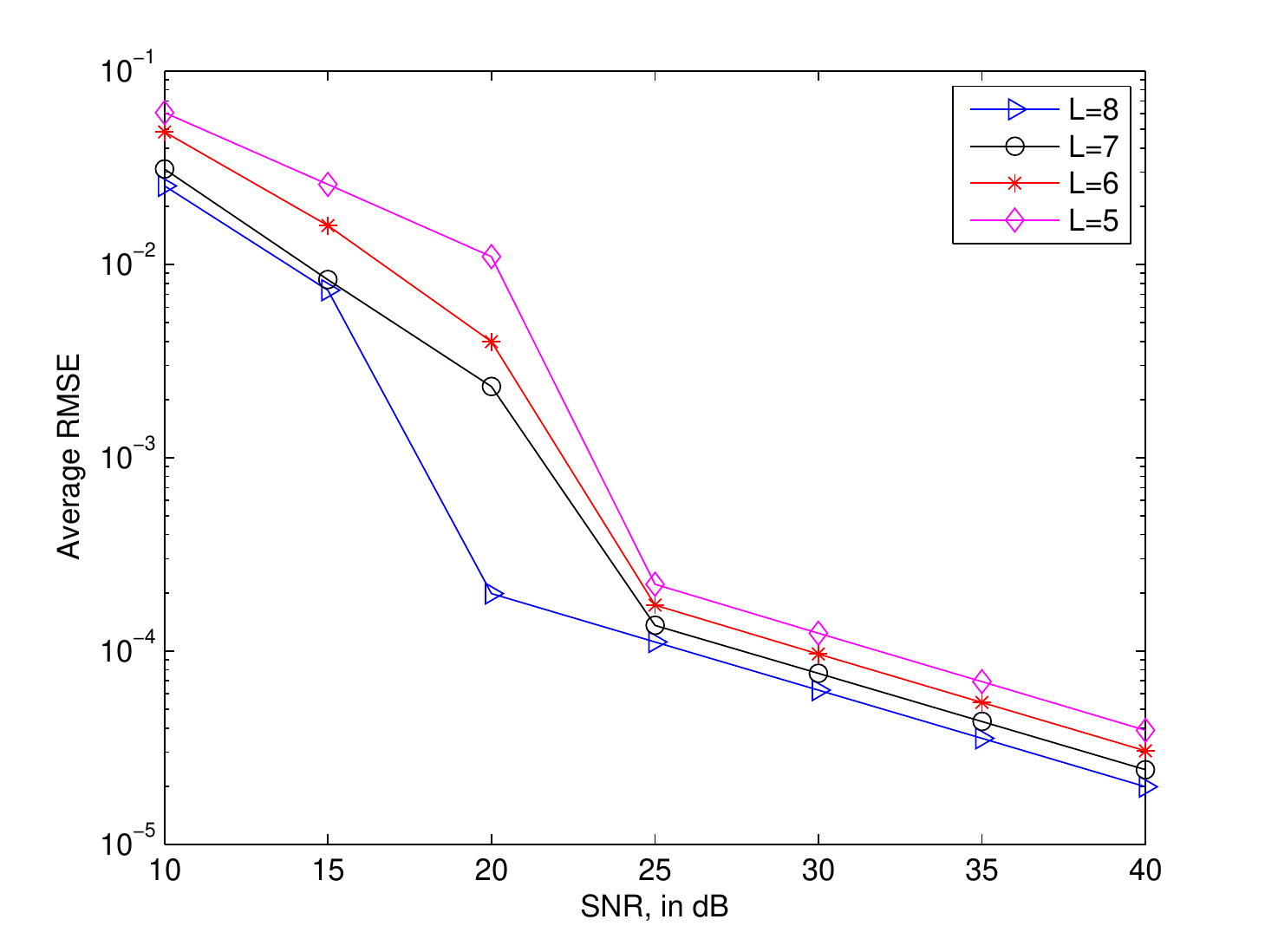}
\caption{The average RMSEs of direction-cosines versus SNR with different $L$ in a four-source scenario with the array geometry in Figure \ref{DT-SS}.}
\label{RMSESNRL}
\end{figure}

\begin{figure*}
\begin{minipage}{2.0in}
  \centering
  \centerline{\epsfig{figure=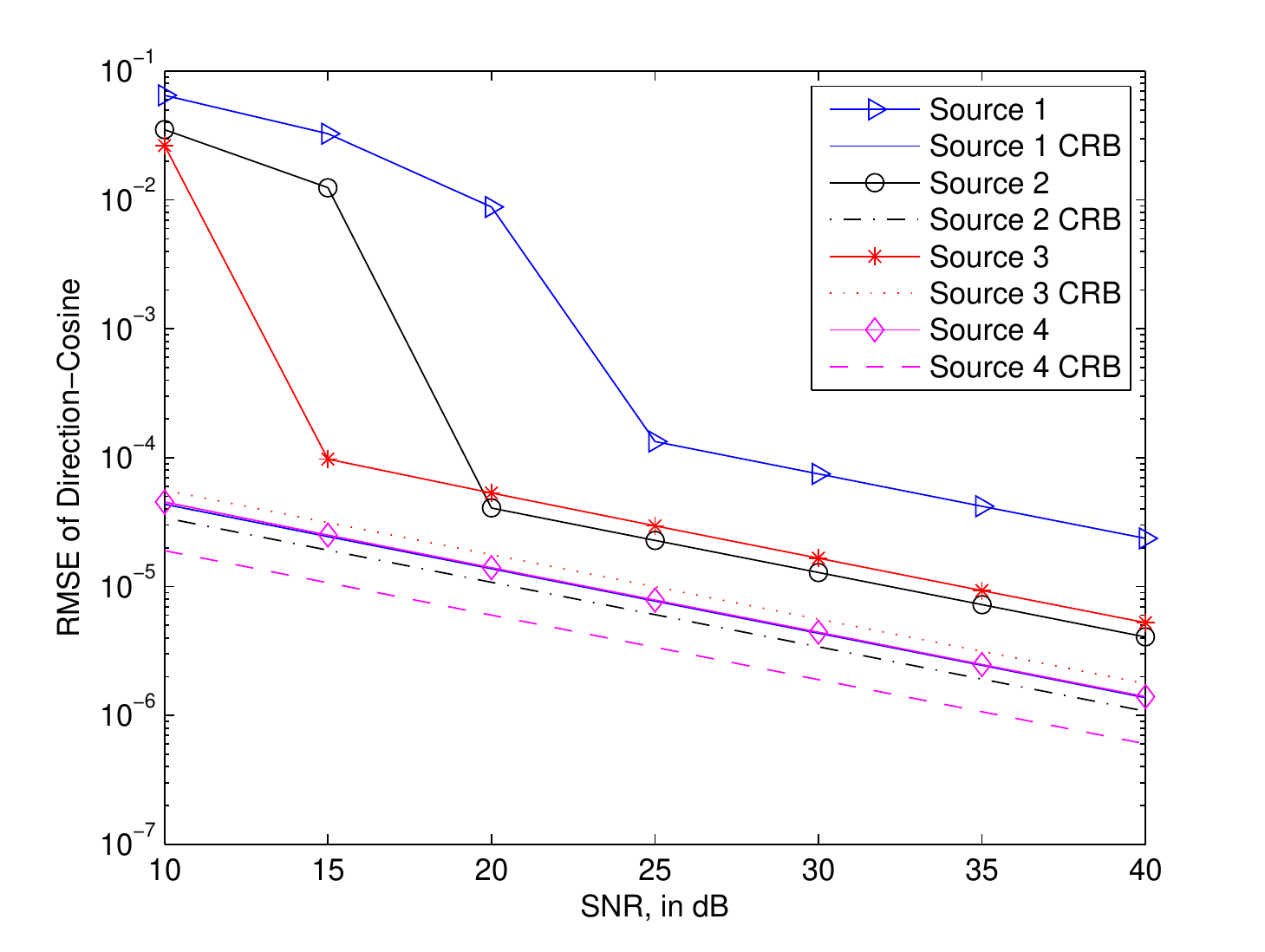,width=5.2cm,height=5cm}}
  \centerline{\footnotesize{(a) array geometry in Figure \ref{EMVS-SS}}}
\end{minipage}
\hspace{0.1in}
\begin{minipage}{2.0in}
  \centering
  \centerline{\epsfig{figure=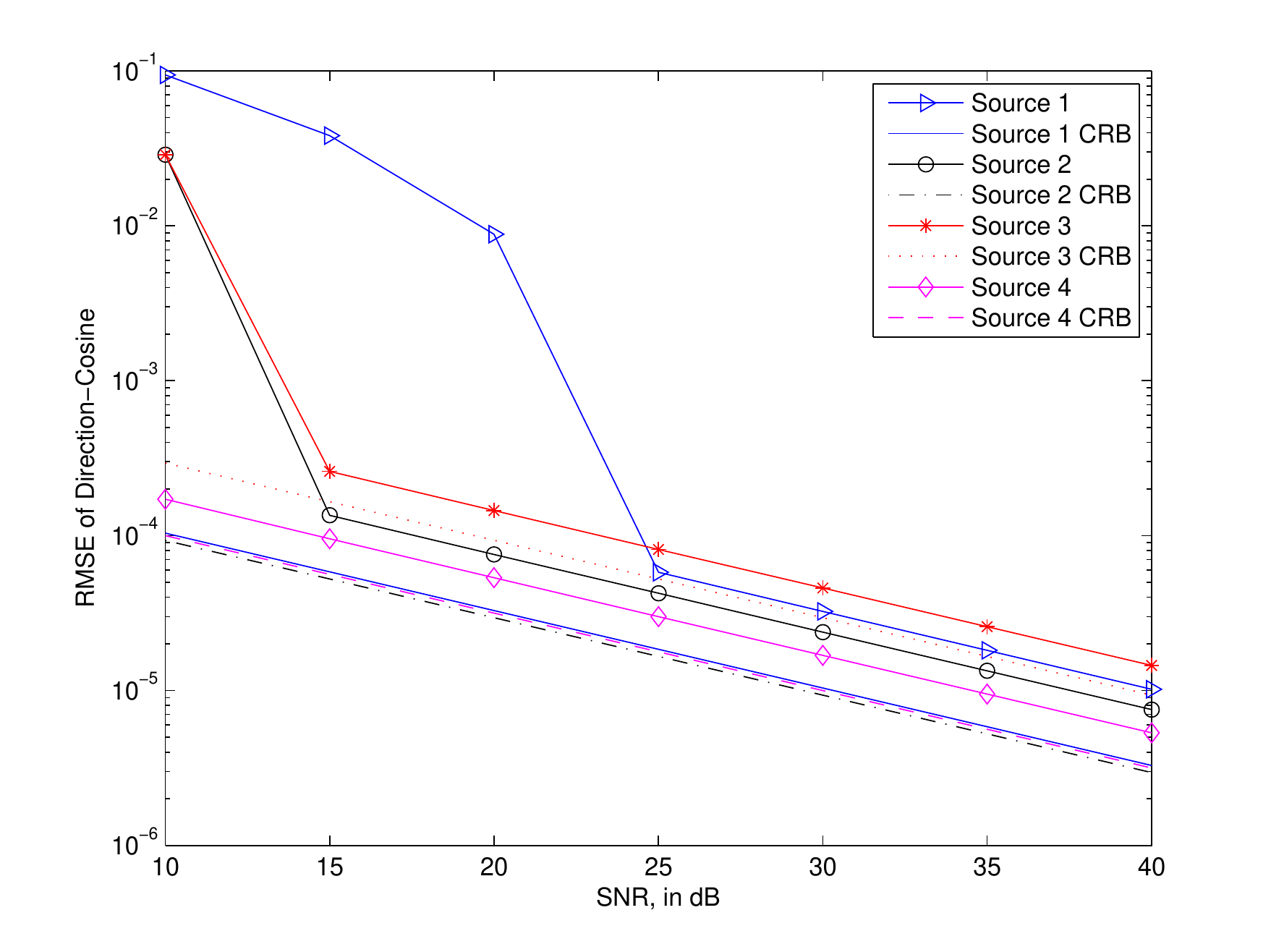,width=5.2cm,height=5cm}}
  \centerline{\footnotesize{(b) array geometry in Figure \ref{pair-SS}}}
\end{minipage}
\hspace{0.1in}
\begin{minipage}{2.0in}
  \centering
  \centerline{\epsfig{figure=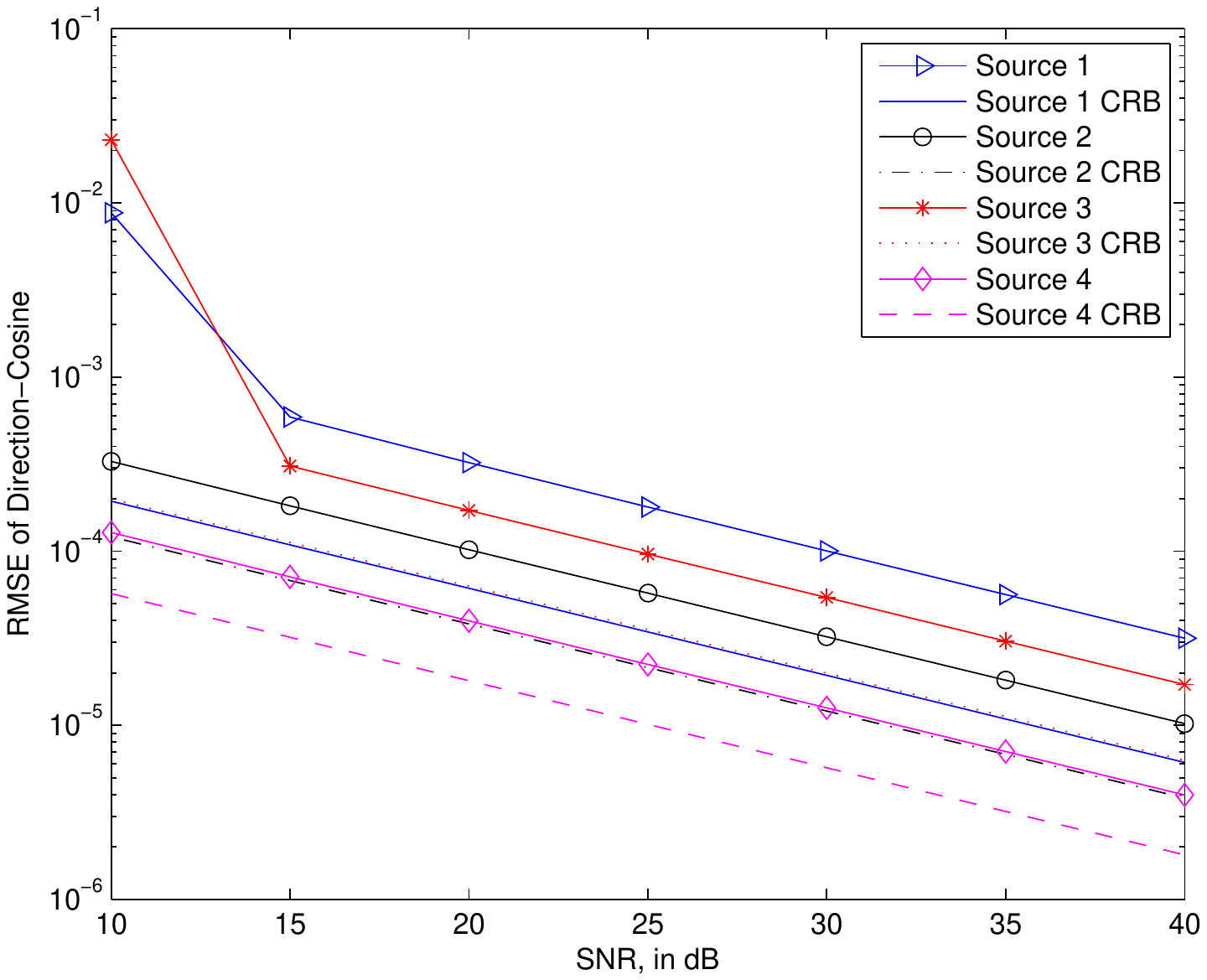,width=5.2cm,height=5cm}}
  \centerline{\footnotesize{(c) array geometry in Figure \ref{DLT-Sparse}}}
\end{minipage}
\caption{Estimation RMSEs of direction-cosines versus SNR in a four-source scenario with the array geometries in Figures \ref{EMVS-SS}-\ref{DLT-Sparse}.}
\label{RMSE-sspair}
\end{figure*}

The proposed algorithm's direction-finding efficacy and extended-aperture capability are demonstrated by Monte Carlo simulations.
The estimates use
$100$ temporal snapshots and $100$ independent
runs.
The root mean square error (RMSE) is utilized
as the performance measure. The RMSE for the direction-cosine is defined as:
\begin{eqnarray}
{\rm RMSE}&=& \sqrt{\frac{1}{100}\sum^{100}_{i=1}\left[({\hat u_x}^i - u_x)^2 + ({\hat u_y}^i - u_y)^2\right]}, \notag
\end{eqnarray}
where $\{{\hat u_x}^i,{\hat u_y}^i\}$ are the estimates of direction-cosines at $i$th run.
Figures \ref{RMSE-dltss} plots the RMSEs of the direction-cosines and polarization parameters versus signal-to-noise ratio (SNR) in a four-source scenario with the bi-sparse arrays proposed in Figures \ref{DT-SS}-\ref{LT-SS}.
The four sources share the same digital frequency $f_1=f_2=f_3=f_4=0.0895$ and
they are correlated with each other by ${\bf s}_2 =-{\bf s}_1, {\bf s}_3 =(1+j){\bf s}_1, {\bf s}_4 =(1-j){\bf s}_1$.
The DOAs and polarizations of the four sources are set as:
$(\theta_{1,1},\theta_{2,1},\theta_{3,1},\theta_{4,1})=(20^{\circ},15^{\circ},45^{\circ},90^{\circ})$,
$(\theta_{1,2},\theta_{2,2},\theta_{3,2},\theta_{4,2})=(53^{\circ},33^{\circ},45^{\circ},-90^{\circ})$,
$(\theta_{1,3},\theta_{2,3},\theta_{3,3},\theta_{4,3})=(81^{\circ},57^{\circ},45^{\circ},90^{\circ})$,
$(\theta_{1,4},\theta_{2,4},\theta_{3,4},\theta_{4,4})=(110^{\circ},70^{\circ},45^{\circ},-90^{\circ})$.
In the simulation, the inter-sensor spacing $d_x=d_y=8\lambda$.
$L=7$ sub-arrays are used and $P=2$.
Figure \ref{RMSE-dltss} clearly demonstrates that all the four correlated sources are accurately resolved by the proposed algorithm.
Figure \ref{RMSESNRL} presents the average RMSEs of the four sources with different $L$ at various SNR.
It can be seen that: a) The RMSEs decrease with the increasing $L$;
b) When SNR increases, this decreasing trend becomes weak.
Figure \ref{RMSE-sspair} plots the corresponding RMSEs of the direction-cosines versus signal-to-noise ratio (SNR) with the array geometries in Figures \ref{EMVS-SS}-\ref{DLT-Sparse} in the same four-source scenario as in Figure \ref{RMSE-dltss}.

\subsection{Aperture Extension Property Analysis}
\begin{figure*}
\begin{minipage}{3.3in}
  \centering
  \centerline{\epsfig{figure=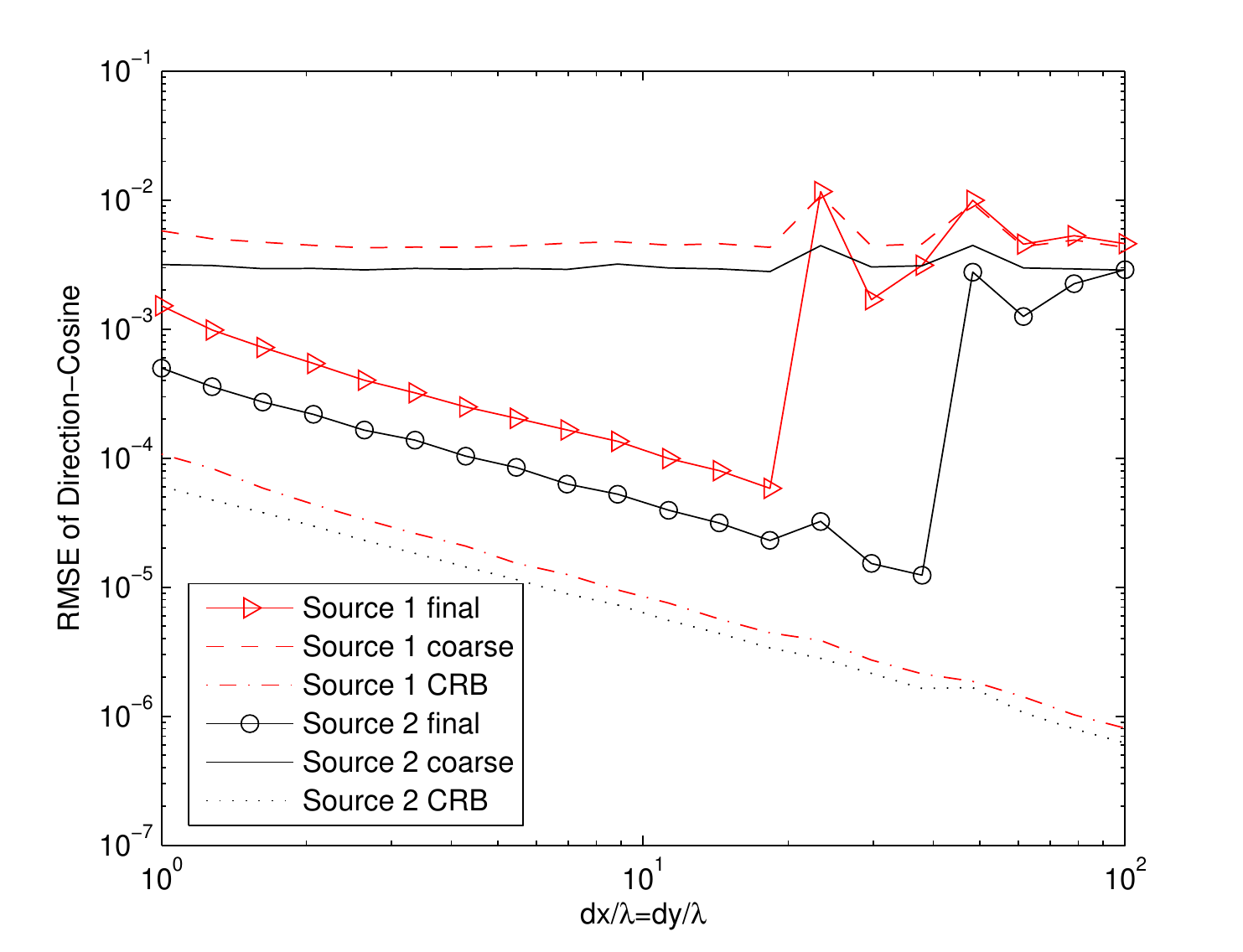,width=9.0cm,height=6cm}}
  \centerline{\footnotesize{(a) array geometry in Figures \ref{DT-SS}-\ref{LT-SS}}}
\end{minipage}
\hspace{0.05in}
\begin{minipage}{3.3in}
  \centering
  \centerline{\epsfig{figure=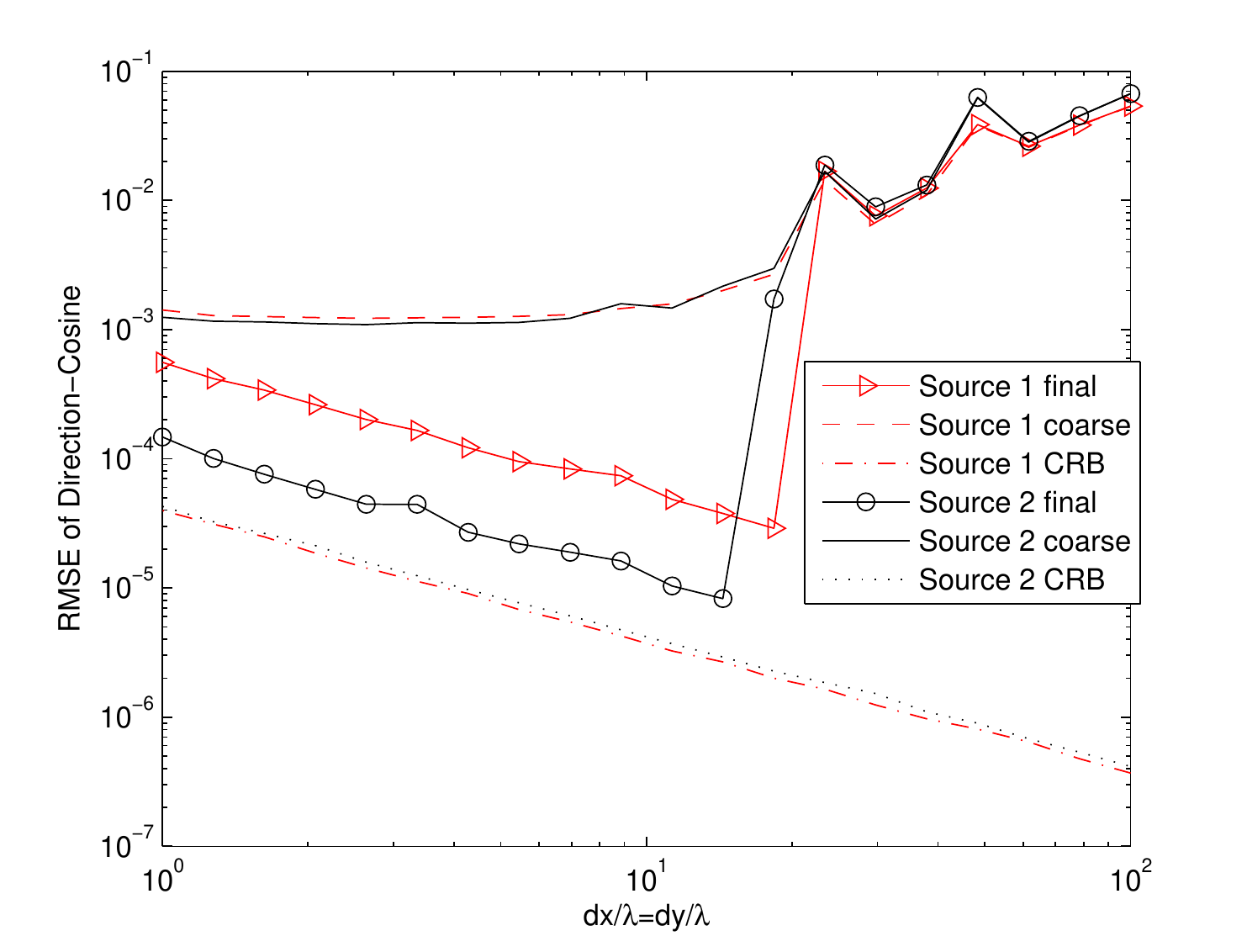,width=9.0cm,height=6cm}}
  \centerline{\footnotesize{(b) array geometry in Figure \ref{EMVS-SS}}}
\end{minipage}
\begin{minipage}{3.3in}
  \centering
  \centerline{\epsfig{figure=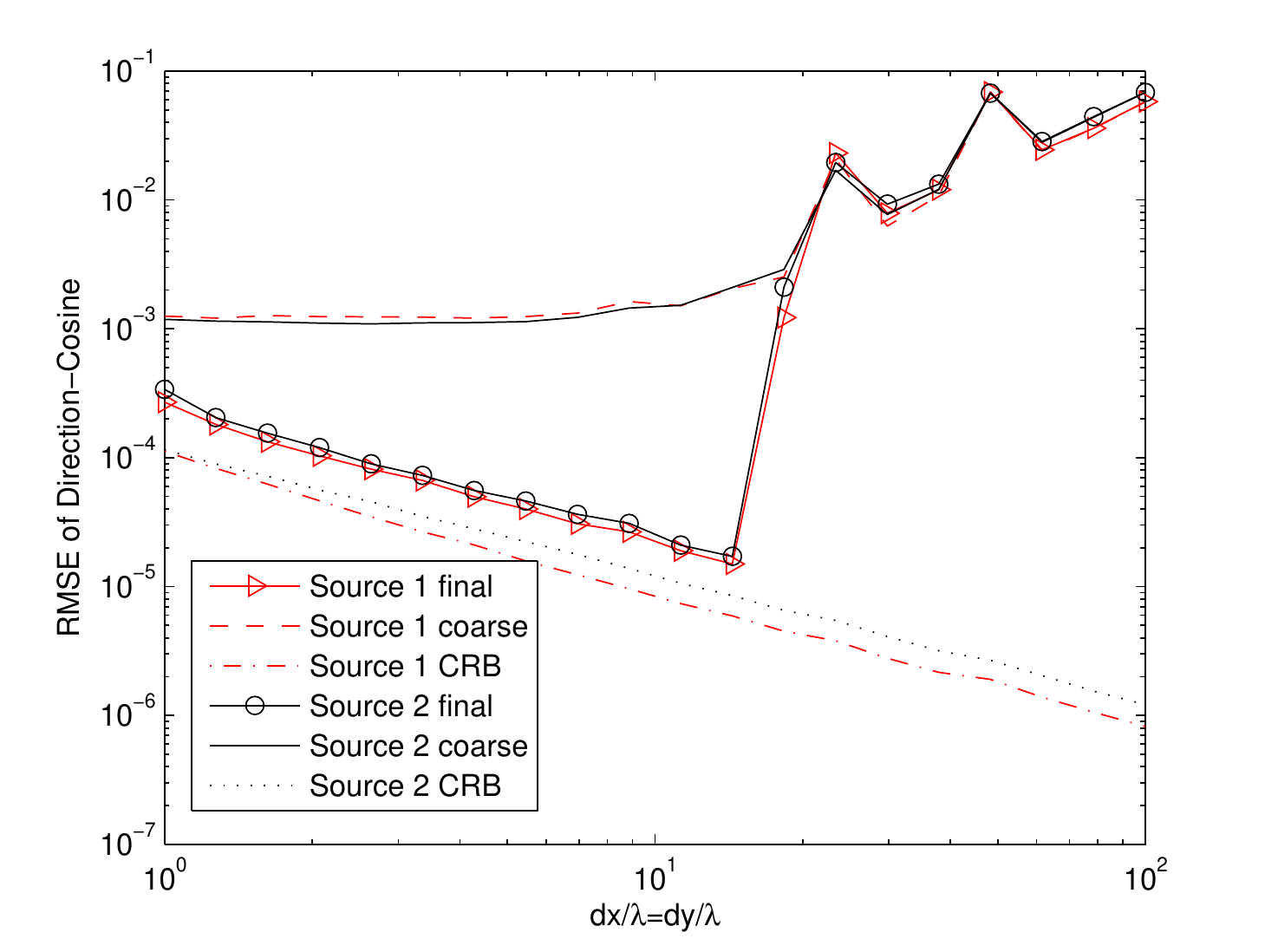,width=9.0cm,height=6cm}}
  \centerline{\footnotesize{(c) array geometry in Figure \ref{pair-SS}}}
\end{minipage}
\hspace{0.05in}
\begin{minipage}{3.3in}
  \centering
  \centerline{\epsfig{figure=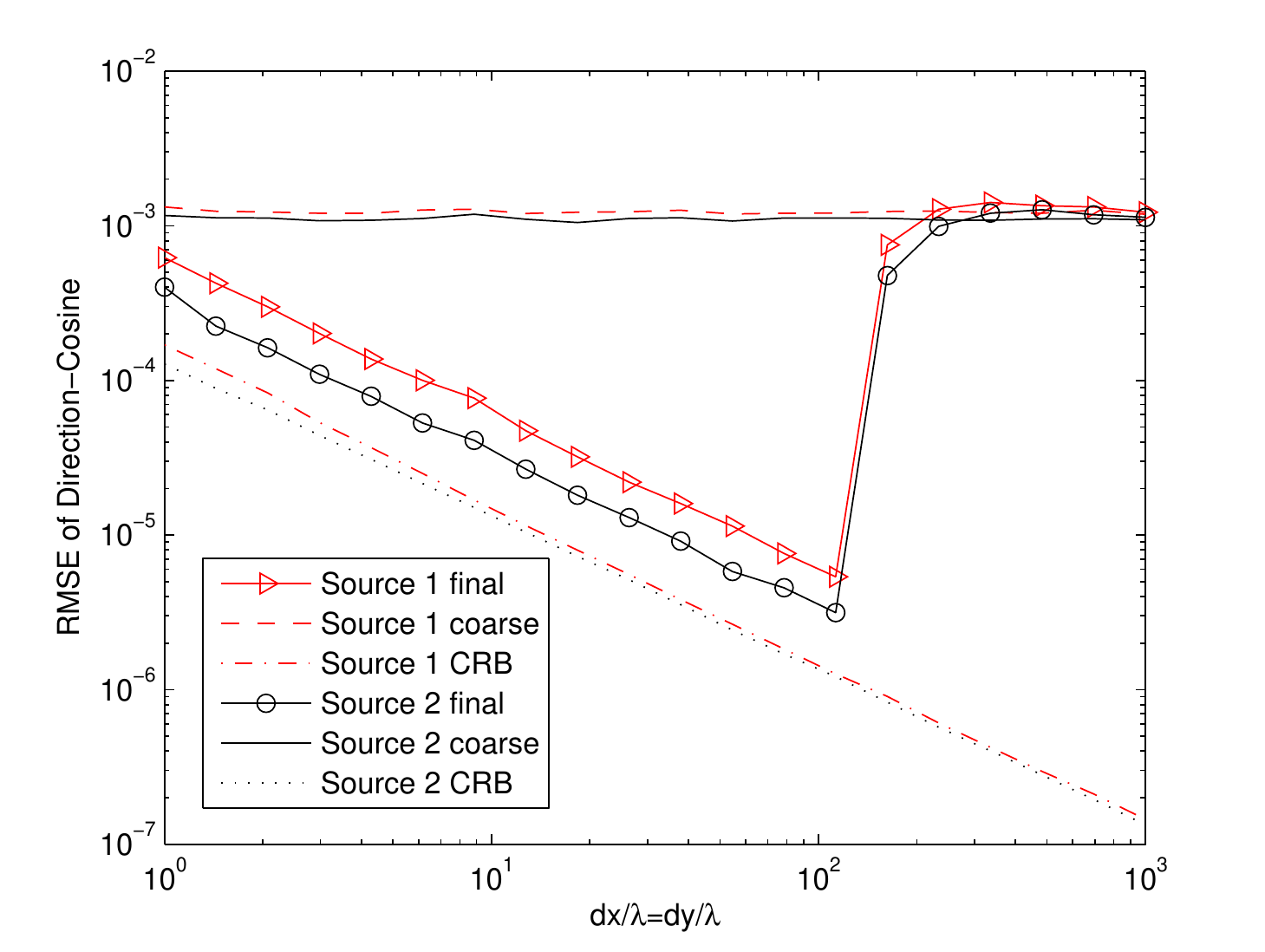,width=9.0cm,height=6cm}}
  \centerline{\footnotesize{(d) array geometry in Figure \ref{DLT-Sparse}}}
\end{minipage}
\caption{The RMSEs of direction-cosines versus inter-sensor spacing $d_x=d_y$ in a two-source scenario, at SNR$=30$dB with array geometries in Figures \ref{DT-SS}-\ref{DLT-Sparse}.}
\label{RMSE_kk}
\end{figure*}

It is well known that the larger the array-aperture, the better the angular resolution.
In order to investigate the aperture extension property of the proposed algorithm, Figure \ref{RMSE_kk} plots the
RMSEs of direction-cosines versus inter-sensor spacing $d_x=d_y$ in a two-source scenario, at SNR$=30$dB.
The two sources are set as ${\bf s}_2 =-{\bf s}_1$, and $(\theta_{1,1},\theta_{2,1},\theta_{3,1},\theta_{4,1})=(20^{\circ},15^{\circ},45^{\circ},90^{\circ})$,
$(\theta_{1,2},\theta_{2,2},\theta_{3,2},\theta_{4,2})=(55^{\circ},40^{\circ},45^{\circ},-90^{\circ})$.
Both the {\em coarse} estimates and the {\em final} estimates of the direction-cosines are plotted in the figure.
It can be seen that the RMSEs of direction-cosines estimated by the proposed algorithm decrease with the increase of inter-sensor spacing and they are close to the Cram\'{e}r-Rao bounds (CRB). For more investigations of the sparse array, please refer to \cite{Zoltowski1SPT0800,Zoltowski2SPT0800}.
It is notable that there is a breakdown phenomenon in Figure \ref{RMSE_kk}. When the inter-sensor spacing $d_x=d_y$ is beyond a specific spacing point, the RMSEs of the final estimates will be the same as the coarse estimates. This is because the coarse estimates will identify the wrong estimation grid at the pre-set SNR and thus it can not be used to disambiguate the fine estimates. For the details of this breakdown phenomenon, please refer to \cite{Zoltowski1SPT0800,Zoltowski2SPT0800}.
It can be found from Figure \ref{RMSE_kk} that this breakdown spacing point exists in all the proposed bi-sparse geometries.
For the array geometry in Figures \ref{DT-SS}-\ref{LT-SS}, it is about $20-40\lambda$.
For the array geometry in Figures \ref{EMVS-SS}-\ref{pair-SS}, it is about $15\lambda-20\lambda$.
For the array geometry in Figure \ref{DLT-Sparse}, it is $100\lambda$.
Thus, the array geometry in Figure \ref{DLT-Sparse} is recommended to be used to form a large-aperture sparse array.
On the other hand, if only the dipoles or loops are used, the array geometries in Figures \ref{DT-SS}-\ref{LT-SS} are recommended to be utilized in order to further reduce the mutual coupling since no antennas are collocated.

\subsection{Comparison with the Polarization Smoothing Algorithm}

\begin{figure*}
\begin{minipage}{2.0in}
  \centering
  \centerline{\epsfig{figure=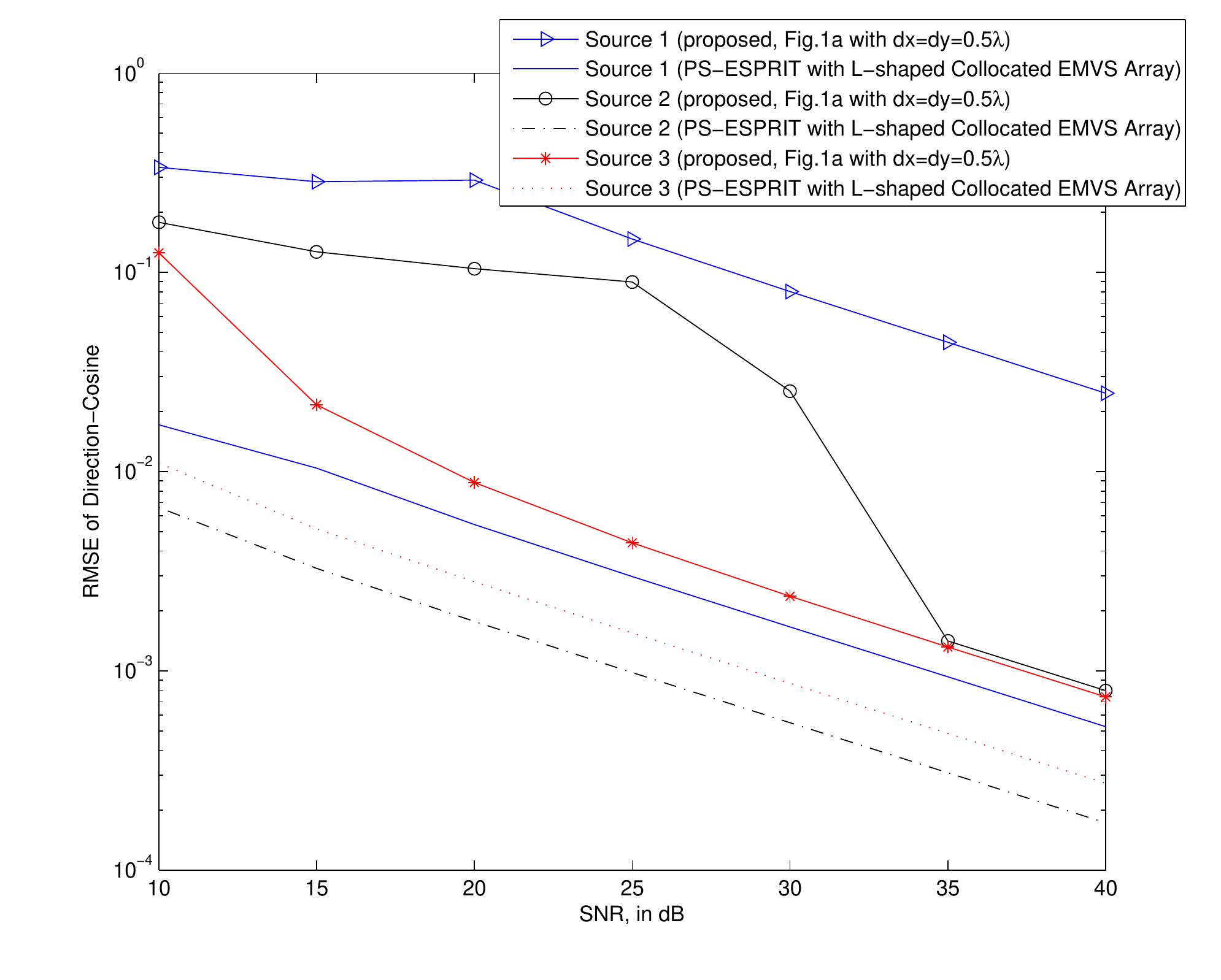,width=5.2cm,height=5cm}}
  \centerline{\footnotesize{(a) $d_x=d_y=0.5\lambda$}}
\end{minipage}
\hspace{0.1in}
\begin{minipage}{2.0in}
  \centering
  \centerline{\epsfig{figure=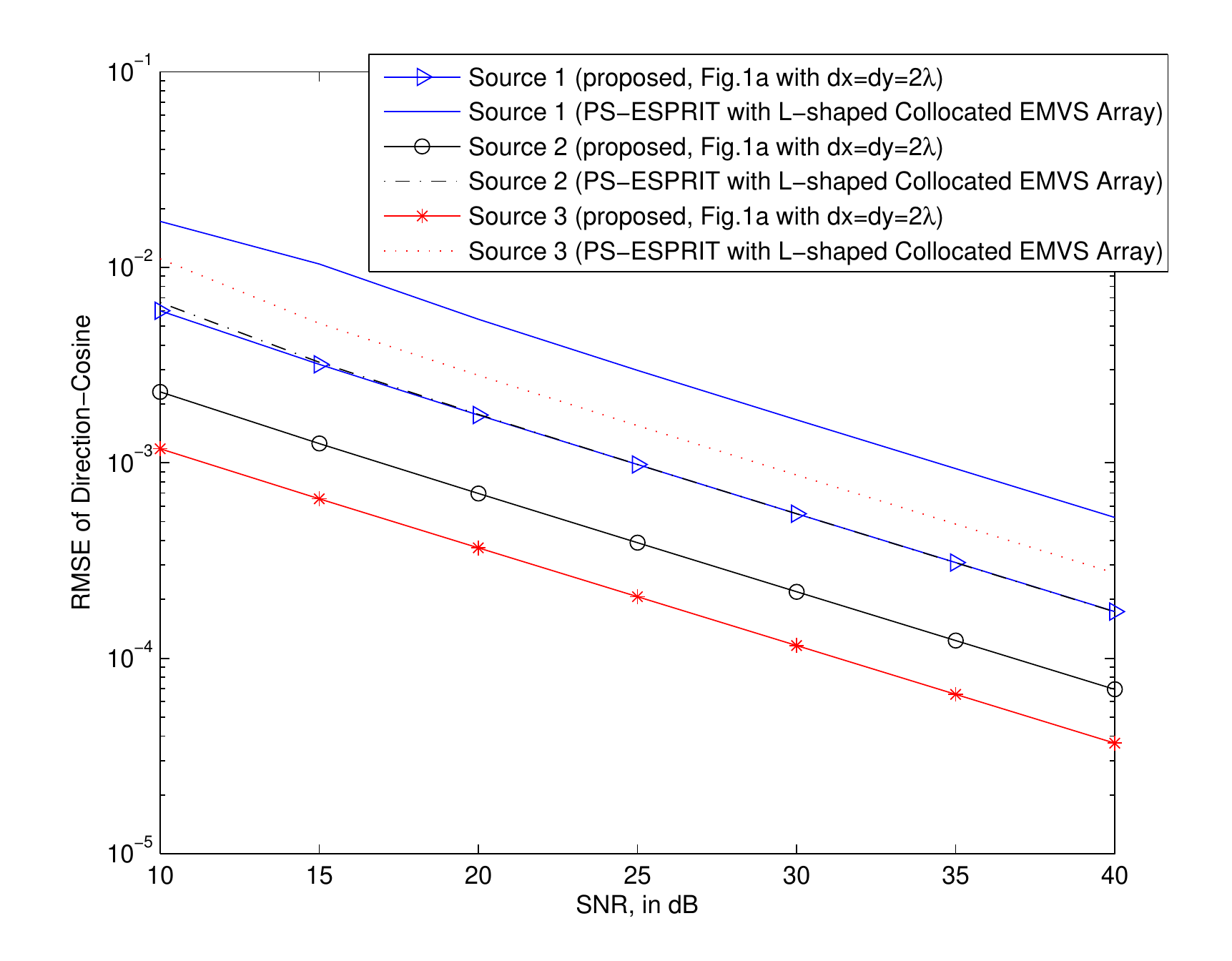,width=5.2cm,height=5cm}}
  \centerline{\footnotesize{(b) $d_x=d_y=2\lambda$}}
\end{minipage}
\hspace{0.1in}
\begin{minipage}{2.0in}
  \centering
  \centerline{\epsfig{figure=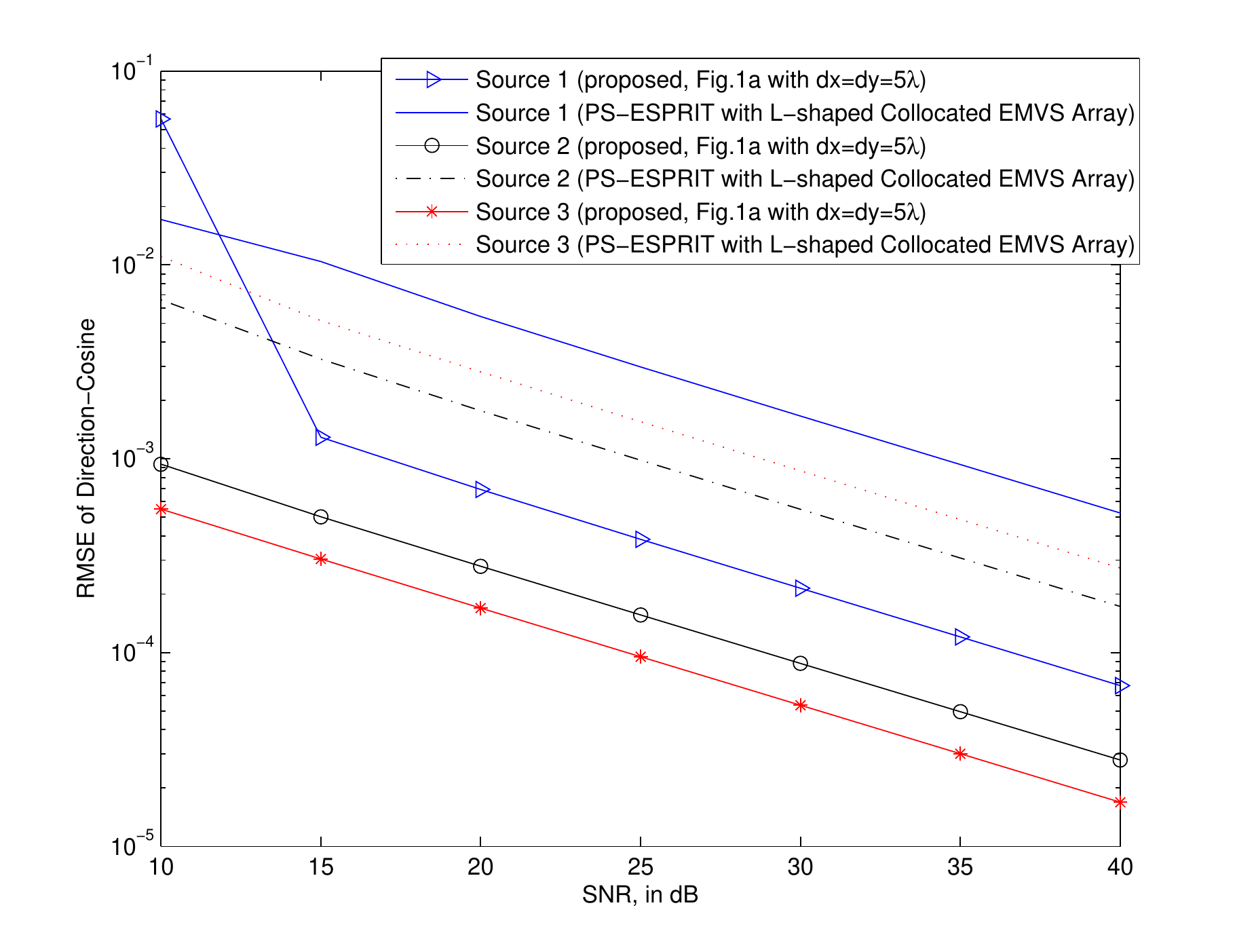,width=5.2cm,height=5cm}}
  \centerline{\footnotesize{(c) $d_x=d_y=5\lambda$}}
\end{minipage}
\caption{Estimation RMSEs of direction-cosines versus SNR in a three-source scenario, the PS-ESPRIT algorithm is adopted in an L-shaped array with $7$ collocated electromagnetic vector-sensors on each leg the the inter-sensor spacing is $\lambda/2$. The proposed algorithm is used in the demonstrated sparse array in Figure \ref{DT-SS} with different inter-sensor spacings.}
\label{compPS}
\end{figure*}

It is worth pointing out that with the demonstrated sparse array in Figures \ref{DT-SS}-\ref{DLT-Sparse}, the PS algorithm can not be used to estimate the two-dimensional DOA of the sources.
In order to compare the performance of the proposed algorithm with the polarization-smoothing algorithm, we adopt an L-shaped array \cite{XuIET-RSN0607} for the estimation with the polarization-smoothing algorithm. In this L-shaped array, $7$ collocated electromagnetic vector-sensors are displaced both on $x$-axis and $y$-axis at $\lambda/2$ inter-sensor spacings. The ESPRIT \cite{RoyASSPT0789} algorithm is incorporated with the PS algorithm to estimate the DOAs of the sources, named as PS-ESPRIT in Figure \ref{compPS}.
Three correlated sources ${\bf s}_2 =-{\bf s}_1, {\bf s}_3 =(1+j){\bf s}_1$, and $(\theta_{1,1},\theta_{2,1},\theta_{3,1},\theta_{4,1})=(15^{\circ},20^{\circ},45^{\circ},90^{\circ})$,
$(\theta_{1,2},\theta_{2,2},\theta_{3,2},\theta_{4,2})=(43^{\circ},53^{\circ},45^{\circ},-90^{\circ})$.
$(\theta_{1,3},\theta_{2,3},\theta_{3,3},\theta_{4,3})=(57^{\circ},81^{\circ},45^{\circ},90^{\circ})$, are used in the simulation.
Figure \ref{compPS} shows the simulation results of the PS-ESPRIT algorithm with the L-shaped array and the results of the proposed algorithm at the
demonstrated bi-sparse array in Figures \ref{DT-SS}-\ref{LT-SS} with various inter-sensor spacing.
It can be seen from part (a) of Figure \ref{compPS} that the performance of the PS-ESPRIT algorithm is better than the proposed algorithm when $d_x=d_y=0.5\lambda$.
However, when $d_x=d_y\ge 2\lambda$, as in parts (b)-(c) of Figure \ref{compPS}, the efficacy of the proposed algorithm is improved significantly and the RMSEs of the proposed algorithm are much lower than their counterparts of the PS-ESPRIT algorithm, but with fewer antennas.
(The L-shaped array needs $13$ six-component electromagnetic vector-sensors (39 dipoles and 39 loops) while the
proposed bi-sparse array only needs $28$ dipoles or $28$ loops.)
Therefore, the proposed scheme can reduce the hardware cost substantially in practical applications.

\subsection{Cross-Correlation Effect of the Sources}

\begin{figure}
\centering
\includegraphics[height=6cm,width=8.0cm]{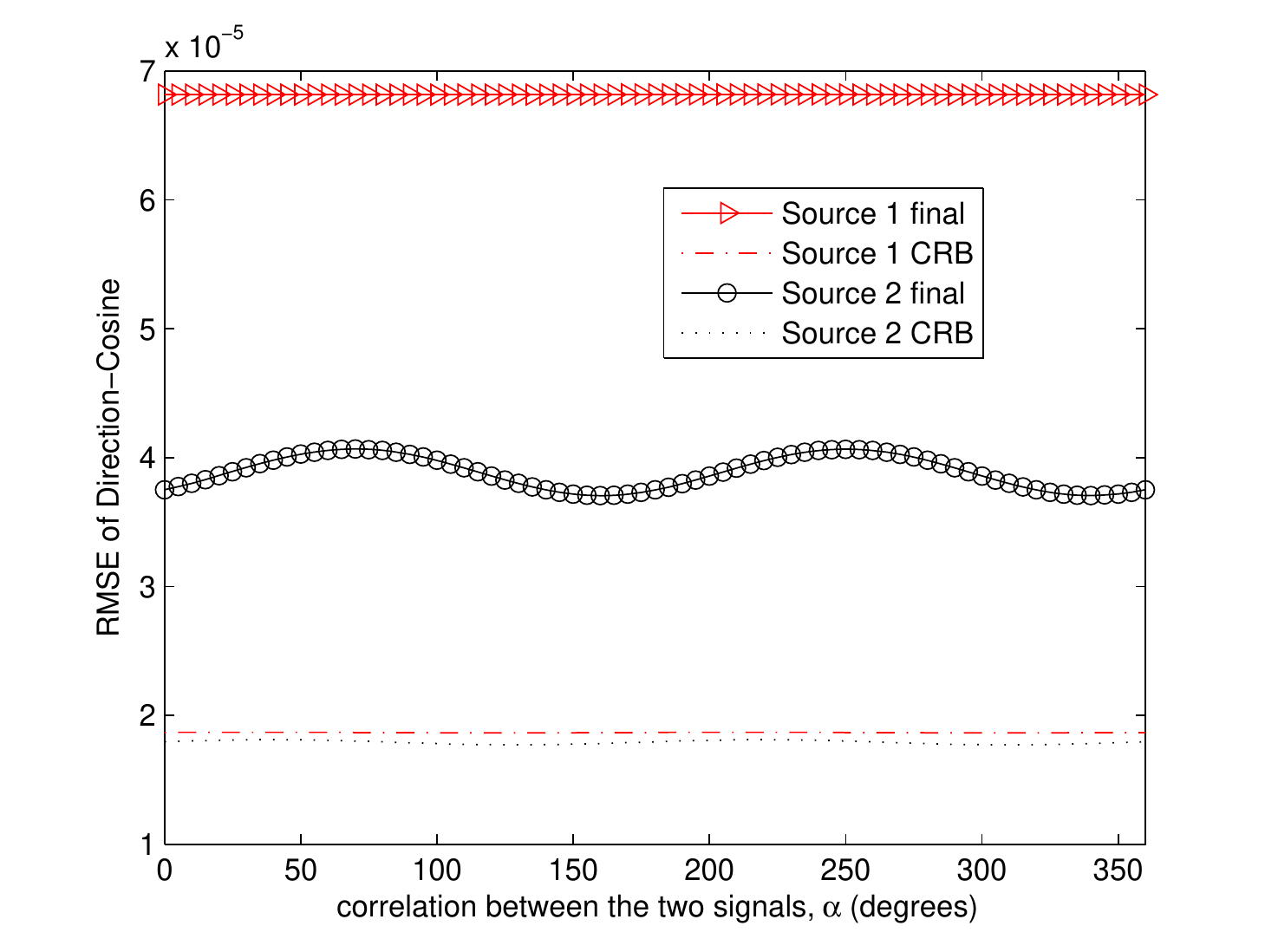}
\caption{The RMSEs of direction-cosines versus the correlation between the two sources with $d_x=d_y=8\lambda$ at SNR$=30$dB, using the array geometry in Figure \ref{DLT-Sparse}.}
\label{corr}
\end{figure}

In order to investigate how the algorithm is affected by the cross-correlation of the sources, two cross-correlated sources with the same amplitude are used in the following simulation, ${\bf s}_2 =e^{j \alpha}{\bf s}_1$, where $\alpha$ denotes the angular difference between the two sources.
Figure \ref{corr} plots the RMSEs of the two sources versus $\alpha$.
The array geometry in Figure \ref{DLT-Sparse} is used as an example in this simulation.
It can be seen that the RMSE of ${\bf s}_1$ remains the same with various $\alpha$, while the RMSE of ${\bf s}_2$ reaches its maximum value at
$\alpha=\{70^{\circ}, 250^{\circ}\}$, and reaches its minimum value at
$\alpha=\{160^{\circ}, 340^{\circ}\}$.
Though the RMSE of ${\bf s}_2$ changes sinusoidally with $\alpha$, it remains in a very small region.
It can be concluded that the proposed algorithm depends slimly on the cross-correlation of different sources.
\footnote{The decorrelation is a priori not uniform whatever the DOA of the sources. Since we use the conventional spatial-smoothing algorithm, this decorrelation should be the same as the generalized scaler sensor arrays  investigated in \cite{ShanASSPT0885,PillaiASSPT0189}.}

\subsection{Comparison of Different Array Geometries}

\begin{figure*}
\begin{minipage}{3.3in}
  \centering
  \centerline{\epsfig{figure=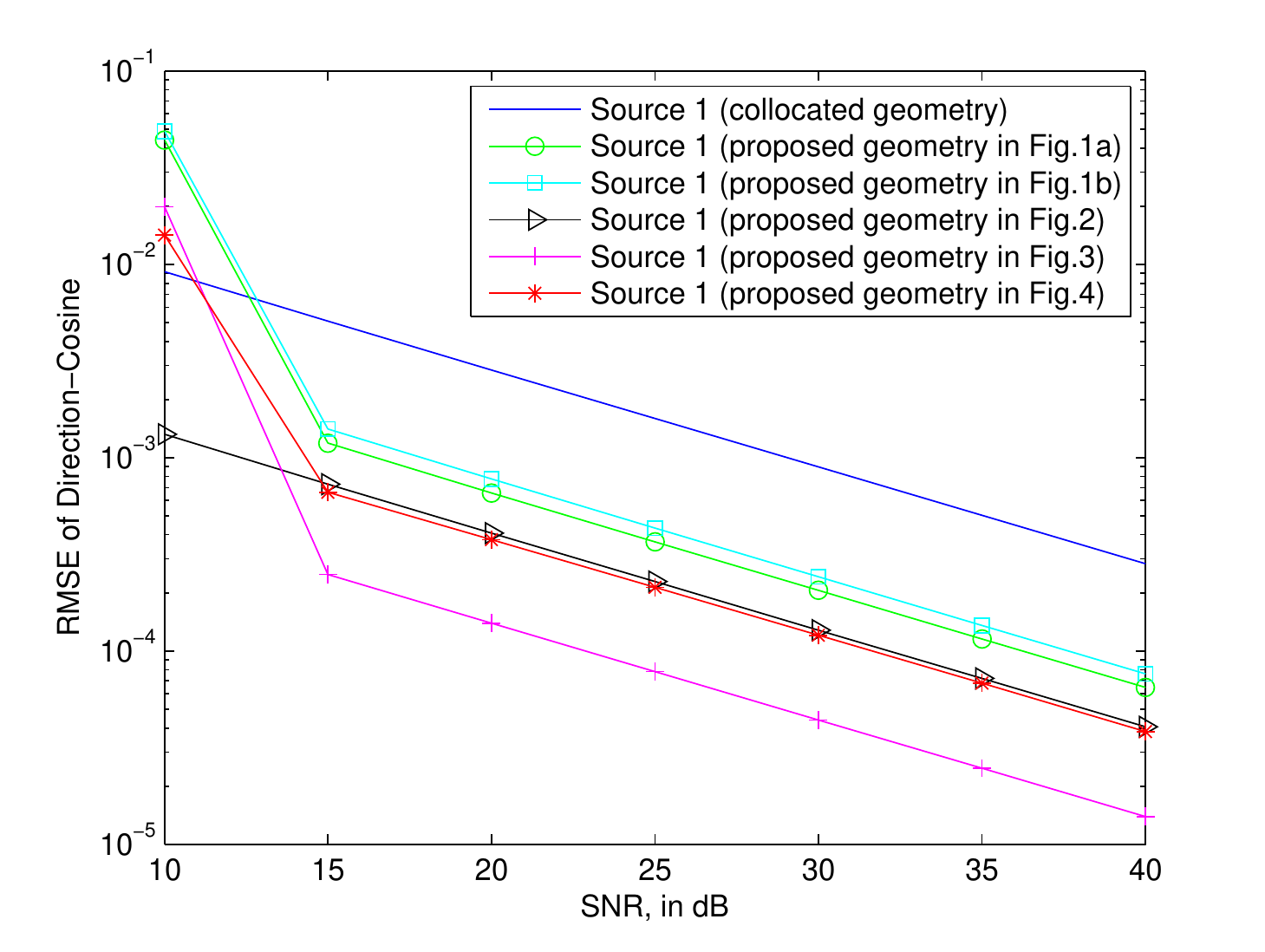,width=9.0cm,height=6cm}}
  \centerline{\footnotesize{(a) source 1}}
\end{minipage}
\hspace{0.1in}
\begin{minipage}{3.3in}
  \centering
  \centerline{\epsfig{figure=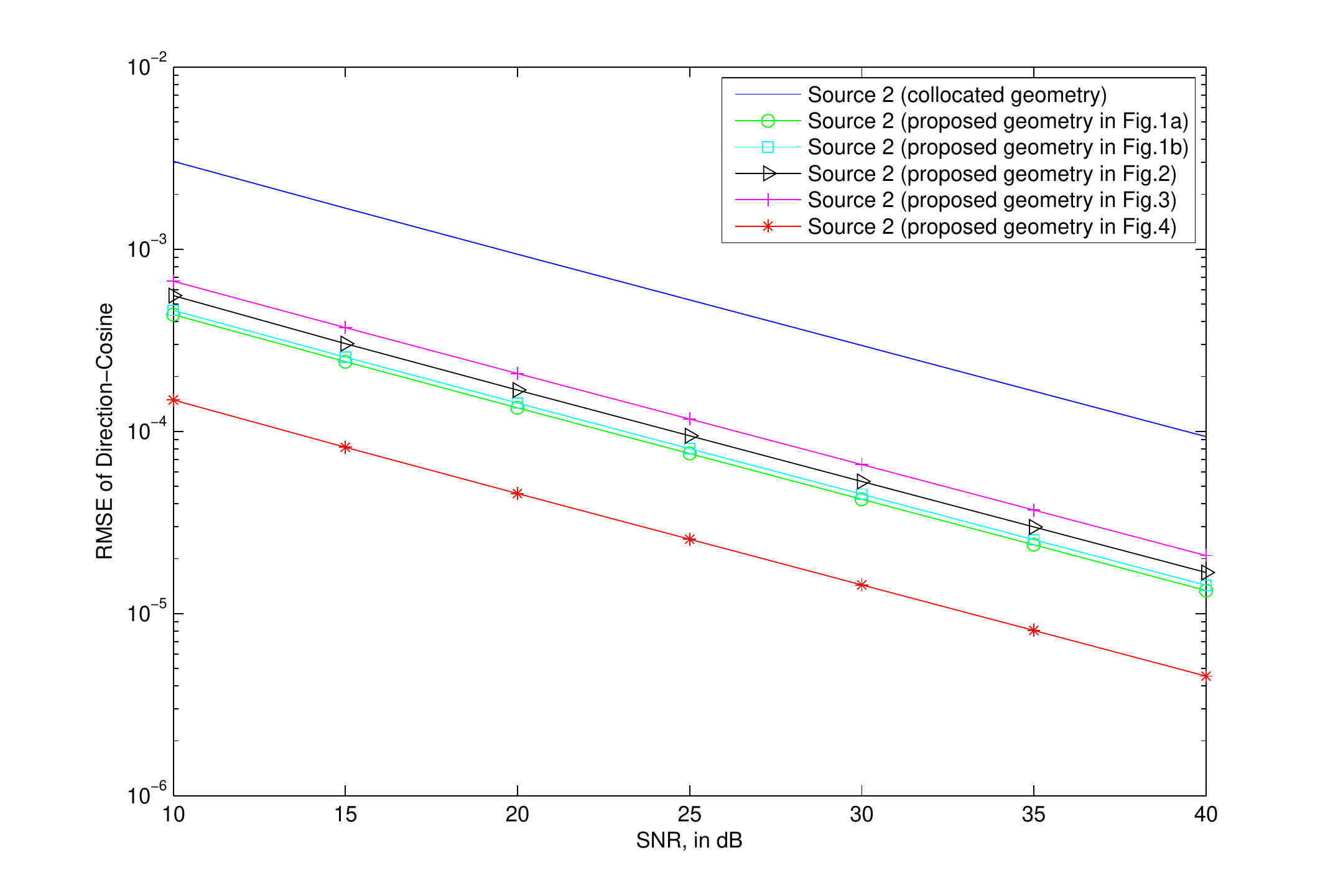,width=9.0cm,height=6cm}}
  \centerline{\footnotesize{(b) source 2}}
\end{minipage}
\caption{The RMSEs of direction-cosines versus SNR for the two sources in a two-source scenario at $d_x=5\lambda$. The collocated geometry denotes that the dipole-triad is collocated with the loop-triad to constitute a collocated electromagnetic vector-sensor.}
\label{Compcol}
\end{figure*}

The ESPRIT-based algorithm is incorporated with the demonstrated sparsely-distributed array-geometry, and unlike the sparse array
investigated in \cite{Zoltowski1SPT0800}, which utilized the collocated electromagnetic vector-sensors,
the novelty in this work is a new bi-sparse array-scheme composed of sparsely-distributed dipoles or loops.
The following simulation will demonstrate how the proposed sparsely-distributed array geometries outperforms the {\em collocated} geometry of the six-component electromagnetic vector-sensor in \cite{LiuJEIT1010,XuIET-RSN0607}.
The proposed algorithm will be adopted to all the six array geometries to estimate the DOAs of the two sources, ${\bf s}_2 =-{\bf s}_1$, and $(\theta_{1,1},\theta_{2,1},\theta_{3,1},\theta_{4,1})=(20^{\circ},15^{\circ},45^{\circ},90^{\circ})$,
$(\theta_{1,2},\theta_{2,2},\theta_{3,2},\theta_{4,2})=(110^{\circ},73^{\circ},45^{\circ},-90^{\circ})$.
The collocated geometry can be obtained from the sparsely-distributed array-geometry in Figure \ref{DLT-Sparse} by setting $d_y=0$.
Figure \ref{Compcol} plots the RMSEs of the two sources in the six different geometries.
In Figure \ref{Compcol}, for the sparsely-distributed array-geometry, we set $d_x=d_y=5\lambda$, and for the collocated geometry, we set $d_x=5\lambda, d_y=0$.
Figure \ref{Compcol} clearly demonstrates that the proposed geometries outperforms the {\em collocated} geometry (particularly when SNR$\ge15$dB) used in \cite{LiuJEIT1010,XuIET-RSN0607}.

Table \ref{Table:summary} summarized the performances of the proposed arrays.
Note that all these arrays are robust to non coherent sources, in which case, the spatial-smoothing algorithm is not required for direction finding and polarization estimation.
The array size depends on the inter-sensor spacings of the antennas. Since the SS-EMVS will have 6 individual dipoles/loops, we argue here that it will have the largest array size.
If the forward/backward spatial-smoothing technique \cite{PillaiASSPT0189} is used, the resolvable coherent source number can be increased.
The direction and polarization
estimation precision for these arrays are similar and the resolution depends on the inter-sensor spacings. If the sources have same polarization states, the arrays can also estimate them correctly, since the proposed algorithm replies on the directions to identify different sources. The polarization is estimated after the arriving angles are determined.

\begin{table}[ht!]
\caption{\small{Summary of the proposed method for the different arrays (Figures \ref{DT-SS}-\ref{DLT-Sparse}). $L$ denotes the number of sub-array unit. For the array size, $1$ denotes the largest and $4$ denotes the smallest. }}
\centering
\begin{tabular}{|c||c|c|c|c|}
\hline Sub-array unit &  Sensor number & $\begin{array}{c} \text{Resolvable coherent}\\ \text{source number}\end{array} $& Array size&$\begin{array}{c} \text{Robust to}\\ \text{non coherent sources}\end{array}$\\
\hline\hline
Dipoles only &  $4L$ & 4 &  2&  Yes\\
\hline
Loops only & $4L$   & 4 & 2 &  Yes\\
\hline
SS-EMVS & $6L$ & 6& 1  & Yes \\
\hline
Pairs & $6L$  & 6 & 3 &  Yes\\
\hline
Triads & $6L$ & 6 & 4  &  Yes\\
\hline\hline
\end{tabular}
\label{Table:summary}
\end{table}

\section{Concluding Remarks}
\label{Sec:con}

Motivated by the sub-array idea, we investigate a spatial-smoothing algorithm to estimate the DOAs and polarizations of coherent sources
based on five bi-sparse arrays, which can be composed of any one of the following five sub-array geometries:
(a) four sparsely spaced  but orthogonally oriented dipoles,
(b) four sparsely spaced  but orthogonally oriented loops,
(c) three spatially spread but orthogonally oriented dipoles and three spatially spread but orthogonally oriented  loops,
(d) three dipole-loop pairs with orthogonal orientations, and
(e) a dipole-triad and a loop-triad.
Both the spacing between two adjacent sub-arrays and the spacing between two adjacent sensors in the sub-array are far larger than a half-wavelength of the incident source.
Hence, these arrays are called ``bi-sparse arrays".
Compared with the recently developed  polarization-smoothing algorithm,
the investigated approach requires no iterative search and can improve the estimation accuracy by the sparse geometries.
Compared with the conventional spatial-smoothing algorithm investigated for the scalar arrays, the proposed approach can offer two-dimensional (closed-form) DOA and polarization estimation using only
one-dimensional spatial-smoothing.
Furthermore, the mutual coupling is reduced efficiently because of this ``bi-sparse" scheme.
The bi-sparse array composed of dipole-triads and loop-triads is recommended to be used to form a large aperture sparse array and the inter-triad spacing can be about $100$ wavelengths.
The bi-sparse arrays composed of only dipoles or only loops are recommended to further reduce the mutual coupling since no antennas are collocated.
For practical applications, the users may need to install these antennas at different locations (e.g, vehicles) to form distributed arrays since they may occupy a large space.
It is worth noting that we assumed the antennas (dipoles/loops) are calibrated well in the proposed arrays. If there are some calibration errors, which will introduce model errors of the array system. This will the future research of the author. Some initial analysis of vector-sensor with model error can be found in  \cite{YuanAEST0412,TamSJ0809}.

The proposed arrays and algorithm have been built to process coherent sources.
In real applications, the sources may be de-correlated, in which case, the proposed arrays will perform well as other polarized antenna arrays (e.g., \cite{WongKT_SPT0111,YuanSPT0312,YuanDLTss}) and this will obviate the requirement of the spatial-smoothing algorithm.
For the case of sources with same polarization, the proposed algorithm can also work well  since the identifiability of different sources is based on their directions of arrival. The proposed algorithm first estimates the arriving angles of the source; then these angles are used to estimate the polarizations of the incident sources associated with the array manifolds.

Looking forward, the identifiability of proposed bi-sparse arrays is an interesting direction. The linear dependence of the steering vectors of different array geometries vary with the polarization states of the incident sources. Though haven't shown in this paper, the research results will be reported elsewhere.

\small
\bibliographystyle{IEEEtran}

\begin{thebibliography}{10}
\providecommand{\url}[1]{#1}
\csname url@samestyle\endcsname
\providecommand{\newblock}{\relax}
\providecommand{\bibinfo}[2]{#2}
\providecommand{\BIBentrySTDinterwordspacing}{\spaceskip=0pt\relax}
\providecommand{\BIBentryALTinterwordstretchfactor}{4}
\providecommand{\BIBentryALTinterwordspacing}{\spaceskip=\fontdimen2\font plus
\BIBentryALTinterwordstretchfactor\fontdimen3\font minus
  \fontdimen4\font\relax}
\providecommand{\BIBforeignlanguage}[2]{{%
\expandafter\ifx\csname l@#1\endcsname\relax
\typeout{** WARNING: IEEEtran.bst: No hyphenation pattern has been}%
\typeout{** loaded for the language `#1'. Using the pattern for}%
\typeout{** the default language instead.}%
\else
\language=\csname l@#1\endcsname
\fi
#2}}
\providecommand{\BIBdecl}{\relax}
\BIBdecl

\bibitem{ShanASSPT0885}
T.-J. Shan, M.~Wax, and T.~Kailath, ``On spatial smoothing for
  direction-of-arrival estimation of coherent signals,'' \emph{IEEE
  Transactions on Acoustics, Speech, and Signal Processing}, vol.~33, no.~8,
  pp. 806--811, August 1985.

\bibitem{PillaiASSPT0189}
S.~U. Pillai and B.~H. Kwon, ``Forward/backward spatial smoothing techniques
  for coherent signal identification,'' \emph{IEEE Transactions on Acoustics,
  Speech, and Signal Processing}, vol.~37, no.~1, pp. 8--15, January 1989.

\bibitem{HuaSPT0992}
Y.~Hua, ``Estimation two-dimensional frequencies by matrix enhancement and
  matrix pencil,'' \emph{IEEE Transactions on Signal Processing}, vol.~40,
  no.~9, pp. 2267--2280, September 1992.

\bibitem{WangAEST0498}
H.~Wang and K.~J.~R. Liu, ``{2-D} spatial smoothing for multipath coherent
  signal separation,'' \emph{IEEE Transactions on Aerospace and Electronic
  Systems}, vol.~34, no.~2, pp. 391--405, April 1998.

\bibitem{WilliamsASSPT0488}
R.~T. Williams, S.~Prasad, A.~K. Mahalanabis, and L.~H. Sibul, ``An improved
  spatial smoothing technique for bearing estimation in a multipath
  environment,'' \emph{IEEE Transactions on Acoustics, Speech, and Signal
  Processing}, vol.~36, no.~4, pp. 425--432, April 1988.

\bibitem{LiSPT1292}
J.~Li, ``Improved angular resolution for spatial smoothing techniques,''
  \emph{IEEE Transactions on Signal Processing}, vol.~40, no.~12, pp.
  3078--3081, December 1992.

\bibitem{LiAPT0393}
------, ``Direction and polarization estimation using arrays with small loops
  and short dipoles,'' \emph{IEEE Transactions on Antennas and Propagation},
  vol.~41, no.~3, pp. 379--387, March 1993.

\bibitem{NehoraiSPT0294}
A.~Nehorai and E.~Paldi, ``Vector-sensor array processing for electromagnetic
  source localization,'' \emph{IEEE Transactions on Signal Processing},
  vol.~42, no.~2, pp. 376--398, February 1994.

\bibitem{WongKT_APT1097}
K.~T. Wong and M.~D. Zoltowski, ``Uni-vector-sensor {ESPRIT} for multi-source
  azimuth, elevation, and polarization estimation,'' \emph{IEEE Transactions on
  Antennas and Propagation}, vol.~45, no.~10, pp. 1467--1474, October 1997.

\bibitem{LuoEURASIP0512}
F.~Luo and X.~Yuan, ``Enhanced ``vector-cross-product" direction-finding using
  a constrained sparse triangular-array,'' \emph{EURASIP Journal on Advances in
  Signal Processing}, vol. 2012:115 doi:10.1186/1687-6180-2012-115, May 2012.

\bibitem{YuanAPT0512}
X.~Yuan, K.~T. Wong, and K.~Agrawal, ``Polarization estimation with a
  dipole-dipole pair, a dipole-loop pair, or a loop-loop pair of various
  orientations,'' \emph{IEEE Transactions on Antennas and Propagation},
  vol.~60, no.~5, pp. 2442--2452, May 2012.

\bibitem{YuanSJ0612}
X.~Yuan, K.~T. Wong, Z.~Xu, and K.~Agrawal, ``Various compositions to form a
  triad of collocated dipoles/loops, for direction finding \& polarization
  estimation,'' \emph{IEEE Sensors Journal}, vol.~12, no.~6, pp. 1763--1771,
  June 2012.

\bibitem{XuIET-MAP0612}
Z.~Xu and X.~Yuan, ``{Cramer-Rao} bounds of angle-of-arrival \& polarisation
  estimation for various triads,'' \emph{IET Microwaves, Antennas \&
  Propagation}, vol.~6, no.~15, pp. 1651--1664, 2012.

\bibitem{YuanAWPL2012}
X.~Yuan, ``Quad compositions of collocated dipoles and loops: for direction
  finding and polarization estimation,'' \emph{IEEE Antennas and Wireless
  Propagation Letters}, vol.~11, pp. 1044--1047, 2012.

\bibitem{WongKT_APT0500}
K.~T. Wong and M.~D. Zoltowski, ``Closed-form direction-finding with
  arbitrarily spaced electromagnetic vector-sensors at unknown locations,''
  \emph{IEEE Transactions on Antennas and Propagation}, vol.~48, no.~5, pp.
  671--681, May 2000.

\bibitem{WongKT_APT0800}
------, ``Self-initiating {MUSIC} direction finding \& polarization estimation
  in spatio-polarizational beamspace,'' \emph{IEEE Transactions on Antennas and
  Propagation}, vol.~48, no.~5, pp. 1235--1245, August 2000.

\bibitem{Zoltowski1SPT0800}
M.~D. Zoltowski and K.~T. Wong, ``{ESPRIT}-based {2D} direction finding with a
  sparse array of electromagnetic vector-sensors,'' \emph{IEEE Transactions on
  Signal Processing}, vol.~48, no.~8, pp. 2195--2204, August 2000.

\bibitem{Zoltowski2SPT0800}
------, ``Closed-form eigenstructure-based direction finding using arbitrary
  but identical subarrays on a sparse uniform rectangular array grid,''
  \emph{IEEE Transactions on Signal Processing}, vol.~48, no.~8, pp.
  2205--2210, August 2000.

\bibitem{YuanSPT0312}
X.~Yuan, ``Estimating the {DOA} and the polarization of a polynomial-phase
  signal using a single polarized vector-sensor,'' \emph{IEEE Transactions on
  Signal Processing}, vol.~60, no.~3, pp. 1270--1282, March 2012.

\bibitem{DeschampsIRE51}
G.~A. Deschamps, ``Geometrical representation of the polarization of a plane
  electromagnetic wave,'' \emph{Proceedings of the IRE}, vol.~39.

\bibitem{LiAPT0592}
J.~Li and J.~R.~T.~Compton, ``Two-dimensional angle and polarization estimation
  using the esprit algorithm,'' \emph{IEEE Transactions on Antennas and
  Propagation}, vol.~40, no.~5, pp. 550--555, May 1992.

\bibitem{WongKT_AEST0401}
K.~T. Wong, ``Direction finding / polarization estimation --- dipole and/or
  loop triad(s),'' \emph{IEEE Transactions on Aerospace and Electronic
  Systems}, vol.~37, no.~2, pp. 679--684, April 2001.

\bibitem{AuYeungAEST0109}
C.~K.~A. Yeung and K.~T. Wong, ``{CRB}: Sinusoid-sources' estimation using
  collocated dipoles/loops,'' \emph{IEEE Transactions on Aerospace and
  Electronic Systems}, vol.~45, no.~1, pp. 94--109, January 2009.

\bibitem{HochwaldSPT0895}
B.~Hochwald and A.~Nehorai, ``Polarimetric modeling and parameter estimation
  with applications to remote sensing,'' \emph{IEEE Transactions on Signal
  Processing}, vol.~43, no.~8, pp. 1923--1935, August 1995.

\bibitem{HoSPT1097}
K.-C. Ho, K.-C. Tan, and B.~T.~G. Tan, ``Efficient method for estimating
  directions-of-arrival of partially polarized signals with electromagnetic
  vector sensors,'' \emph{IEEE Transactions on Signal Processing}, vol.~45,
  no.~10, pp. 2485--2498, October 1997.

\bibitem{HoSPT1099}
K.-C. Ho, K.-C. Tan, and A.~Nehorai, ``Estimating directions of arrival of
  completely and incompletely polarized signals with electromagnetic vector
  sensors,'' \emph{IEEE Transactions on Signal Processing}, vol.~47, no.~10,
  pp. 2845--2852, October 1999.

\bibitem{WongKT_AEST0101}
K.~T. Wong, ``Blind beamforming/geolocation for wideband-{FFH}s with unknown
  hop-sequences,'' \emph{IEEE Transactions on Aerospace and Electronic
  Systems}, vol.~37, no.~1, pp. 65--76, January 2001.

\bibitem{KoAEST0702}
C.~C. Ko, J.~Zhang, and A.~Nehorai, ``Separation and tracking of multiple
  broadband sources with one electromagnetic vector sensor,'' \emph{IEEE
  Transactions on Aerospace and Electronic Systems}, vol.~38, no.~3, pp.
  1109--1116, July 2002.

\bibitem{MironSPT0406}
S.~Miron, N.~L. Bihan, and J.~I. Mars, ``{Quaternion-MUSIC} for vector-sensor
  array processing,'' \emph{IEEE Transactions on Signal Processing}, vol.~54,
  no.~4, pp. 1218--1229, April 2006.

\bibitem{HurtadoAEST0407}
M.~Hurtado and A.~Nehorai, ``Performance analysis of passive low-grazing-angle
  source localization in maritime environments using vector sensors,''
  \emph{IEEE Transactions on Aerospace and Electronic Systems}, vol.~43, no.~2,
  pp. 780--789, April 2007.

\bibitem{LeBihanSPT0907}
N.~L. Bihan, S.~Miron, and J.~Mars, ``{MUSIC} algorithm for vector-sensors
  array using biquaternions,'' \emph{IEEE Transactions on Signal Processing},
  vol.~55, no.~9, pp. 4523--4533, September 2007.

\bibitem{XuAEST1008}
Y.~Xu, Z.~Liu, K.~T. Wong, and J.~Cao, ``Virtual-manifold ambiguity in
  hos-based direction-finding with electromagnetic vector-sensors,'' \emph{IEEE
  Transactions Aerospace and Electronic Systems}, vol.~44, no.~4, pp.
  1291--1308, October 2008.

\bibitem{HurtadoSPM0109}
M.~Hurtado, J.-J. Xiao, and A.~Nehorai, ``Target estimation, detection, and
  tracking,'' \emph{IEEE Signal Processing Magazine}, vol.~42, no.~1, pp.
  42--52, January 2009.

\bibitem{SunCISP09}
L.~Sun, G.~Ou, and Y.~Lu, ``Vector sensor cross-product for direction of
  arrival estimation,'' \emph{International Congress on Image and Signal
  Processing}, pp. 1--5, 2009.

\bibitem{XiaoSPT0209}
J.-J. Xiao and A.~Nehorai, ``Optimal polarized beampattern synthesis using a
  vector antenna array,'' \emph{IEEE Transactions on Signal Processing},
  vol.~57, no.~2, pp. 576--587, February 2009.

\bibitem{GongSP0509}
X.~Gong, Z.-W. Liu, Y.-G. Xu, and M.~I. Ahmad, ``Direction-of-arrival
  estimation via twofold mode-projection,'' \emph{Signal Processing}, vol.~89,
  no.~5, pp. 831--842, May 2009.

\bibitem{SunAPSEC10}
L.~Sun, B.~Li, Y.~Lu, and G.~Ou, ``Distributed vector sensor cross product
  added with {MUSIC} for direction of arrival estimation,'' \emph{Asia-Pacific
  International Symposium on Electromagnetic Compatibility}, pp. 1354--1357,
  2010.

\bibitem{SunCISP10}
L.~Sun, B.~Li, Y.~Wang, G.~Ou, and Y.~Lu, ``Distributed reduced vector sensor
  for direction of arrival and polarization state estimation,''
  \emph{International Congress on Image and Signal Processing}, pp. 3935--3939,
  2010.

\bibitem{GongSP0411}
X.~Gong, Z.-W. Liu, and Y.-G. Xu, ``Direction finding via biquaternion matrix
  diagonalization with vector-sensors,'' \emph{Signal Processing}, vol.~91,
  no.~4, pp. 821--831, April 2011.

\bibitem{GongIET-SP0411}
------, ``Regularised parallel factor analysis for the estimation of
  direction-of-arrival and polarisation with a single electromagnetic
  vector-sensor,'' \emph{IET Signal Processing}, vol.~5, no.~4, pp. 390--396,
  2011.

\bibitem{GuoSPT0711}
X.~Guo, S.~Miron, D.~Brie, S.~Zhu, and X.~Liao, ``A {CANDECOMP/PARAFAC}
  perspective on uniqueness of {DOA} estimation using a vector sensor array,''
  \emph{IEEE Transactions on Signal Processing}, vol.~59, no.~7, pp.
  3475--3481, July 2011.

\bibitem{YuanICASSP2012}
X.~Yuan, ``Polynomial-phase signal source-tracking using an electromagnetic
  vector-sensor,'' \emph{International Conference on Acoustics, Speech, and
  Signal Processing (ICASSP)}, pp. 2577--2580, March 2012.

\bibitem{RahamimSPT1104}
D.~Rahamim, J.~Tabrikian, and R.~Shavit, ``Source localization using vector
  sensor array in a multipath environment,'' \emph{IEEE Transactions on Signal
  Processing}, vol.~52, no.~11, pp. 3096--3103, November 2004.

\bibitem{XuIET-RSN0607}
Y.~Xu and Z.~Liu, ``Polarimetric angular smoothing algorithm for an
  electromagnetic vector-sensor array,'' \emph{IET Radar, Sonar \& Navigation},
  vol.~1, no.~3, pp. 230--240, June 2007.

\bibitem{HeAEST0110}
J.~He, S.~Jiang, J.~Wang, and Z.~Liu, ``Polarization difference smoothing for
  direction finding of coherent signals,'' \emph{IEEE Transactions on Aerospace
  and Electronic Systems}, vol.~46, no.~1, pp. 469--480, January 2010.

\bibitem{GongAEST0711}
X.-F. Gong, Z.-W. Liu, and Y.-G. Xu, ``Coherent source localization: Bicomplex
  polarimetric smoothing with electromagnetic vector-sensors,'' \emph{IEEE
  Transactions on Aerospace and Electronic Systems}, vol.~47, no.~3, pp.
  2268--2285, July 2011.

\bibitem{XuJCIC0504}
Y.~Xu and Z.~Liu, ``Simultaneous estimation of {2-D} doa and polarization of
  multiple coherent sources using an electromagnetic vector sensor array,''
  \emph{Journal of China Institute of Communications}, vol.~2, no.~5, pp.
  28--38, May 2004.

\bibitem{LiuJEIT1010}
Z.~Liu, J.~He, and Z.~Liu, ``Extended aperture-based {DOA} estimation of
  coherent sources using a electromagnetic vector-sensor array,'' \emph{Journal
  of Electronics \& Information Technology}, vol.~32, no.~10, pp. 2511--2515,
  October 2010.

\bibitem{LiuEURASIP2011}
------, ``Computationally efficient {DOA} and polarization estimation of
  coherent sources with linear electromagnetic vector-sensor array,''
  \emph{EURASIP Journal on Advances in Signal Processing}, vol. 2011, no.
  Article ID 490289, pp. 1--10, 2011.

\bibitem{WongKT_SPT0111}
K.~T. Wong and X.~Yuan, ```{V}ector cross-product direction-finding' with an
  electromagnetic vector-sensor of six orthogonally oriented but spatially
  non-collocating dipoles / loops,'' \emph{IEEE Transactions on Signal
  Processing}, vol.~59, no.~1, pp. 160--171, January 2011.

\bibitem{YuanDLTss}
X.~Yuan, ``Joint {DOA} and polarization estimation with sparsely distributed
  and spatially non-collocating dipole/loop triads,'' \emph{arXiv:1308.0072},
  2013.

\bibitem{TanSP93}
K.-C. Tan, G.-L. Oh, and M.~H. Er, ``A study of the uniqueness of steering
  vectors in array processing,'' \emph{Signal Processing}, vol.~34, no.~3, pp.
  245--256, 1993.

\bibitem{HoSP95}
K.-C. Ho, K.-C. Tan, and W.~Ser, ``An investigation on number of signals whose
  directions-of-arrival are uniquely determinable with an electromagnetic
  vector sensor,'' \emph{Signal Processing}, vol.~47, no.~1, pp. 41--54, 1995.

\bibitem{HochwaldTSP0196}
B.~Hochwald and A.~Nehorai, ``Identifiability in array processingmodels with
  vector-sensor applications,'' \emph{IEEE Transactions on Signal Processing},
  vol.~44, no.~1, pp. 83--95, January 1996.

\bibitem{TanTSP1296}
K.-C. Tan, K.-C. Ho, and A.~Nehorai, ``Linear independence of steering vectors
  of an electromagnetic vector sensor,'' \emph{IEEE Transactions on Signal
  Processing}, vol.~44, no.~12, pp. 3099--3107, December 1996.

\bibitem{HoTAP1198}
a.~K.-C.~T. K.-C.~Ho and B.~T.~G. Tan, ``Linear dependence of steering vectors
  associated with tripole arrays,'' \emph{IEEE Transactions on Antennas and
  Propagation}, vol.~46, no.~11, pp. 1705--1711, November 1998.

\bibitem{XuTAES1008}
Y.~Xu, Z.~Liu, K.~T. Wong, and J.~Cao, ``Virtual-manifold ambiguity in
  hos-based direction-finding with electromagnetic vector-sensors,'' \emph{IEEE
  Transactions on Aerospace and Electronic Systems}, vol.~44, no.~4, pp.
  1291--1308, October 2008.

\bibitem{YuanSSP2011}
X.~Yuan, ``{Cramer-Rao Bound} of the direction-of-arrival estimation using a
  spatially spread electromagnetic vector-sensor,'' \emph{IEEE Statistical
  Signal Processing Workshop}, pp. 1--4, June 2011.

\bibitem{YuanAWPL2013}
------, ``Spatially spread dipole/loop quads/quints: For direction finding and
  polarization estimation,'' \emph{IEEE Antennas and Wireless Propagation
  Letters}, vol.~12, pp. 1081--1084, 2013.

\bibitem{HeIET-RSN0509}
J.~He and Z.~Liu, ``Computationally efficient two-dimensional
  direction-of-arrival estimation of electromagnetic sources using the
  propagator method,'' \emph{IET Radar, Sonar and Navigation}, vol.~3, no.~5,
  pp. 437--448, 2009.

\bibitem{MarcosSP0395}
S.~Marcos, A.~Marsal, and M.~Benidir, ``The propagator method for source
  bearing estimation,'' \emph{Signal Processing}, vol.~42, no.~2, pp. 121--138,
  March 1995.

\bibitem{RoyASSPT0789}
R.~Roy and T.~Kailath, ``{ESPRIT} - estimation of signal parameters via
  rotational invariance techniques,'' \emph{IEEE Transactions on Acoustics,
  Speech, and Signal Processing}, vol.~37, no.~7, pp. 984--995, July 1989.

\bibitem{HeDSP0509}
J.~He and Z.~Liu, ``Computationally efficient {2D} direction finding and
  polarization estimation with arbitrarily spaced electromagnetic vector
  sensors at unknown locations using the propagator method,'' \emph{Digital
  Signal Processing}, vol.~19, no.~5, pp. 491--503, May 2009.

\bibitem{YuanAEST0712}
X.~Yuan, ``Coherent source direction-finding using a sparsely-distributed
  acoustic vector-sensor array,'' \emph{IEEE Transactions on Aerospace and
  Electronic Systems}, vol.~48, no.~3, pp. 2710--2715, July 2012.

\bibitem{SeeSPA03}
C.-M. See and A.~Nehorai, ``Source localization with distributed
  electromagnetic component sensor array processing,'' \emph{International
  Symposium on Signal Processing \& Its Applications}, vol.~1, pp. 177--180,
  2003.

\bibitem{SeeSSP03}
------, ``Source localization with partially calibrated distributed
  electromagnetic component sensor array,'' \emph{IEEE Workshop on Statistical
  Signal Processing}, pp. 441--444, 2003.

\bibitem{VanTrees02}
H.~L.~V. Trees, \emph{Detection, Estimation, and Modulation Theory, Part IV:
  Optimum Array Processing}.\hskip 1em plus 0.5em minus 0.4em\relax New York,
  U.S.A.: Wiley, 2002.

\bibitem{YuanIET-RSN0712}
X.~Yuan, ``{Cramer-Rao} bounds of direction-of-arrival and distance estimation
  of a near-field incident source for an acoustic vector-sensor: {Gaussian}
  source and polynomial-phase source,'' \emph{IET Radar, Sonar and Navigation},
  vol.~6, no.~7, pp. 638--648, July 2012.

\bibitem{YuanAEST0412}
------, ``Direction-finding with a misoriented acoustic vector sensor,''
  \emph{IEEE Transactions on Aerospace and Electronic Systems}, vol.~48, no.~2,
  pp. 1809--1815, April 2012.

\bibitem{TamSJ0809}
P.~K. Tam and K.~T. Wong, ``Cramer-rao bounds for direction finding by an
  acoustic vector-sensor under non-ideal gain-phase responses, non-collocation,
  or non-orthogonal orientation,'' \emph{IEEE Sensors Journal}, vol.~9, no.~8,
  pp. 969--982, August 2009.

\end{thebibliography}

\clearpage
\newpage

\end{document}